%% file: mast.tex
\PassOptionsToPackage{table}{xcolor}
\documentclass[twoside,11pt,final]{entics} 
\usepackage{enticsmacro}
\usepackage{graphicx}
\usepackage[all]{xy}
\usepackage{url}
\def\conf{MFPS 2025}
\volume{5}			
\def\volu{5}			
\def\papno{8}			
\def\mydoi{16879}
\def\lastname{Fiore, Kammar, Moser, and Staton} 
\def\runtitle{Modular abstract syntax trees (MAST)} 
\def\copynames{M.P.~Fiore, O.~Kammar, G.~Moser, and S.~Staton}

\newcommand\authoremail[2]{
  \thanks[#1]{Email: \href{mailto:#2} {\texttt{\normalshape
        #2}}}}

\input{defs}

\newenvironment{noop}{}{}

\newboolean{finalversion}
\setboolean{finalversion}{true}
\ifthenelse{\boolean{finalversion}}%
 {%
}{%
}

\begin{document}

\begin{frontmatter}
  \title{Modular Abstract Syntax Trees (MAST):\\
    Substitution Tensors with Second-class Sorts}
  \author{Marcelo P. Fiore\thanksref{d}\thanksref{marceloemail}}
  \author{Ohad Kammar\thanksref{a}\thanksref{ohademail}}  
  \author{Georg Moser\thanksref{b}\thanksref{georgemail}} 
  \author{Sam Staton\thanksref{c}\thanksref{samemail}} 
  \address[d]{Department of Computer Science and Technology\\
    University of Cambridge\\ England}
  \authoremail{marceloemail}{marcelo.fiore@cl.cam.ac.uk}%
  \address[a]{Laboratory for Foundations of Computer Science\\
     School of Informatics\\
     University of Edinburgh\\
     Scotland}
   \authoremail{ohademail}{ohad.kammar@ed.ac.uk} %
   \address[b]{Theoretical Computer Science\\Department of Computer Science\\
    University of Innsbruck\\Austria}
  \authoremail{georgemail}{georg.moser@uibk.ac.at}
  \address[c]{Department of Computer Science\\University of Oxford\\England}
  \authoremail{samemail}{sam.staton@cs.ox.ac.uk}
\begin{abstract}
\input{abstract}
\end{abstract}
\end{frontmatter}

\input{introduction}
\input{tutorial}

\input{case-studies}
\input{technical-development}

\input{related-work}
\input{conclusion}
\input{acks}
\bibliographystyle{./entics}
\bibliography{ref}
\begin{techreport}
\appendix
\input{case-studies-appendix}
\input{bicat-app}
\input{monoidal-cat}
\input{representation-appendix}
\end{techreport}
\end{document}

%% file: defs.tex
\usepackage{wrapfig}
\usepackage{suffix}
\usepackage{amsmath}
\usepackage{mathtools}
\usepackage{ebproof}
\usepackage{extarrows}
\usepackage{okexplain}
\usepackage{scalerel}
\usepackage{multicol}

\usepackage{todonotes}
\usepackage[notext,upint]{stix}

\usepackage{quiver}
\usepackage{tikz,tikz-cd}
\usetikzlibrary{decorations.pathmorphing}
\newcommand{\citepos}[1]{\citeauthor{#1}'s~\citeyearpar{#1}}
\WithSuffix\newcommand\citepos*[2]{\citeauthor{#1}'s #2~\citeyearpar{#1}}
\WithSuffix\newcommand\citepos-[2]{\citeauthor{#1}#2~\citeyearpar{#1}}

\usepackage{contour}
\usepackage[normalem]{ulem}

\newcommand\diagram[1]{\begin{center}
    \diagram*{#1}
\end{center}}
\WithSuffix\newcommand\diagram*[1]{\begin{tabular}[c]{@{}c@{}}\includegraphics{#1.mps}\end{tabular}}

\newcommand\exlabel[1]{\label{ex:#1}}
\newcommand\exref[1]{Ex.~\ref{ex:#1}}
\WithSuffix\newcommand\exref*[1]{\ref{ex:#1}}
\newcommand\exrefs{Exs.~}
\newcommand\applabel[1]{\label{app:#1}}
\newcommand\appref[2][??]{%
\ifthenelse{\boolean{finalversion}}%
  {Appendix~#1}%
  {Appendix~\ref{app:#2}%
}%
}
\WithSuffix\newcommand\appref*[2][??]{%
\ifthenelse{\boolean{finalversion}}%
  {#1}%
  {\ref{app:#2}}%
}
\newcommand\thmlabel[1]{\label{thm:#1}}
\newcommand\thmref[1]{Thm~\ref{thm:#1}}
\newcommand\lemmalabel[1]{\label{lemma:#1}}
\newcommand\lemmaref[1]{Lemma~\ref{lemma:#1}}
\newcommand\applemmaref[2][??]{%
\ifthenelse{\boolean{finalversion}}%
  {Lemma~#1}%
  {Lemma~\ref{lemma:#1}}%
}
\newcommand\appeqref[2][??]{%
\ifthenelse{\boolean{finalversion}}%
  {(#1)}%
  {\eqref{#2}}%
}
\WithSuffix\newcommand\lemmaref*[1]{\ref{lemma:#1}}
\newcommand\deflabel[1]{\label{def:#1}}
\newcommand\defref[1]{Def~\ref{def:#1}}
\newtheorem{lem*}[thm]{Lemma}
\newcommand\setproposition{}
\newcounter{ignorecounter}
\newcommand\ignore[1]{}

\newtheorem{propositionExplicit}[ignorecounter]%
  {Proposition~\setproposition{}\ignore}
\newenvironment{proposition@}[1]{%
  \renewcommand\setproposition{#1}%
  \begin{propositionExplicit}%
}{%
  \end{propositionExplicit}%
}
\newtheorem{lemmaExplicit}[ignorecounter]%
  {Lemma~\setproposition{}\ignore}
\newenvironment{lemma@}[1]{%
  \renewcommand\setproposition{#1}%
  \begin{lemmaExplicit}%
}{%
  \end{lemmaExplicit}%
}
\newtheorem{theoremExplicit}[ignorecounter]%
  {Theorem~\setproposition{}\ignore}
\newenvironment{theorem@}[1]{%
  \renewcommand\setproposition{#1}%
  \begin{theoremExplicit}%
}{%
  \end{theoremExplicit}%
}
\newcommand\proplabel[1]{\label{prop:#1}}
\newcommand\propref[1]{Prop.~\ref{prop:#1}}
\WithSuffix\newcommand\propref*[1]{\ref{prop:#1}}
\WithSuffix\newcommand\thmref*[1]{\ref{thm:#1}}
\newcommand\subseclabel[1]{\label{subsec:#1}}
\newcommand\subsecref[1]{\S\ref{subsec:#1}}
\newcommand\appsubsecref[2][??]{%
\ifthenelse{\boolean{finalversion}}%
  {\S#1}%
  {\S\ref{subsec:#2}
  }%
}
\newcommand\seclabel[1]{\label{sec:#1}}
\newcommand\secref[1]{Sec.~\ref{sec:#1}}
\WithSuffix\newcommand\secref*[1]{\ref{sec:#1}}
\newcommand\figlabel[1]{\label{fig:#1}}
\newcommand\figref[1]{Fig~\ref{fig:#1}}
\newcommand\Figref[1]{Figure~\ref{fig:#1}}
\WithSuffix\newcommand\figref*[1]{\ref{fig:#1}}

\makeatletter
\newcommand\newmvar[3][]{%
  \newcommand#2[1][1]{#1{\ifcase ##1\or #3\or\else@ctrerr\fi}}%
}

\newenvironment{longexample}{\begin{example}\upshape}%
{\end{example}}

\makeatother

\newcommand\pair[2]{\seq{#1,#2}}
\WithSuffix\newcommand\pair-[2]{\seq-{#1,#2}}
\newcommand\triple[3]{\seq{#1,#2,#3}}
\newcommand\coseq[2][]{\bracks{#2}_{#1}}
\newcommand\seq[2][]{\parent{#2}_{#1}}
\WithSuffix\newcommand\seq-[2][]{(#2)_{#1}}
\newcommand\parent[1]{\left(#1\right)}
\newcommand\bracks[1]{\left[#1\right]}
\newcommand\gdefinedby{\Coloneq}
\newcommand\gor{\mathrel{\vert}}
\newcommand\definedby{\coloneq}
\newcommand\defines{\mathrel{=:}}
\newcommand\set[1]{\left\{#1\right\}}
\newcommand\suchthat{\middle\vert}
\WithSuffix\newcommand\suchthat*{\middle\vert\normalcolor}

\newcommand\isomorphic\cong
\newcommand\xfrom\xleftarrow
\newcommand\monic\rightarrowtail
\newcommand\xto\xrightarrow
\makeatletter
\def\xepifill@{%
  \arrowfill@\relbar\relbar\twoheadrightarrow}
\providecommand{\xepi}[2][]{%
	\ext@arrow 0099\xepifill@{#1}{#2}}
\def\xmonicfill@{%
  \arrowfill@\righttail\relbar\rightarrow}
\providecommand{\xmonic}[2][]{%
	\ext@arrow 0099\xmonicfill@{#1}{#2}}
\def\xtotofill@{%
  \arrowfill@\Relbar\Relbar\rightrightarrows}
\providecommand{\xtoto}[2][]{%
	\ext@arrow 0099\xtotofill@{#1}{#2}}
\makeatother

\newcommand\unitVal{\seq{\mathvisiblespace}}
\WithSuffix\newcommand\++{\mathbin{\mathchoice%
  {\plus\mspace{-9mu}\plus}%
  {\plus\mspace{-9mu}\plus}%
  {\plus\mspace{-9mu}\plus}%
  {\plus\mspace{-9mu}\plus}%
  }%
}
\newcommand\plus{+}

\let\subset\subseteq
\newcommand\finsubset{\subset_{\mathrm{fin}}}
\newcommand\leftadjointto\dashv
\newcommand\tensor\otimes
\newcommand\inv[1]{{#1}^{-1}}
\WithSuffix\newcommand\inv-[1]{#1^{-1}}
\newcommand\const[1]{\underline{#1}}
\newcommand\abst[1]{\lambda#1.}
\newcommand\bmul\olessthan

\newcommand\id[1][]{\mathsf{id}_{#1}}
\newcommand\compose\circ
\newcommand\terminal{\mathbb 1}
\newcommand\initial{\mathbb 0}
\newcommand\projection{\boldsymbol\pi}
\newcommand\Category[1]{\mathbf{#1}}
\newcommand\Set{\Category{Set}}
\newcommand\Cat[1]{\mathcal{#1}}
\newcommand\SmallCat[1]{\mathbb{#1}}
\newmvar\SC{\SmallCat A \or \SmallCat B \or \SmallCat C}
\newcommand\Prof{\Category{Prof}}
\newcommand\rmul{\otimes}
\newcommand\zmul{\oslash}
\newcommand\fix{{\boldsymbol\mu}}

\newmvar\mF{F\or G\or K}
\newmvar\nt{\alpha\or\beta\or\gamma}
\newcommand\PSh{\mathop{\Category{PSh}}\nolimits}
\WithSuffix\newcommand\PSh+[1]{\PSh{\parent{#1}}}
\newcommand\opposite[1]{{#1}^{\mathsf{op}}}
\WithSuffix\newcommand\opposite-[1]{#1^{\mathsf{op}}}
\newcommand\yoneda[1][]{\mathbf{y}_{#1}}

\newcommand\mast{\textsc{mast}}
\newcommand\Mast{\textsc{Mast}}
\newcommand\lamcalc{$\lambda$-calculus}
\newcommand\CBVFull{Call-by-Value \lamcalc}
\newcommand\CBPVFull{Call-by-Push-Value}

\newcommand\CBV{\textsc{cbv}}
\newcommand\CBVSort[1][]{\textsc{cbv}_{#1}}
\newcommand\CBN{\textsc{cbn}}
\newcommand\CBPV{\textsc{cbpv}}

\newmvar\gsort{\mathbf{R}\or\mathbf{S}\or\mathbf{T}}

\newcommand\Fst[1][]{\mathsf{Fst}_{#1}}
\newcommand\Snd[1][]{\mathsf{Snd}_{#1}}
\WithSuffix\newcommand\Fst-{\mathsf{Fst}}
\WithSuffix\newcommand\Snd-{\mathsf{Snd}}
\newcommand\fst[1][]{\mathop{\mathsf{fst}}\nolimits_{#1}}
\newcommand\snd[1][]{\mathop{\mathsf{snd}}\nolimits_{#1}}
\newcommand\Sort[1][]{\mathsf{Sort}_{#1}}
\WithSuffix\newcommand\Sort*{\Sort[]}

\newcommand\restrict[2]{\left.#1\right\rvert_{#2}}
\newcommand\rest[1]{\restrict{#1}{\fst}}
\newmvar\sort{s\or r\or t}

\newcommand\Fioreetal{Fiore et al.}
\WithSuffix\newcommand\Fioreetal*{Fiore, Plotkin, and Turi}

\newcommand\Param[1]{\textsf{#1}}
\newcommand\Base{\Param{Base}}
\newcommand\SimpleType[1][]{\Param{SimpleType}}
\newcommand\CbvType[1][]{\Param{\CBV{}Type}_{#1}}
\newmvar\base{\beta}
\newmvar\ty{A\or B\or C}
\newcommand\FUN[2]{#1 \mathrel{\syntax{\boldsymbol\rightarrow}} #2}
\newmvar\Context{\Gamma\or \Delta \or \Xi}
\newmvar\M{M \or N \or K \or L}
\newcommand\N{\M[2]}
\newmvar\mV{V \or W}
\newcommand\types\vdash
\newcommand\syntax[1]{{\color{olive}#1}}
\newcommand\syntaxBegin{\global\color{olive}}
\newcommand\syntaxClose{\normalcolor}
\newcommand\sabst[1]{\syntax{\boldsymbol\lambda} #1.}
\newcommand\sapply[2]{#1 \syntax{\boldsymbol @} #2}

\newcommand\Comp{\mathop{\mathsf{comp}}\nolimits}
\newcommand\Val{\mathop{\mathsf{val}}\nolimits}
\newcommand\val{\mathop{\syntax{\mathsf{val}}}\nolimits}
\newmvar\V{V\or W}

\newcommand\newSig[2]{\newcommand{#1}{\mathop{\mathsf{#2}}\nolimits}}
\newSig\ValSig{ValOp}
\newSig\AppSig{AppOp}
\newSig\AbsSig{LamOp}
\newSig\CBVSig{CbvOps}
\newcommand\Variant[1]{\left\lBrace#1\right\rBrace}
\newcommand\Record[1]{\left\lParen#1\right\rParen}
\newcommand\sRecord[1]{%
  \syntaxBegin%
    \left\lParen
  \syntaxClose%
  #1%
  \syntaxBegin%
  \right\rParen%
  \syntaxClose}
\newcommand\sVariant[1]{%
  \syntaxBegin%
  \left\lBrace%
  \syntaxClose%
  #1
  \syntaxBegin
    \right\rBrace%
  \syntaxClose}

\newcommand\RenCat[1]{#1_{\types}}
\WithSuffix\newcommand\RenCat*[1]{\RenCat{\parent{#1}}}
\newmvar\mP{P\or Q\or L\or K} 
\newcommand\mQ{\mP[2]}
\newcommand\Vars[1][]{\mathbb{V}_{#1}}
\usepackage[bbsymbols]{dsserif}
\newcommand\List{\mathop{\mathsf{List}}\nolimits}
\newmvar\ren{\rho\or\sigma\or\zeta}
\newcommand\from\leftarrow
\newcommand\Neut[1][]{{\mathbb V}_{#1}}
\newcommand\mNeut[1][]{{\mathbb I}_{#1}}
\newcommand\kNeut[1][]{{\mathbb J}_{#1}}
\newcommand\Syn[1]{#1\textrm{\textbf{-}}\mathbf{Struct}}
\newcommand\gast{\mathbb s}
\newcommand\SndSyn[2][\snd]{#2\textrm{\textbf{-}}\mathbf{Struct}_{#1}}
\newcommand\Lam[1][]{\bbLambda^{#1}}
\newcommand\AST[1][]{\mathbb{S}^{#1}}
\newcommand\Env[2]{#1^{\mathrm{Env}}_{#2}}
\newcommand\op{\mathrm{op}}
\newcommand\At{\mathbin{\mathsf{@}}}
\newcommand\OnlyAt[1]{\mathop\looparrowright_{#1}}
\newcommand\Shift{\mathbin\triangleright}
\newcommand\var[1][]{\mathop{\mathsf{var}}\nolimits_{#1}}
\newcommand\Sig[1][]{\mathbf{O}}

\newcommand\rEnv[1]{\Env{#1}}
\newcommand\qsim{\sim}
\newmvar\thin{\tau\or\theta}
\newcommand\ellunit{\boldsymbol\ell}
\newcommand\runit{\mathbf{r}}
\newcommand\assoc{\mathbf{a}}

\newcommand\curry{\mathsf{curry}}
\newcommand\uncurry{\mathsf{uncurry}}
\newcommand\eval{\mathsf{eval}}

\newcommand\carrier[1]{\underline{#1}}

\newmvar\Monoid{\mathbf M\or \mathbf N\or \mathbf L}
\newmvar\Action{\mathbf A\or \mathbf B\or \mathbf C}
\newmvar\SubStruct{\mathbf S\or \mathbf Q\or \mathbf P}
\newcommand\monprod\bullet
\newcommand\monunit{\mathsf e}
\newcommand\monact\bullet
\newcommand\sem[1]{\left\lBrack #1 \right\rBrack}
\newcommand\bigsem[1]{\stretchleftright{\lBrack}{#1}{\rBrack}}
\newmvar\Model{\mathbf M \or\mathbf N}
\newcommand\smashrel[1]{\mathrel{\smash{#1}}}
\newmvar\msubst{\theta\or\sigma\or\xi}

\newcommand\catMon{\Cat V}
\newcommand\catAct{\Cat A}
\newcommand\catSkew{\Cat S}
\newcommand\ract{\otimesrhrim}

\newcommand\Pointed[1]{\Pointed-{{#1}}}
\WithSuffix\newcommand\Pointed-[1]{#1^{\bullet}}
\newcommand\ptensor{\Pointed\tensor}
\newcommand\paction{\PointedAct\tensor}
\newcommand\pract{\PointedAct\ract}
\newcommand\pinitial[1][\Neut]{\Pointed{#1}}
\newmvar\pA{A\or B\or C}
\newcommand\pB{\pA[2]}

\newcommand\PointedAct[1]{\PointedAct-{{#1}}}
\WithSuffix\newcommand\PointedAct-[1]{#1_{\bullet}}
\newcommand\strength[1][]{\mathop{\mathsf{str}}\nolimits^{#1}}
\newcommand\pstrength[1]{\mathop{\mathsf{str}}\nolimits^{\PointedAct{#1}}}
\newcommand\prmul{\PointedAct\rmul}

\newcommand\MonoidCat{\mathop{\Category{Monoid}}}
\newcommand\ActionCat[1]{\mathop{#1\textbf{-}\Category{Action}}}

\newcommand\encode[2]{\left\lAngle#1,#2\right\rAngle}
\newcommand\roll{\mathsf{roll}}
\newcommand\fold{\mathsf{fold}}

\newcommand\Free[1][]{\mathbf{F}_{#1}}

\newcommand\Hole{\mathbf H}
\newcommand\ghole{\mathbb h}
\newmvar\mhole{?m\or?k\or?h}

\newcommand\meta[1][]{\mathop{?}\nolimits^{#1}}
\newcommand\metaenv[1][]{\sem{-}^{#1}_{\mathsf{meta}}}

\newcommand\alacarte{\'a la carte}
\newcommand\Extension{\mathbf{Ext}}
\newmvar\frag{\varepsilon\or\xi\or\zeta}
\newmvar\ext{\mathrm{ext}}

\usepackage[table]{xcolor}
\usepackage{tcolorbox}
\newtcbox{\myfigbox}{%
  on line, arc=1pt, outer arc=1pt,colframe=black,
  colback=white,
  boxsep=0pt,
  left=1pt,right=1pt,top=1pt,bottom=1pt,%
  boxrule=.5pt,bottomrule=.5pt,toprule=.5pt}

\newcommand\syncat[1]{\mspace{-25mu}\synname{#1}}
\newcommand\synname[1]{\qquad\text{#1}}

\newenvironment{syntaxDef}[1][]{%
\(
  \rowcolors{100}{white}{white}%
  \begin{array}[t]{#1l@{}l@{\,}*2{l@{}}@{\,}l}
}{
\end{array}
\)%
}

\newcommand\RecordCase[3]{%
  \mathop{\syntax{\mathbf{case}}}\nolimits%
  #1%
  \mathbin{\syntax{\mathbf{of}}}%
  \sseq{#2}\mathrel{\syntax{\To}}#3
  }

\newcommand\Clause[1]{%
  #1\mathrel{\syntax{\To}}
}

\newcommand\sset[1]{
 \syntaxBegin\left\{
  \syntaxClose #1
  \syntaxBegin\right\}
  \syntaxClose
}

\newcommand\VariantCase[2]{%
  \mathop{\syntax{\mathbf{case}}}\nolimits%
  #1%
  \mathbin{\syntax{\mathbf{of}}}%
  \sset{#2}
  }

\newcommand\LetName{\syntax{\mathbf{let}}}
\WithSuffix\newcommand\:={\mathrel{\syntax{\boldsymbol{=}}}}

\WithSuffix\newcommand\Let*[1]{%
  \mathop{\syntax{\mathbf{let}}}\nolimits%
  #1
  \mathrel{\syntax{\mathbf{in}}}}

\newcommand\ForLoop[3]{
  \mathop{\syntax{\mathbf{for}}}\nolimits%
  #1\:=#2 \mathbin{\syntax{\mathbf{do}}}
       #3}
\newcommand\LetRecName{\syntax{\mathbf{let\ rec}}}
\newcommand\LetRec[1]{
  \mathop{\LetRecName{}}\nolimits%
  #1
  \mathrel{\syntax{\mathbf{in}}}}

\newcommand\Roll{\syntax{\mathbf{roll}}}
\newcommand\Unroll{\syntax{\mathbf{unroll}}}
\newcommand\Fold[2]{%
  \mathop{\syntax{\mathbf{fold}}}\nolimits
  #1
  \mathop{\syntax{\mathbf{by}}}
  \sset{#2}
}
\WithSuffix\newcommand\Fold*[3]{%
  \mathop{\syntax{\mathbf{fold}}}\nolimits
  #1
  \mathop{\syntax{\mathbf{by}}}
  \sset{\Clause{#2} #3}
}
\newcommand\literal[1]{\syntax{#1}}
\newcommand\Suc{\mathop{\syntax{1+}}\nolimits}
\newcommand\Done{\mathop{\syntax{\mathbf{done}}}\nolimits}
\newcommand\Cont{\mathop{\syntax{\mathbf{continue}}}\nolimits}

\newcommand\sseq[2][]{\syntaxBegin\parent{\syntaxClose#2\syntaxBegin}\syntaxClose_{#1}}
\newcommand\Cons[1]{\syntax{#1}}
\newcommand\sof{\syntax{\colon}}
\newcommand\Subst\bracks

\newcommand\To\Rightarrow

\newcommand\naturals{\mathbb{N}}
\newcommand\Nat{\syntax\naturals}

\newmvar\TySig{\mathbf{L}\or\mathbf{R}}

\newcommand\BaseTySig{\mathop{\carrier{\Base}}\nolimits}
\newcommand\NatTySig{\mathop{\mathsf{NatTy}}\nolimits}
\newcommand\FunTySig{\mathop{\mathsf{FunTy}}\nolimits}
\newcommand\FieldLabel{\mathsf{Label}}
\newcommand\RowSig{\mathop{\mathsf{Row}}\nolimits}
\newcommand\RecordTySig{\mathop{\mathsf{RecordTy}}\nolimits}
\newcommand\VariantTySig{\mathop{\mathsf{VariantTy}}\nolimits}
\newcommand\EmptyRecordTySig{\mathop{\mathsf{EmptyRecord}}\nolimits}
\newcommand\NatConstTySig{\mathop{\mathsf{NatConstVariant}}\nolimits}
\newcommand\FullCBVTySig{\mathop{\mathsf{FullCBV}}\nolimits}

\newcommand\BaseSig{\mathop{\mathsf{Base}}\nolimits}
\newcommand\SeqSig{\mathop{\mathsf{Seq}}\nolimits}
\newcommand\FunSig{\mathop{\mathsf{Fun}}\nolimits}
\newcommand\RecordSig{\mathop{\mathsf{Record}}\nolimits}
\newcommand\VariantSig{\mathop{\mathsf{Variant}}\nolimits}
\newcommand\NatSig{\mathop{\mathsf{Nat}}\nolimits}
\newcommand\Maybe{\mathsf{Maybe}}
\newcommand\WhileSig{\mathop{\mathsf{While}}\nolimits}
\newcommand\RecSig{\mathop{\mathsf{Rec}}\nolimits}

\newcommand\Type{\mathsf{Type}}

\newcommand\Id{\mathsf{Id}}
\newcommand\rto{\nvrightarrow}

\newcommand\RecReq{\mathop{\mathsf{RecNeed}}}

\newcommand\seqFrag{\mathrm{seq}}
\newcommand\funFrag{\mathrm{fun}}

\newcommand\newop[2]{\newcommand{#1}{\mathop{\mathrm{#2}}\nolimits}}
\newcommand{\bindsymbol}{\mathchoice%
  {\rcurvyangle\mspace{-7mu}\joinrel\rcurvyangle\mspace{-4mu}\joinrel\Relbar}%
  {\rcurvyangle\mspace{-7mu}\joinrel\rcurvyangle\mspace{-4mu}\joinrel\Relbar}%
  {\rcurvyangle\mspace{-3mu}\joinrel\rcurvyangle\mspace{-1mu}\joinrel\Relbar}%
  {\rcurvyangle\mspace{-3mu}\joinrel\rcurvyangle\mspace{-1mu}\joinrel\Relbar}%
}

\newcommand\Monad[1][]{\mathrm{T}_{#1}}

\newcommand{\bind}[1][]{\mathrel{\bindsymbol}_{#1}}
\newop{\return}{return}
\WithSuffix\newcommand\overline*[1]{\parent{\overline{#1}}}
\newcommand\keep{\mathop\ltcir\nolimits}
\newcommand\bindkeep{\bind\joinrel\ltcir}

\newcommand\SeqCompatProofObligationB{($*$)}
\newcommand\AppCompatProofObligation{($**$)}
\newcommand\AssocProofObligation{($***$)}

\newcommand\rTo{\nVrightarrow}

\newcommand\ssToArrow[5]{%
\arrow[""{name=#1, anchor=center, inner sep=0},"\textstyle{#2}",
      "\times"{marking}, curve={#3},
      from=#4, to=#5]%
}
\WithSuffix\newcommand\ssToArrow'[5]{%
\arrow[""{name=#1, anchor=center, inner sep=0},"\textstyle{#2}"',
      "\times"{marking}, curve={#3},
      from=#4, to=#5]%
}
\newcommand\ToArrow[6][]{%
  \arrow[""{name=#2, anchor=center, inner sep=0}, "{#3}"#1,
    curve={#4}, from=#5, to=#6]%
}
\WithSuffix\newcommand\ToArrow'[6][]{%
  \arrow[""{name=#2, anchor=center, inner sep=0}, "{#3}"'#1,
    curve={#4}, from=#5, to=#6]%
}
\newcommand\rtoArrow[5]{%
\arrow[""{name=#1, anchor=center, inner sep=0},"\textstyle{#2}",
      "\shortmid"{marking}, curve={#3},
      from=#4, to=#5]%
}
\WithSuffix\newcommand\rtoArrow'[5]{%
\arrow[""{name=#1, anchor=center, inner sep=0},"\textstyle{#2}"',
      "\shortmid"{marking}, curve={#3},
      from=#4, to=#5]%
}
\newcommand\rToArrow[5]{%
\arrow[""{name=#1, anchor=center, inner sep=0},"\textstyle{#2}",
      "\shortmid\mspace{-10mu}\shortmid"{marking}, curve={#3},
      from=#4, to=#5]%
}
\WithSuffix\newcommand\rToArrow'[5]{%
\arrow[""{name=#1, anchor=center, inner sep=0},"\textstyle{#2}"',
      "\shortmid\mspace{-10mu}\shortmid"{marking}, curve={#3},
      from=#4, to=#5]%
}

\newcommand\smalltensor{{
\mathchoice%
  {\scriptstyle\tensor}%
  {\scriptstyle\tensor}%
  {\scriptscriptstyle\tensor}%
  {\scriptscriptstyle\tensor}%
}}
\newcommand\smallract{{
\mathchoice%
  {\scriptstyle\ract}%
  {\scriptstyle\ract}%
  {\scriptscriptstyle\ract}%
  {\scriptscriptstyle\ract}%
}}
\newcommand\smallodot{{
\mathchoice%
  {\scriptstyle\odot}%
  {\scriptstyle\odot}%
  {\scriptscriptstyle\odot}%
  {\scriptscriptstyle\odot}%
}}
\newcommand\smallpract{{
\mathchoice%
  {\scriptstyle\PointedAct\ract}%
  {\scriptstyle\PointedAct\ract}%
  {\scriptscriptstyle\PointedAct\ract}%
  {\scriptscriptstyle\PointedAct\ract}%
}}

\newcommand\tfrom{\from\joinrel\mathrel{\smalltensor}}
\newcommand\afrom{\from\joinrel\mathrel{\smallract}}

\newcommand\dfrom{\from\joinrel\mathrel{\smallodot}}
\newcommand\pfrom{\from\joinrel\mathrel{\smallpract}}
\newcommand\rfrom\tfrom
\newcommand\texp[2]{#1 \tfrom #2}
\newcommand\aexp[2]{#1 \afrom #2}

\newenvironment{proofNoQED}{\begin{pf}}{\end{pf}}

\newcommand\Commute[2]{%
  \arrow["{=}"{description}, draw=none, from=#1, to=#2]
}

\newcommand\CommuteDefBy[2]{%
  \arrow["{\defines}"{description}, draw=none, from=#1, to=#2]
}

\newcounter{junk}

\newcommand\ops{\mathrm{ops}}
\let\altphi\phi
\let\phi\varphi
\let\varphi\altphi
\let\altphi\undefined
\newcommand\gSig{\Sig_{\ghole}}
\newcommand\gsem[1]{\sem-^{\ghole}}



%% file: abstract.tex
We adapt Fiore, Plotkin, and Turi's treatment of abstract syntax with
binding, substitution, and holes to account for languages with
second-class sorts. These situations include programming calculi such
as the Call-by-Value $\lambda$-calculus (CBV) and Levy's
Call-by-Push-Value (CBPV). Prohibiting second-class sorts from
appearing in variable contexts changes the characterisation of the abstract
syntax from monoids in monoidal categories to actions in actegories.
We reproduce much of the development
through bicategorical arguments. We apply the resulting theory by
proving substitution lemmata for varieties of CBV.

%% file: introduction.tex
\section{Introduction}
\Fioreetal*'s~\cite{fiore-plotkin-turi:ast} mathematical foundations
for abstract syntax with binding and substitution
possess several unique properties. It is based on Goguen et
al.'s~\cite{goguen-et-al:initial-algebra-semantics} initial algebra
semantics, and as such provides an abstract interface that supports
modularity and extensibility.  It accommodates both syntactic
representations and their semantic models in one semantic setting.
Its multi-sorted extension supports intrinsically-typed
representation: the abstract syntax also encodes simply-typed
constraints, ensuring only well-typed syntax trees are expressed.
It supports context-aware \emph{holes}, called
\emph{metavariables}.  Despite its mathematical sophistication, it
lends itself to formalisation and computational
representation~\cite{allais-et-al:type-and-scope-safe-syntax-conf,allais-et-al:type-and-scope-safe-syntax-journal,fiore-szamozvancev:formal-soas,crole-representational-adequacy-hybrid}.
This approach is robust to generalisation, e.g.~to
polymorphic~\cite{fiore-hamana:mpats}, and
dependently-typed~\cite{fiore:soas} settings.

This work concerns
another such generalisation: support for second-class sorts~\cite{arkor-mcdermott:abstract-clones-for-abstract-syntax}, as
employed by common calculi such as the \CBVFull{} and Levy's
\CBPVFull~\cite{levy:book,levy:cbpv-conference}. Typically we consider separate syntax for values and for computations, but variables in the language only stand for values, leaving the sorts of computations `second-class' w.r.t.~substitution. Supporting the
syntactic needs of these calculi is essential for the applicability of
this theory to formalisation and computational representation of programming languages.  After all, common programming languages and their syntactic
representations are overwhelmingly Call-by-Value.

We make these ideas and motivation precise through a comprehensive
case-study of a \CBVFull{} (\CBV) with records/products,
variants/coproducts, and a simple inductive type of natural numbers,
with various flavours of recursion: structural, unbounded, and general
recursion. \figref{cbv syntax} presents the raw terms of this
calculus.  Each type $\ty$ has \emph{two} sorts of terms associated
with it: values $\mV[1]$ and unrestricted terms $\M[1]$. We can embed
values into unrestricted terms through the term constructor $\val$,
and use them wherever we may use a term.  Values enjoy a first-class
status w.r.t.~binding and substitution: we only have value variables and may
only substitute values for variables. This distinction partitions the
abstract syntax into first-class sorts for values and second-class
sorts for computations. Our contribution is to modify the classical theory
to allow this distinction between these two classes of sorts.

As a concrete yard-stick for this new theory, consider fragments of
the calculus. For example, a fragment involving only functions and
records, but not natural numbers. Each such fragment makes a different
set of semantic demands on its class of models. For example, functions
require certain exponentials, natural numbers require a parameterised
natural number object, recursion requires parameterised fixed-points,
etc. So long as we consider a fragment in isolation, we can simply
aggregate all the required structure into one definition, define the
semantic interpretation function, and establish its basic properties
directly. However, once we consider multiple fragments simultaneously,
we quickly need a modular representation of the syntax, its semantics,
and their properties. A concrete example of a property that typically
requires tedious reproof is the semantic substitution lemma, which
states that the semantics of a substituted term is the semantics of
the term composed with the semantics of the subsitution:
$\sem{\M{}[\msubst]} = \sem\M\compose \sem\msubst$. Without
a modular theory encompassing both syntax and semantics, proving this
lemma requires a separate tedious inductive argument for each
fragment. The theory we develop here will allow us to prove it
modularly for each fragment of the calculus, and combine these results
together to $2^7 = 128$ different substitution lemmata for $128$
different languages in \lemmaref{substitution}.

This problem is a semantic counterpart to Wadler's Expression
Problem\footnote{Philip Wadler, \textit{The Expression Problem}, Java
Genericity mailing list,
1998:\\ \url{https://homepages.inf.ed.ac.uk/wadler/papers/expression/expression.txt} .%
}. We do not purport to solve the Expression Problem, which
concern the design and implementation of abstract syntax and its
evaluator in an existing programming language. We review the immediate
literature and draw some connections in the Related Work
\secref{related work}.  Our approach is closest to Swierstra's
\alacarte{} solution to the Expression Problem. Like him, we use coproducts of
signature functors~\cite{swierstra:data-types-a-la-carte-journal}.

\begin{figure}
  \input{cbv-raw-syntax}
  \caption{Syntax of \CBV{}}
  \figlabel{cbv syntax}
\end{figure}

The classical theory phrases the universal property of
capture-avoiding substitution over the abstract syntax as follows. The
abstract syntax is a presheaf indexed by sorts and simply-typed
contexts. Both the syntax and its semantic models form algebras for
signature functors over this presheaf category, which allow them to be
aware of binding constructs. Both the syntax and its semantic models
also form monoids w.r.t.~a monoidal tensor product whose unit is the
presheaf of variables. This tensor product is the input to
capture-avoiding substitution, pairing a value with its closure,
subject to the structural properties of substitution. Tensoring with a
\emph{pointed} presheaf then provides closures that may contain
syntactic or semantic representations of variables.
The algebra structure and the substitution monoid must be
compatible. Phrasing this compatibility requires a \emph{pointed
strength} for the signature functor. It is this strength that yields
the theory its modularity.  It explains how to propagate
variable-aware closures under each syntactic construct, independently
from all the other language constructs and their semantics.

\subsection*{Contribution}
Generalising the theory to account for second-class sorts involves
refining these ingredients further. We generalise compatible monoids
to compatible \emph{actions}. In this sense, we extend Fiore and
Turi's~\cite{fiore-turi:name-and-value-passing} treatment of
\emph{value-passing} process calculis~(cf.~Related Work
\secref{related work}).  The terms of first-class sorts and their
semantics still form a monoid, allowing us to collapse towers of
substitutions into first-class terms or denotations. This monoid
\emph{acts} on terms of second-class sorts and their semantics, but
those no longer form a monoid, but an \emph{action}.
Just as monoidal categories are the
natural categorical setting for monoids, \emph{actegories} are the
natural categorical setting for actions. We dub the resulting theory
\emph{Modular Abstract Syntax Trees} (\mast). This simple change
propagates throughout the theory and requires reformulating it from
start to finish. Fortunately, the following relatively recent
development makes our work straightforward.

\emph{Bicategorical foundations.} Fiore, Gambino, Hyland and
Winskel~\cite{fiore-et-al:relative-pseudomonads-etc} characterised the
substitution tensor product as $1$-cell composition in a
Kleisli bicategory over \emph{profunctors}. While technical, this
bicategory captures the structure of the tensor as it operates on
heterogeneously sorted structures: one set of sorts for syntactic
classes, and one set of sorts for the context of bound variables. In
the same way that monoidal categories come from one object
bicategories, actegories come from a choice of two objects in a
bicategory. This observation avoids the many calculations direct
treatments incur.

\emph{Evaluation.} To evaluate \mast{}, we study
fragments of the \CBV{} calculus in \figref{cbv syntax}. \Mast{}'s modularity allows us
to formulate succinctly their syntax and semantics, deriving
$128$ substitution lemmata.

\emph{Skew monoidal structure.}  In an earlier version of this work,
we observed that the presheaves indexed by both kinds of sorts and by contexts of
first-class sorts have a substitution tensor with an associator and unitors.
Unfortunately, the left unitor is
not invertible. This failure means the tensor is \emph{left-skew}
monoidal~\cite{szlachanyi:skew-monoidal-categories-and-bialgebroids}.
Moreover, it is \emph{right-unital} and \emph{associative}, meaning
both the right-unitor and associator are invertible. Adapting the
theory is straightforward, but requires redeveloping it completely.
Fiore and
Szamozvancev's~\cite{fiore-szamozvancev:formal-soas,fiore-szamozvancev:soas-theory}
recent skew monoidal theory of abstract syntax has been influential to
our work. Skewness in their development has a different
source. Nonetheless, we reused their results verbatim thanks to our
shared categorical abstractions. While our recourse to actegories and
actions avoids the need for skew monoidal structure, we can connect
the two approaches. We prove that sufficiently well-behaved actegories
induce skew monoidal categories, and moreover the category of actions
in the actegory is isomorphic to the category of monoids in the skew
monoidal category. When working concretely, the skew structure
provides a uniform list of proof-obligations, avoiding the need to split
the structure into the components of its first-class and second-class sorts.

\emph{Alternative.}
Instead of re-developing the theory, in practice
one can take a concrete semantic structure $\Model$ with second-class
sorts, and complete it to obtain a model $\overline\Model$ of the
classical theory, in which the second-order sorts can appear in
contexts. We are not interested in this alternative. First, it leads
to unnecessarily complicated models (i.e., the hypothetical $\overline\Model$) which
contain formal/syntactic extensions needed to account for variables
of second-class sorts that programs will never exhibit. Second, we
would expect proponents of this alternative to prove this kind of
completion is always possible. Doing so would require to either
empirically complete all known models, or, more satisfyingly, define a
completion construction $\Model \mapsto \overline\Model$. In this
work, we define models for second-class sorts, i.e., the structure for
$\Model$, and moreover show they have the same abstract theory as
those with first-class sorts. This development lays the foundation for
such wholesale completion construction $\Model \mapsto
\overline\Model$, which we leave to further work.

\emph{Paper structure.} While the theory is technical, applying it
concretely involves few self-contained ingredients. Sections
\secref*{sorting and structures}--\secref*{compatible monoids} cover
these ingredients in a tutorial style. We accompany each concept with
motivating explanations and concrete examples (more than $25$ examples overall). This
development culminates in the Special Representation
\thmref{representation} which characterises the presheaf of abstract
syntax with holes in terms of $\Sig$-actions.
This theorem enables us to prove dozens of substitution
lemmata.  \secref{case
  studies} reports on a case study: extensions to \CBV{}.
\secref{technical development} provides a detailed technical outline
of the main ingredients in the bicategorical development and presents
the General Representation \thmref{general representation} for O-actions.
\thmref{general representation} implies the Special
Representation \thmref{representation} directly, but abstracts away
from the concrete syntactic details and technicalities of presheaves
using skew monoidal products and actions.
\secref{related work} surveys closely
related work.
\secref{conclusion} concludes.
\begin{mfpspaper}
We prepared an accompanying technical report~\cite{mast:techreport},
which is not published as part of this article.  It contains the same
contents of this article expanded with Appendices~A--D, which include
further technical details.
\end{mfpspaper}
\begin{techreport}
Appendices~\appref*{case-studies}--\appref*{representation} include further
technical details.
\end{techreport}
\appref[A]{case-studies} provides omitted proofs concerning
our case study. \appref[B]{skew bicats} expands on our bicategorical development,
including: the full definition of a bicategory; how to obtain monoidal
actions from bicategories; and the existence of the closed structure
for substitution tensors. \appref[C]{skew monoidal cats} provides the technical
details behind our development of monoidal categories, including:
actegories, strong functors; and connections to skew monoidal
categories. \appref[D]{representation} gives the full proof for the
General Representation \thmref{general representation} for \mast{},
based on parameterised initiality as outlined in Szamozvancev's
thesis~\cite{szamozvancev:thesis}. We will refer to these Appendices
throughout this article.

%% file: cbv-raw-syntax.tex
  \begin{tabular}{@{}ll@{}}
  \begin{syntaxDef}
    \ty[1], \ty[2], \ty[3] &{}\gdefinedby & & \syncat{type}\\
    & & \base & \synname{base} \\
    &{}\gor& \FUN{\ty[1]}{\ty[2]} & \synname{function}\\
    &{}\gor& \sRecord{C_i : \ty_i \suchthat* i \in I}
                                  & \synname{record ($I$ finite)}\\
    &{}\gor& \sVariant{C_i : \ty_i \suchthat* i \in I}
                                  & \synname{variant ($I$ finite)}\\
    &{}\gor& \Nat
                                  & \synname{natural number}\\
  \end{syntaxDef}
  &
  \begin{syntaxDef}
    \mV[1],\mV[2]
    &{}\gdefinedby& &\syncat{value}\\
    &{}    & x & \synname{variable}\\
    &{}\gor& \sabst {x : \ty}\M & \synname{function abstraction}\\
    &{}\gor& \sseq{\Cons{C_i} \sof \V_i \suchthat* \syntax i \in I}
                          & \synname{record constructor}\\
    &{}\gor& \ty.\Cons{C_i}\ \V & \synname{variant constructor}\\
    &{}\gor& \literal n  & \synname{number literal}\\
  \end{syntaxDef}
  \end{tabular}
  \begin{syntaxDef}
    \M[1],\M[2],\M[3],\M[4]
    &{}\gdefinedby& &\syncat{term}\\
    &{}    & \val \mV & \synname{value}\\
    &{}\gor& \Let*{x_1\:=\M_1; \ldots; x_n\:=\M_n}\N & \synname{sequencing}\\
    &{}\gor& \sapply \M\N & \synname{function application}\\
    &{}\gor& \sseq{\Cons{C_1} \sof \M_1, \ldots, \Cons{C_n}\sof \M_n}
                          & \synname{record constructor}\\
    &{}\gor& \RecordCase\M{\Cons{C_1}x_1, \ldots, \Cons{C_n}x_n}\N & \synname{record pattern match}\\
    &{}\gor& \ty.\Cons{C_i} \M & \synname{variant constructor}\\
    &{}\gor& \VariantCase\M{\Clause{\Cons{C_i}x_i} \M_i \suchthat* i \in I}\N & \synname{variant pattern match}\\
    &{}\gor& \Unroll \M \gor \Roll \M & \synname{number (de)constructor}\\
    &{}\gor& \Fold* \M x{\M[2]} & \synname{bounded iteration}\\
    &{}\gor& \ForLoop i\M\N & \synname{unbounded iteration}\\
    &{}\gor& \LetRec{
    f_1 \Context_1 \sof \ty_1 \:= \M_1;
    \ldots;
    f_n \Context_n \sof \ty_n \:= \M_n}
      \N
      &\synname{recursion}
  \end{syntaxDef}

%% file: tutorial.tex
\input{sorting-systems}
\input{combinators}
\input{tensors}
\input{signature-functors}
\input{compatibility}

%% file: sorting-systems.tex
\section{Heterogeneous sorting systems and structures}
\seclabel{sorting and structures}
The first component in the theory specifies the available sorts of
syntactic classes of interest. The
first-class sorts, unlike the second-class sorts, can appear in contexts and binding
positions. Formally, a \emph{heterogeneous sorting system} $\gsort =
\seq{\Fst[\gsort], \Snd[\gsort], \Sort[\gsort], \fst[\gsort],\snd[\gsort]}$ consists of:
\begin{itemize}
\item a set $\Fst[\gsort]$ whose elements are the \emph{first-class sorts};
\item a set $\Snd[\gsort]$ whose elements are the \emph{second-class sorts};
\item a coproduct diagram: \(
  \Fst[\gsort] \xto{\fst[\gsort]} \Sort[\gsort] \xfrom{\snd[\gsort]} \Snd[\gsort]
\)
\end{itemize}
We denote by
$b = \coprod_{i\in I}(c_i : a_i)$
the coproduct diagram with apex $b$ and injections:
$\seq[i \in I]{c_i : a_i \to b}$. So, for
a sorting system $\gsort$, we have $\Sort[\gsort] = (\fst[\gsort] :
\Fst[\gsort])\amalg(\snd[\gsort] : \Snd[\gsort])$.  We call the apex
$\Sort[\gsort]$ the
\emph{total} set of sorts.  We omit the sorting system $\gsort$ and
write $\Fst$, $\Snd$, and $\Sort$ when $\gsort$ is unambiguous.
When we don't specify a set for the apex, we will use the disjoint union
and its injections as the apex.
A
\emph{homogeneous} sorting system is a sorting system which only has
first class sorts: $\Snd[\gsort] = \emptyset$ so that $\fst[\gsort] :
\Fst[\gsort] \isomorphic \Sort[\gsort]$.  In this way, \mast{}
generalises the classical theory from homogeneous sorting systems to
heterogeneous ones.

\begin{longexample}
  The homogeneous sorting system $\CBN$ for the \emph{call-by-name}
  \lamcalc{} has the simple types as first-class sorts and no
  second-class sorts, i.e., $\Sort[\CBN] \definedby \Fst[\CBN]
  \definedby \SimpleType$.  The sorting system $\CBV$ for the
  \emph{call-by-value} \lamcalc{} has both a first-class sort for
  \emph{$\ty$-values} and a second-class sort for
  \emph{$\ty$-computations} for each simple type $\ty \in
  \SimpleType$:
  \[
  \Sort[\CBV] \definedby
  \set{\ty, \Comp\ty \suchthat \ty \in \SimpleType}
  = (\Val \definedby (\abst \ty\ty) : \SimpleType) \amalg (\Comp : \SimpleType)
  \]
  Here, the total set of sorts includes two versions of each simple
  type $\ty$: an untagged version, representing the sort of
  $\ty$-values, and a tagged version $\Comp \ty$ representing the sort
  of $\ty$-computations. We take the inclusion $\val \ty \definedby
  \ty$ as the left coproduct injection, and the tagging function
  $\Comp$ as the right injection.

  The sorting systems $\CBN$ and $\CBV$ codify
  that in \CBV{} only values may be substituted for variables, and
  \CBN{} does not make this distinction and all expressions can be
  substituted for variables.
\end{longexample}

\begin{longexample}
  Each sorting system $\gsort$ restricts to a homogeneous
  system $\rest\gsort \definedby
  \pair{\Fst[\gsort]}{\emptyset}$.
\end{longexample}

Every set of first-class sorts determines a category of contexts and
renamings in the following way.  Given a set $S$, we write $\RenCat S
\definedby \List S$ for the set of finite sequences $\Context \in
\RenCat S$ with elements in $S$ which we call the \emph{$S$-sorted
contexts}. We will need to refer to positions in a given context. For
example, the third position in the context $\Context \definedby {}
[\ty[1], \FUN{\ty[1]}{\ty[2]}, \ty[2]]$ has sort $\ty[2]$.  To
simplify such references, we will label these positions with distint
meta-level variable names, and omit the brackets when other symbols
delimit the context unambiguously.  For example, we will write
$\Context \definedby{} x : \ty[1], f : \FUN{\ty[1]}{\ty[2]}, y :
\ty[2]$, and refer to the third position by $y$. This presentation
makes it seem as if we use a nominal representation of binding,
whereas in fact we use a nameless (i.e., de
Bruijn~\cite{de-bruijn:nameless-dummies}) representation, indexing
context positions by ordinals.  We thus refer to the
set $\Vars\Context$ of positions in a given context $\Context$ as the
set of its \emph{variables}, and will use the meta-level labels to
refer to its elements. We write $(x : a) \in \Context$ when the
element $a$ appears in position $x \in \Vars\Context$.

 A \emph{renaming} $\ren : \Context[1] \to \Context[2]$ is a function
 $-[\ren] : \Vars\Context[1] \from \Vars\Context[2]$ satisfying
 $(y[\ren] : \sort) \in \Context[1] \impliedby (y : \sort)\in
 \Context[2]$. The choice of direction (\underline{from} $\Context[1]$
 \underline{to} $\Context[2]$) is a matter of taste. One mnemonic for
 our choice is the fictional typing judgement $\Context[1] \types \ren :
 \Context[2]$. Renamings compose as functions in opposite order with
 the identity function acting as the identity renaming, and
 collect into a category $\RenCat S$ of \emph{contexts} and \emph{renamings}
 between them.

 \begin{longexample}
   The following is a renaming from the context to itself:
   \[
   \ren : \Context \definedby [x : \base_1,
     f : \FUN{\base_1}{\base_2}, g : \FUN{\base_1}{\base_2}, y : \base_1]\to
   \Context
   \qquad
   x\bracks\ren \definedby x
   \quad
   f\bracks\ren \definedby g
   \quad
   g\bracks\ren \definedby f
   \quad
   y\bracks\ren \definedby x
   \]
   Thus a renaming may permute the order, e.g., permuting the
   variables in positions $f$ and $g$, identify some variables, e.g.,
   $x$ and $y$, or weaken the context, e.g., $y$ is not in the image.
 \end{longexample}

 \begin{longexample}
 Each category $\RenCat S$ has chosen finite products. The terminal object
 is the empty context $\terminal \definedby []$ with the unique
 renaming $\unitVal : \Context \to \terminal$ being the
 empty function $\Vars\Context \from \Vars{}[]$. The product of two
 contexts is their concatenation, with the left/right projection
 sending the $i^{\text{th}}$ position from the left/right to the
 $i^{\text{th}}$ position to the left/right. Letting
 $\Context[1] \definedby [x_1 : \sort_1, \ldots, x_n : \sort_n]$
 and $\Context[2] \definedby [x_{n+1} : \sort_{n+1}, \ldots, x_{n+m} : \sort_{n+m}]$:
 \[
 \Context[1]\++\Context[2]
   \definedby
       [x_1 : \sort_1, \ldots, x_{n+m} : \sort_{n+m}]
       \qquad
       \Context[1] \xfrom{x_i[\projection_1] := x_i}
       \Context[1] \++ \Context[2]
       \xto{x_{n+i} \defines x_{i}[\projection_2]}
       \Context[2]
 \]
 \end{longexample}

 Let $\gsort$ be a sorting system.
 Consider the set $\Sort[\gsort]$ as a discrete category.
 An \emph{$\gsort$-structure} $\mP$
 is a presheaf $\mP \in \PSh{\parent{\Sort[\gsort]\times
     \RenCat*{\Fst[\gsort]}}}$, i.e., a functor $\mP : \Sort[\gsort]
 \times \opposite-{\RenCat*{\Fst[\gsort]}} \to \Set$
 indexed by sorts and contexts and contravariant in renamings.  If $I$ is a set,
 an \emph{$I$-family} $\mF$ is a presheaf over the discrete category
 induced by $I$, and an $\gsort$-family is a family over
 $\Sort[\gsort]\times \RenCat*{\Fst[\gsort]}$. Thus an $\gsort$-family
 $\mF$ assigns to each sort $\sort \in \Sort[\gsort]$ and context
 $\Context \in \RenCat*{\Fst[\gsort]}$ a set $\mF_{\sort}\Context$. So an
 $\gsort$-structure $\mP$ is an $\gsort$-family equipped with a
 functorial action:
\[
-[\ren]_{\mP} : \mP_{\sort}\Context \from \mP_{\sort}\Context[2]
\qquad
p[\id]_{\mP} = p
\qquad
(q[\ren[2]])[\ren] = q[\ren[2]\compose \ren]
\tag{$\sort\in\Sort*, p \in \mP_{\sort}\Context, q \in \mP_{\sort}\Context[3]$, and $\Context \xto\ren \Context[2] \xto{\ren[2]}\Context[3]$}
\]
Let $\Syn\gsort$ be the category of $\gsort$-structures and natural transformations $\nt : \mP \to \mQ$ between them.
This category is our central semantic domain. Through it we will
define constructions and universal properties for most other concepts.
Each $\gsort$-structure $\mP$ amounts to two presheaves:
\[
\rest\mP \in \PSh+{\Fst \times \opposite-{\RenCat{\Fst-}}}
\qquad
\restrict\mP\snd \in \PSh+{\Snd \times \opposite-{\RenCat{\Fst-}}}
\]
which we call the \emph{first-class} and \emph{second-class fragments}
of $\mP$, respectively. In a homogeneous sorting system, the
second-class fragment of every presheaf trivialises to the empty
functor from the empty category to $\Set$, and we will identify such
presheaves with their first-class fragment.  In this way, \mast{}
presheaves generalise the homogeneous presheaves of the classical
theory.\footnote{The other trivialising case, $\Sort[\gsort] \definedby
\emptyset \amalg \Snd$, yields $\Syn\gsort \isomorphic \Set^{\Snd}$,
and so \mast{} generalises multi-sorted algebra, without accounting
for equational presentations and their varieties. We will not make
use of this fact further.}

\begin{longexample}
  Let $\Sort = \Fst$ be a \emph{homogeneous} sorting system, i.e., a set of
  first-class sorts. The presheaf of
  \emph{variables} $\Neut \in \Syn\Fst$ is given by
  \(
  \Neut[\sort]\Context \definedby \set{x \suchthat (x : \sort) \in \Context}
  \) equipped with renaming
\(
  x[\ren]_{\Neut} \definedby x[\ren]
  \).
\end{longexample}

\begin{longexample}
  Let $\Lam[\CBV]$ be the $\CBV$-structure comprising the values
  and terms of the \CBV{} \lamcalc:
  \[
  \Lam[\CBV]_{\Val\ty}\Context \definedby \set{\V \suchthat \Context \types \V : \Val \ty}
  \qquad
  \Lam[\CBV]_{\Comp\ty}\Context \definedby \set{\M \suchthat \Context \types \M : \Comp \ty}
  \qquad
  X[\ren]_{\Lam[\CBV]} \definedby X[\ren]
  \]
  Its functorial action is given by the usual, syntactic, definition of renaming.
  Its first-class fragment supports a natural transformation from the homogeneous
  presheaf of variables:
  \(
  \var : \Neut \to \restrict{\Lam[\CBV]}\Val
  \).
\end{longexample}

Let $\Cat C$ be a category with chosen finite products. Every
functor $F : S \to \Cat C$ extends to a product-preserving functor
$\Env F{} : \RenCat S \to \Cat C$ via:
\[
\Env F{\Context} \definedby \prod_{(x : \sort) \in \Context} F_{\sort}
\qquad
\Env F\ren \definedby \seq[{(x : \sort) \in \Context[2]}]{\Env F{\Context[1]}
  \xto{\projection_{x[\ren]}}
    F_{\sort}}
  :
  \Env F{\Context[1]} \to
  \Env F{\Context[2]}
\tag{$\ren : \Context[1] \to \Context[2]$}
\]
The product-preservation condition provides the \emph{concatenation} operation $(\++) : \Env\mP{\Context[1]}\times \Env\mP{\Context[2]} \xto\isomorphic \Env{\mP}{\Context[1]\++\Context[2]}$.
The typical case is $\Cat C \definedby
\PSh{\RenCat*{\Fst[\gsort]}}$. By considering each homogeneous
presheaf $\mP \in \Syn\Fst$ as a
functor $\Fst[\gsort] \to \PSh{\RenCat*{\Fst[\gsort]}}$, we form the functor
$\Env\mP{} \in \RenCat*{\Fst[\gsort]} \to \PSh{\RenCat*{\Fst[\gsort]}}$. We call the elements $e \in
\Env\mP\Context\Context[2]$, the \emph{$\mP$-valued
$\Context$-environments} in context $\Context[2]$. They are are
$\Context[1]$-indexed tuples of $\mP\Context[2]$ elements of the
appropriate sorts, and can represent both semantic and syntactic substitutions.
As with coproducts, we write $b = \prod_{i \in I}(c_i : a_i)$ for the product
diagram $\seq[i \in I]{b \xto{c_i} a_i}$, and $\seq[i \in I]{c_i : u_i}$ for the
tuple whose $i$-th component is $u_i$.

\begin{longexample}
  The following environment
  is in $\Env{(\Lam[\CBV])}{[a : \base, b : \base, c : \FUN\base\base]}[x : \base, f : \FUN\base\base]$:
  \[
  \triple {a : x}{b : x}{c : \sabst {z : \base}{\sapply f{(\sapply f z)}}}
  \]
\end{longexample}

\begin{remark}
  We chose a nameless representation for contexts since it simplifies
  some concrete aspects in the development. As in the classical
  theory, this choice is not essential. For example, we can
  represent contexts nominally as a list pairing variable names
  and sorts. To define $\Neut$, we must
  disambiguate variables with the same concrete
  name. These different choices give equivalent small categories of
  contexts and therefore equivalent categories of presheaves. All of
  our concepts are defined by universal properties, and so the same
  development can be carried out in any of those representations.
\end{remark}

%% file: combinators.tex
\section{Signature combinators}
One defining feature of the classical theory is how it
deconstructs signatures with binding~\cite{aczel1978:general-church-rosser,plotkin:illative}
into smaller components. Formally, and following categorical logic and
Goguen's initial algebra approach to semantics, we use endofunctors
$\Sig : \Syn\gsort \to \Syn\gsort$ to represent signatures. Each
element $\op \in (\Sig X)_{\sort}\Context$ represents a
language constructs of sort $\sort$. The presheaf $X$
represents the sub-terms $\op$ may use and the context
$\Context$ represents the free variables in scope.  Representing
signatures with endofunctors enables some degree of modularity.  For
example, the coproduct of signature functors $\Sig_1\amalg\Sig_2$
gives terms in which we can take operators from either $\Sig_1$ or
$\Sig_2$. This decomposition allows us to study each operator on its own
and combine their properties modularly. In doing so, we abstract from
sorting systems and their categories of structures. Thus we define these
combinators on more abstract presheaf categories.

\subsection*{Application, restriction, and extension}
We often want to project out one or more subterms, or only define a
language construct in a specific collection of sorts. For example, in
the simplest case we project out or define in a single sort:
\begin{align*}
(\At \sort_0)
  &: \Syn\gsort \to \PSh{\RenCat*{\Fst[\gsort]}}
  &  X \At\sort_0 &\definedby X_{\sort_0} &
  \tag{$\sort_0 \in \Sort\gsort$}\\
  \OnlyAt{\sort_0}
  & : \Syn\gsort \from \PSh{\RenCat*{\Fst[\gsort]}}
  & (\OnlyAt{\sort_0}Y)_{\sort} &\definedby \begin{cases}
    \sort = \sort_0: & Y \\
    \text{otherwise:}& \initial
  \end{cases}
\end{align*}
\begin{longexample}
  The \CBV{} inclusion of $\ty$-values into $\ty$-computations is:
  \(
  \ValSig_{\ty} X \definedby \OnlyAt{\Comp\ty} (X\At{\Val \ty})
  \).
\end{longexample}
In a more general form of these combinators we restrict to, and extend along,
a function $f : I \to J$ between sets, for any small category $\SC$,
obtaining an adjoint pair of combinators $\OnlyAt f \leftadjointto (\restrict{}f)$:
\begin{align*}
(\restrict {}f)
  &: \PSh+{J\times\SC} \to \PSh+{I \times \SC}
  &  (\restrict Xf)_{i} &\definedby X_{fi} &
  \tag{$f : I \to J$ in $\Set$}\\
  \OnlyAt{f}
  & : \PSh+{J\times\SC} \from \PSh+{I \times \SC}
  & (\OnlyAt{f}Y)_{j} &\definedby \coprod_{i \in \inv f[j]}Y_i
\end{align*}
\begin{longexample}
  Taking $\fst : \Fst[\gsort] \to \Sort[\gsort]$ recovers the
  combinator $(\rest{})$. We recover the combinators
  $(\At{\sort_0})$ and $\OnlyAt{\sort_0}$ using the constant
  function $\const \sort_0 \definedby (\abst {\sort[2]}\sort_0) : \terminal \to \Sort[\gsort]$.
\end{longexample}

\subsection*{Products and coproducts}
As in categorical logic and initial algebra semantics, the
bread-and-butter combinators are products and coproducts. Products
allow us to express $n$-ary syntactic constructs. Coproducts allow us
to combine signatures into larger signatures. We will use both in many
different settings, and so define them in their utmost generality as
$\prod_I,\coprod_I : \Cat C^I \to \Cat C$, where $\Cat C$ has
$I$-indexed products or coproducts, as appropriate.

\begin{longexample}\exlabel{valsig}
  We combine the value inclusions in one functor: $\ValSig X \definedby
  \coprod_{\ty \in \SimpleType}(\val_{\ty}: \ValSig_{\ty} X)$.
\end{longexample}

\begin{longexample}\exlabel{appsig}
  Application in \CBV{} has the signature:
  \[
  \AppSig X \definedby \coprod_{\ty,\ty[2] \in \SimpleType}\parent{
  (\sapply{}{}) : \OnlyAt{\Comp\ty[2]} \parent{\parent{X \At \Comp(\FUN{\ty[1]}{\ty[2]})}\times (X \At\Comp\ty[1])}}
  \]
\end{longexample}

\subsection*{Scope shift}
We use the following scope shift operation to express operators
that bind variables:
\begin{align*}
\Context\Shift &: \Syn\gsort \to \Syn\gsort
& (\Context\Shift X)_{\sort}\Context[2]  &\definedby
  X_{\sort}(\Context[2] \++ \Context)
\tag{$\Context \in \RenCat*{\Fst[\gsort]}$}
\end{align*}

\textbf{Aside.}
  In the classical single-sorted and homogeneous theory, scope
  shift by one variable is presheaf exponentiation
  by the presheaf of variables. In our setting,
  $(\Context\Shift) = (\yoneda[\Context] \multimap)$, where
  $\yoneda : \RenCat*{\Fst[\gsort]} \to \PSh{\RenCat*{\Fst[\gsort]}}$ is the
  Yoneda embedding and
  $(G\multimap) : \Syn\gsort \to
  \Syn\gsort$, for a single-sorted presheaf $G \in
  \PSh+{\RenCat*{\Fst[\gsort]}}$,
  is the right adjoint
  to the functor $(G\odot) : \Syn\gsort \to
  \Syn\gsort$ given by: $(G\odot\mP)_{\sort}\Context := G\Context
  \times \mP_{\sort}\Context$.

\begin{longexample}\exlabel{abstraction sig}
  For abstraction: $\displaystyle\AbsSig X \definedby
  \coprod_{\ty[1],\ty[2] \in \SimpleType}\parent{(\sabst{x : \ty}) : \OnlyAt{\FUN{\ty[1]}{\ty[2]}}
    [x : \ty]\Shift X\At \Comp\ty[2]}$.
\end{longexample}

Combining these examples, we have the full \CBV{} signature:
\(
\CBVSig \definedby \AbsSig \amalg \ValSig \amalg \AppSig
\).
The presheaf of syntax is then the initial algebra $\Lam[\CBV] = \fix
X . \parent{(\CBVSig X)\amalg \Neut}$.
Using the same methodology, one can mechanically translate, e.g.,
Aczel's~\cite{aczel1978:general-church-rosser,plotkin:illative} \emph{binding
signatures} which express a wide class of syntax.

%% file: tensors.tex
\section{Substitution tensors}
The substitution tensor imposes semantic invariants on the input for
syntactic or semantic substitution operations, which will be of the
form $-[-]_{\mP} : \mP \rmul \rest\mP \to \mP$ where $\mP$ is a
$\gsort$-structure standing for the abstract syntax or its semantics.
Let $R,S,T$ be sets, and $\mP \in \PSh+{R\times\RenCat S}$, $\mQ \in
\PSh+{S\times\RenCat T}$ be two heterogeneous presheaves. Their
\emph{heterogeneous substitution tensor} is the heterogeneous presheaf:
\[
\mP\tensor \mQ \in \PSh+{R\times\RenCat T}
\qquad
(\mP\tensor\mQ)_{\sort[2]}\Context[1] \definedby
\int^{\Context[2] \in \RenCat S} \mP_{\sort[2]}\Context[2] \times \Env\mQ{\Context[2]}\Context
\definedby
\parent{\coprod_{\Context[2] \in \RenCat S} \mP_{\sort}\Context[2]\times
\Env\mQ{\Context[2]}\Context}/(\qsim)
\]
This definition involves the $\mQ$-environment functor $\Env\mQ{} : S
\to \PSh+{\RenCat T}$ given by $\Env\mQ{} \definedby \prod_{(x :
  \sort) \in \Context[2]} P_{\sort}\Context$.  The coend's quotienting
relation $(\qsim)$ is the least equivalence relation generated by
relating the triples:
\[
\triple{\Context[2]_1}{t[\ren]_{\mP}}{e} \qsim
\triple{\Context[2]_2}{t}{e_{-[\ren]}}
\tag{$\ren : \Context[2]_1 \to \Context[2]_2$, $t \in \mP_{\sort}\Context[2]_2$, $e \in \rEnv\mQ{\Context[2]_1}\Context$}
\]
As we explained in the introduction, these identification represent invariants
for substitution:
\begin{longexample}\exlabel{substitution tensor motivation}
  Consider the syntax presheaf $\Lam[\CBV]$. For brevity, we use the
  vernacular ($k\,y$), rather than elaborate ($\sapply{(\val k)}{(\val
    y)}$), syntax.  Consider the following equivalences in
  $\Lam[\CBV]\rmul\Lam[\CBV]$:
  \begin{itemize}
  \item Assigning the same value to different variables ($f,g$ below)
    relates to renaming the two variables to one $(f,g \mapsto h)$. Formally,
    taking $f[\ren] \definedby h, g[\ren] \definedby h$, $t \definedby
    \sabst{x}f(g\,x)$, and
    $e \definedby \seq{f: k\,y,g: k\,y}$
    witnesses:
    \begin{multline*}
    [\sabst{x}f(g\,x)
      , \seq{f: k\,y
            ,g: k\,y}]_{[f,g : \FUN\base\base]}
    =
    [\sabst{x}h(h\,x)
      , \seq{h: k\,y}]_{[h : \FUN\base\base]}
    \\\in (\Lam[\CBV]\rmul \Lam[\CBV])_{\Val{(\FUN\base\base)}}[k : \FUN\base\base, y : \base]
    \end{multline*}
\item Weakening the context by unused variables relates to projecting
  only the used variables. Formally, taking $z[\ren] \definedby z$, $t
  \definedby \sabst{x}z$, and $e \definedby \seq{f: \sabst xx, z : y}$
  witnesses:
  \[
    [\sabst{x}z
      , \seq{f: \sabst xx
        ,z: y}]_{[f : \FUN\base\base, z : \base]}
    =
    [\sabst{x}z
      , \seq{z: y}]_{[z : \base]}
    \tag*{$\in (\Lam[\CBV]\rmul \Lam[\CBV])_{\Val\parent{\FUN\base\base}}[y : \base]$}
  \]
  \item Permuting the environment relates to permuting the variables. Formally, taking $f[\ren] \definedby g, g[\ren]\definedby f$, $t \definedby \sabst{x}g(f\,x)$, and $e \definedby \seq{f: \sabst xx
            ,g: k}$ witnesses:
    \begin{multline*}
    [\sabst{x}f(g\,x)
      , \seq{f: \sabst xx
            ,g: k}]_{[f,g : \FUN\base\base]}
    =
    [\sabst{x}g(f\,x)
      , \seq{f: k
            ,g: \sabst xx}]_{[f,g : \FUN\base\base]}
    \\\in (\Lam[\CBV]\rmul \Lam[\CBV])_{\Val\parent{\FUN\base\base}}[k : \FUN\base\base]
    \end{multline*}
  \end{itemize}
\end{longexample}
These examples are representative in the following sense.
We can represent every renaming $\ren : \Context[2]_0 \to
\Context[2]_2$ as the composition: \( \ren:
\Context[2]_0 \xmonic{i} \Context[2]_1 \xto{\thin} \Context[2]_2 \),
where: $i : \Context[2]_0 \monic \Context[2]_1$ is a renaming with a
surjective action on variables, and $\thin$ is a \emph{thinning}, a
renaming with an injective and relative-order-preserving action on
variables. Permuting the variables and then repeatedly
identifying some, but not necessarily all, adjacent variables of the
same sort generates all renamings with surjective actions.
Repeatedly thinning out a variable generates all thinnings. Therefore
$(\qsim)$ is the smallest equivalence relation that contains the
following three identifications (cf.~\exref{substitution tensor motivation}):
\begin{itemize}
\item Identifying two variables vs.~environments containing the same
  value in their positions.
\item Weakening by a thinning vs.~projecting according to a thinning.
\item Permuting two variables in the term vs.~permuting the values
  in their positions.
\end{itemize}

The substitution tensor product has the following \emph{left/right
unitors} and \emph{associator} maps:
\begin{align*}
  \ellunit &: \Neut \rmul P \xto{\isomorphic} P &
  \runit   &: P \rmul \Neut \xto{\isomorphic} P &
  \assoc &: (P \rmul Q)\rmul L \xto\isomorphic P \rmul (Q \rmul L)\\
  \ellunit &[x,e]_{\Context[2]} \definedby e_x &
  \runit   &[p,\ren]_{\Context[2]} \definedby p[\tilde\ren] &
  \assoc &\bracks{[p,q]_{\Context[2]_1}, e}_{\Context[2]_2} \definedby
  \bracks{p, \seq[{(x : \sort) \in \Context[2]_1}]{\bracks{q_x, e}_{\Context[2]_2}}}_{\Context[2]_1}
\end{align*}
While these \emph{mediators} satisfy familiar-looking pentagon and triangle laws, they
are best understood as part of a bicategory whose 0-cells are small
categories, 1-cells are generalised heterogeneous preshaves, and
2-cells are natural transformations between them. Fiore, Gambino,
Hyland and Winskel~\cite{fiore-et-al:relative-pseudomonads-etc}
introduced and investigated this bicategory, and we expand on this
bicategorical perspective in \subsecref{bicat dev}.

Let $\gsort$ be a heterogeneous sorting system. The substitution
tensor then provides a two-argument functor $(\tensor) : \Syn\gsort \times
\Syn{\Fst[\gsort]} \to \Syn\gsort$. We understand it abstractly through
actions. Formally, a \emph{monoidal action} $\pair\catMon\catAct$
consists of:
\begin{itemize}
\item A monoidal category $\catMon = \seq{\carrier\catMon, \tensor,
  \mNeut, \assoc, \ellunit, \runit}$, we will typically write $\runit' \definedby \inv\runit$; and
\item A $\catMon$-actegory $\catAct = \seq{\carrier\catAct, \ract,
  \assoc, \runit}$, i.e., a functor and isomorphisms, natural in $a
  \in \carrier\catAct$ and $x,y \in \carrier\catMon$:
  \[
  (\ract) : \catAct \times \catMon \to \catAct
  \qquad
  \assoc_{a,x,y} : (a \ract x) \ract y \to a \ract (x \tensor y)
  \qquad
  \runit_{a} : a \ract \mNeut \to a
  \]
  satisfying the following equations:
  \begin{center}
 \diagram*{action-pentagon}\qquad
 \diagram*{action-triangle}
  \end{center}
\end{itemize}

Take $\SndSyn\gsort \definedby
\PSh+{\Snd[\gsort]\times\RenCat*{\Fst[\gsort]}}$ as the heterogeneous structures
with second-class sorts only:

\begin{theorem}\thmlabel{structures are actions}
  Each heterogeneous sorting system $\gsort$
  gives a monoidal action $\pair{\Syn{\Fst[\gsort]}}{\SndSyn\gsort}$:
  \begin{itemize}
  \item The monoidal category of
  homogeneous $\Fst[\gsort]$-structures equipped with substitution tensors,
  the presheaf of variables, and their associator and unitors:
  \[
  \Syn\Fst = \seq{\PSh+{\Fst[\gsort]\times{\RenCat*{\Fst[\gsort]}}},
    (\tensor), \Neut, \assoc, \ellunit, \runit}
  \]
  \item The $\Fst[\gsort]$-actegory of
  the second-class-sorted structures equipped with substitution tensors and mediators:
  \[
  \SndSyn\gsort = \seq{\PSh+{\Snd[\gsort]\times{\RenCat*{\Fst[\gsort]}}},
    (\tensor), \assoc, \runit}
  \]
  \end{itemize}
\end{theorem}

The theorem is a direct consequence of the bicategorical
development~\cite{fiore-et-al:relative-pseudomonads-etc}
(cf.~\propref{actions from two 0-cells}).
The monoidal category is the one from
the classical theory. Identifying the actegory structure is the
technical innovation in \mast{}.  The projections into the first-class
and second-class components form an isomorphism of categories
$\pair{(\rest{})}{(\restrict{}{\snd})} : \Syn\gsort\xto\isomorphic
\Syn\Fst \times \SndSyn\gsort$. We will use its inverse in the sequel:
\[
\encode-- :
\Syn\Fst \times \SndSyn\gsort
\xto\isomorphic
\Syn\gsort
\qquad
\rest{\encode\mP\mQ} = \mP
\quad
\restrict{\encode\mP\mQ}\snd = \mQ
\quad
\mP[3] = \encode{\rest{\mP[3]}}{\restrict{\mP[3]}\snd}
\]
Since each monoidal category acts on
itself, and actegories have componentwise products, we can and will extend the
action, through this isomorphism, to an action of the homogeneous structures on all $\gsort$-structures:
$(\ract) : \Syn\gsort \times \Fst[\gsort] \to \Syn\gsort$, satisfying:
\(
\rest{\parent{\mP\ract\mQ}} = \rest\mP\tensor\mQ
\).
However, the Representation \thmref{representation} in the sequel needs the
action from \thmref{structures are actions}, not the extended action $(\ract)$.

We will use the following two well-known constructions to combine actegories compositionally:
\begin{example}
  Every monoidal category $\catMon$ is a $\catMon$-actegory
  through its own monoidal tensor, i.e., taking:
  \(
  (a \ract b) \definedby (a \tensor b)
  \).
  The actegory axioms are a subset of the monoidal ones.
\end{example}
\begin{example}
  Given a sequence of
  $\catMon$-actegories
  $\seq[i]{\Cat A_i, (\ract^i), \Neut^i, \assoc^i, \runit^{i}{}}$,
  we define the \emph{product $\catMon$-actegory}
   componentwise as follows, validating the axioms componentwise:
  \[
  (\ract) : (\prod_{i \in I}\Cat A_i)\times \Cat C \to\prod_{i \in I}\Cat A_i
  \qquad
  (\vec x \ract a) \definedby \seq[i]{x_i \ract^i a}
  \qquad
  \assoc \definedby \seq[i]{\assoc^i}
  \qquad
  \runit \definedby \seq[i]{\runit^{i}}
  \]
\end{example}

Let $\pair{\catMon}{\catAct}$ be a monoidal action.
A  \emph{$\pair\catMon\catAct$-action} $\pair\Monoid\Action$ consists of:
\begin{itemize}
\item A $\catMon$-monoid $\Monoid$, i.e., a triple
  $\triple{\carrier\Monoid}{\monprod}{\monunit}$ consisting of an
  object $\carrier \Monoid \in \carrier{\catMon}$ and two morphisms
  $\mNeut \xto\monunit \carrier\Monoid \xfrom{(\monprod)}
  \carrier\Monoid \tensor \carrier\Monoid$ satisfying the following
  three equations:
\begin{center}       \hfill
  \diagram*{mon-left}\hfill
  \diagram*{mon-assoc}\hfill
  \diagram*{mon-right}\hfill~%
\end{center}
\item An $\Monoid$-action $\Action$ in $\catAct$, i.e., a pair
  $\pair{\carrier\Action}{\monact}$ consisting of an
  object $\carrier \Action \in \carrier{\catAct}$ and a morphism
  $\monact : \carrier\Action \ract \carrier\Monoid \to \carrier\Action$
  satisfying the following
  two equations:
\begin{center}       \hfill
  \diagram*{act-assoc}\hfill
  \diagram*{act-right}\hfill~%
\end{center}
\end{itemize}

Generalising the classical theory, we
use monoids and actions with respect to the substitution tensors
to axiomatise syntactic and semantic capture-avoiding substitution operations.
We therefore use the more suggestive notation $\Monoid =
\triple{\carrier\Monoid}{-[-]}{\var}$ for monoids in $\Syn\Fst$
and $\Action = \pair{\carrier\Action}{-[-]}$ for their actions in $\SndSyn\gsort$,
which we call \emph{substitution actions}.

\begin{example}
  The first-class fragment of the syntax presheaf $\Lam[\CBV]$ has a
  $\CBV$-monoid structure given by capture-avoiding simultaneous
  substitution on values $[\mV,\msubst]_{\Context} \mapsto
  \mV{}[\msubst]$. The inclusion of variables in values is the map $\var :
  \Neut[\ty] \xto{x\mapsto x} (\Lam[\CBV])_{\ty}$. This monoid
  acts on the second-class fragment through capture-avoiding
  substitution on computations $[\M,\msubst]_{\Context} \mapsto
  \M{}[\msubst]$. The five axioms become the familiar substitution
  lemmata:
  \[
  x\bracks{\msubst}
  =
  \msubst_x
  \qquad
  \parent{\mV{}\bracks{\msubst[1]}}\bracks{\msubst[2]}
  =
  \mV{}\bracks{\msubst[1][\msubst[2]]}
  \qquad
  \mV{}\bracks{\id}
  =
  \mV
  \qquad
  \parent{\M{}\bracks{\msubst[1]}}\bracks{\msubst[2]}
  =
  \M{}\bracks{\msubst[1][\msubst[2]]}
  \qquad
  \M{}\bracks{\id}
  =
  \M
  \]
  We will derive capture-avoiding substitution and its meta-theory
  from the Representation \thmref{representation}.
\end{example}

\begin{example}\exlabel{pure cbv model}
  Let $\Cat C$ be a Cartesian-closed category with chosen finite
  products $\prod$ and exponentials $b^a$. Every choice of
  interpretation $\sem \base$ for the base types equips $\Cat C$ with a
  $\CBV$-action structure by first extending the interpretation to
  types and contexts, and then defining the carrier presheaf by:
  \[
  \sem{\FUN{\ty[1]}{\ty[2]}} \definedby \sem{\ty[2]}^{\ty[1]}
  \qquad
  \sem{\Context} \definedby \prod_{(x : \ty) \in \Context}\sem\ty
  \qquad
  \carrier\Model_{\ty}\Context \definedby \Cat C(\sem\Context, \sem\ty)
  \qquad
  \carrier\Action_{\Comp\ty}\Context \definedby \Cat C(\sem\Context, \sem\ty)
  \]
  Define the unit by projection
  \(
  \parent{\var[{\ty[2];\Context}] y} :  \sem{\Context} =
  \prod_{(x : \ty) \in \Context}\sem\ty \xto{\projection_y}
  \sem{\ty[2]}
  \);
  and substitution by composition:
  \[
  \parent{\sem{\Context[2]} \smashrel{\xto f} \sem\ty}\bracks{
    \seq[{(y : \ty[2]) \in \Context[2]}]{\sem{\Context} \smashrel{\xto{\msubst_y}} \sem{\ty[2]}}
  } \definedby\parent{\sem\Context
    \smashrel{\xto{\seq[{(y : \ty[2]) \in \Context[2]}]{\msubst_y}}}
  \prod\nolimits_{(y : \ty[2]) \in \Context[2]} \sem{\ty[2]}
  =
  \sem{\Context[2]} \smashrel{\xto f} \sem\ty}
  \]
  We explore the richer strong-monad models for \CBV{} in our case study in
  \secref{case studies}.
\end{example}

The isomorphism $\encode-- : \Syn\Fst\times \SndSyn\gsort
\xto\isomorphic \Syn\gsort$ lets us re-package substitution actions in
terms of $\gsort$-structures. A \emph{substitution structure}
$\SubStruct =
\triple{\carrier\SubStruct}{-[-]_{\SubStruct}}{\var[\SubStruct]}$ is
an $\gsort$-structure $\carrier\SubStruct$ equipped with morphisms
$\var[\SubStruct] : \Vars \to \rest{\carrier\SubStruct}$ and $-[-] :
\carrier\SubStruct \tensor \rest{\carrier\SubStruct} \to
\carrier\SubStruct$ forming a monoid
$\triple{\rest\SubStruct}{\var}{(\rest-)[-]_{\SubStruct}}$ and an
action of it
$\pair{\restrict\SubStruct\snd}{(\restrict-\snd)[-]_{\SubStruct}}$. I.e.,
this monoid and its action form a substitution action. Thus we will
speak of substitution actions when we refer to a pair of
structures---a homogeneous monoid and its action on a heterogeneous
structure over second-class sorts. We will speak of substitution structures when
we want to emphasised the combined heterogeneous structure over both
first-class and second-class sorts.

%% file: signature-functors.tex
\section{Signature functors}
\subseclabel{signature functors}

So far we have specified functors $\Sig$, their algebras, and
actions. We can equip the abstract syntax with this structure, and its
denotational semantics with this structure. We can even characterise
the abstract syntax as an initial algebra $\pair{\fix X.
  (\Sig X)\amalg \encode\Vars\initial}{\roll}$ for the signature,
and the denotational semantics as the unique $(\Sig\amalg
\encode\Vars\initial$-algebra homomorphism. This characterisation,
however, does not account for neither the syntactic nor semantic
substitution lemmata. The missing ingredient is a structural map, a
tensorial strength for the signature functor. The strength specifies how to
avoid unintended capture when moving under each operator in the
signature.  More precisely, it specifies how to change an environment
containing the free variables in scope when we propagate it to the
arguments of each language construct. It is this information, phrased
w.r.t.~arbitrary environments, that gives \mast{} much of its
modularity. The overall signature functor is a coproduct of signature
functors, each with their own strength. Each strength determines how
to substitute through each language construct independently from the
other language constructs.  To define the strengths, the classical
theory uses pointed tensors and their actions, which we adapt to
heterogeneous structures in this section.

To describe scope changes, the classical theory uses a point-free way to consider
those homogeneous prehseaves $\mP \in \Syn\Fst$ that can, moreover,
encode variables $\sem{x : \sort}_{\mP} \in
\mP_{\sort}[x : \sort]$.  Since we have that $\Neut[\sort] = \yoneda[{[x:\sort]}]
\in \PSh+{\Fst[\gsort]}$, the Yoneda lemma represents $\seq[\sort
  \in {\Fst[\gsort]}]{\sem{x : s}_{\mP}}$ via a natural transformation:
\[
\var : \Neut \to \mP
\qquad
\var[{\sort;\Context}]x \definedby
\sem{x:\sort}_{\mP}\bracks{[x:\sort]\xfrom{\projection_x}\Context}\in\mP_{\sort}\Context
\qquad
\sem{x:\sort}_{\mP} \definedby \var[{\sort;[x:\sort]}]x
\]

In \mast{} we follow the same technique. We start by organising these
presheaves into a monoidal category of their own.
Let $\catMon = \seq{\carrier{\catMon}, (\tensor), \mNeut, \assoc,
  \ellunit, \runit}$ be a monoidal category. Recall the
\emph{coslice} category $\Pointed{\carrier{\catMon}} \definedby \mNeut/\carrier{\catMon}$:
\begin{wrapfigure}[3]{r}{2cm}
\diagram{pointed-homomorphism}
\end{wrapfigure}
\begin{itemize}
\item \emph{pointed objects} $A$: pairs $
  \pair{\carrier A}{\var[A] : \mNeut \to A}$ consisting of an object
  $\carrier A$ in $\carrier{\catMon}$ and a $\carrier{\catMon}$-arrow $\var[A] : \mNeut \to A$
   called the \emph{point}; and
\item arrows $f : A \to B$ are point-preserving $\carrier{\catMon}$-morphisms $f : \carrier A
  \to \carrier B$ (cf.~diagram on right).
\end{itemize}

The forgetful functor $\carrier- : \Pointed{\carrier\catMon} \to \carrier\catMon$
sending pointed objects and morphisms to their underlying objects and
morphisms is faithful. The tensor product $(\tensor)$ lifts along
$\carrier-$ to the following \emph{pointed} tensor:
\[
(\ptensor) : \Pointed{\carrier\catMon}
\times \Pointed{\carrier\catMon} \to \Pointed{\carrier\catMon}
\quad
\carrier{A \ptensor B} : \carrier A \tensor \carrier B
\quad
\var[A \ptensor B] \definedby \mNeut \xto{\runit'} \mNeut\tensor \mNeut
\xto{\var[A]\tensor \var[B]} \carrier A \tensor \carrier B
\quad
(A \xto f A')\ptensor(B \xto g B') \definedby f\tensor g
\]
This definition for $f \ptensor g$ relies on a simple
point-preservation argument (cf.~\appsubsecref[C.1]{monoidal proofs: pointed tensors}).

The initial pointed object is given by $\mNeut$ equipped with its
identity: $\pinitial[\mNeut] \definedby \pair\mNeut{\id[\mNeut]} \in
\Pointed{\carrier\catMon}$. The unique point-preserving morphism to any pointed
object $A$ is given by the point: $\coseq{} \definedby \var[A]:
\pinitial[\mNeut] \to A$.

The pointed objects inherit the monoidal structure from $\catMon$
(see \appsubsecref[C.1]{monoidal proofs: pointed tensors} for the proof):
\begin{proposition}[Fiore~\cite{fiore:soas}]\proplabel{pointed tensor}
  Every monoidal category $\catMon$ yields a
  monoidal category of pointed objects
  \(
  \Pointed\catMon \definedby \seq{\Pointed{\carrier\catMon}, (\ptensor), \pinitial[\mNeut], \assoc, \ellunit, \runit}
  \).
  I.e., the mediators preserve points, hence lift to
  $\Pointed\catMon$.
\end{proposition}

\begin{example}
  Let $\pA \in \Pointed{\Syn\gsort}$ be a pointed structure. Its point
  $\var[\pA]$ is uniquely determined, through the Yoneda lemma, by the
  tuple of variable interpretations $\seq[\sort \in \Fst]{\sem{x
  : \sort}_{\pA} \definedby (\var[\pA])_{\sort;[x:\sort]}x}$.  Given
  any other pointed structure $\pB$, the variable interpretation for
  $\pA\ptensor\pB$ is $\sem{x : \sort}_{\pA\ptensor\pB}
  = \bracks{\sem{x}_{\pA}, \seq{x : \sem{x}_{\pB}}}_{[x : \sort]}$.
\end{example}

\begin{example}\exlabel{pointed action}
  If $\Cat A$ is a $\catMon$-actegory, then the monoidal
  category of pointed objects $\Pointed\catMon$ acts on $\Cat A$ by:
  \(
  (a \mathbin{\PointedAct\ract} A) \definedby (a \ract \carrier A)
  \).
  The $\Pointed\catMon$-actegory axioms hold straightforwardly. We
  denote this $\Pointed\catMon$-actegory by $\PointedAct{\Cat A}$.
  In particular, we have an $\Pointed{\Syn\gsort}$-actegory
  $\PointedAct{\Syn\gsort}$ given by the $\gsort$-structures.
\end{example}

Let $\Cat A, \Cat B$ be two $\catMon$-actegories,
and consider any functor $F : \carrier{\Cat A} \to \carrier{\Cat
  B}$. A \emph{tensorial strength} for $F$ from $\Cat A$ to $\Cat B$ is a
natural transformation $\strength : (F a) \ract_{\Cat A} b \to F (a
\ract_{\Cat B} b)$ satisfying the two axioms:
\[
\diagram*{strength-pentagon}\qquad
\diagram*{strength-triangle}
\]
A \emph{strong functor} $F : \Cat A \to \Cat B$ is a functor
$\carrier F : \carrier{\Cat A} \to \carrier{\Cat B}$ equipped with a
strength, $\strength_F$, for $\carrier F$ from $\Cat A$ to $\Cat B$.

The strong functors w.r.t.~the pointed monoidal structure are central
to the modularity of both the classical theory and \mast{}. As we will
soon see, the strength of the signature functor will provide the
structure that allows us to propagate syntactic and semantic
substitutions through each term constructor.

\begin{definition}
An \emph{$\gsort$-signature functor} is an
$\Pointed{\Syn\gsort}$-strong functor:
\(
{\Sig : \PointedAct{\Syn\gsort}
\to \PointedAct{\Syn\gsort}
}
\).
\end{definition}

Each signature functor
explains, through its strength, how to propagate substitutions to its
subterms:
\begin{longexample}\exlabel{cbv abst val app}
  The strength for the abstraction signature~(\exref{abstraction sig}):
  \begin{gather*}
  \displaystyle\AbsSig \definedby
  \Variant{(\sabst{x : \ty}) : \OnlyAt{\FUN{\ty[1]}{\ty[2]}}
    [x : \ty]\Shift X\At \ty \suchthat \ty[1],\ty[2] \in \SimpleType}
  \\
  \strength[\AbsSig]_{\mP,\pA;\FUN{\ty[2]}{\ty[3]}, \Context}
  \bracks{\sabst{x : \ty}t, e}_{\Context[2]} \definedby
  \sabst{x : \ty[2]}\bracks{t, e\bracks{\Context \xfrom{\projection_1} \Context\++[x : \ty[2]]}\++(\var[\pA])_{\Context \++ [x : \ty[2]]}x}_{\Context[2]\++[x : \ty[2]]}
  \end{gather*}
  Thus, to propagate the environment $e$ under the binder, we \emph{weaken} it.
  The strengths for the other \CBV{} signature functors from \exrefs\exref*{valsig} and~\exref*{appsig} propagate the environment as it is:
  \[
  \strength[\ValSig]\bracks{\val t, e}_{\Context[2]} \definedby \val\bracks{t,e}_{\Context[2]}
  \qquad
  \strength[\AppSig]\bracks{\sapply{t_1}{t_2}, e}_{\Context[2]} \definedby
  \sapply{\bracks{t_1,e}_{\Context[2]}}
         {\bracks{t_2,e}_{\Context[2]}}
  \]
  These rules are identical to the syntactic rules one may use to
  define capture-avoiding substitution, but crucially the environment
  $e$ is taken from any pointed presheaf, not just the presheaf of
  abstract syntax. It is this additional generality that allows the
  classical theory and \mast{} to combine signatures modularly, as
  well as account for both syntactic and semantic substitution.
\end{longexample}

It is straightforward to show the transformations in the previous
example are strengths from first principles. However, this fact follows
compositionally. We will spend the remainder of this section building
up the machinery to derive this strength compositionally.

\begin{example}
  Recall the \emph{scope shift} combinator $(\Shift\Context{}) :
  \Syn\gsort \to \Syn\gsort$ that we use to describe binding
  constructs, such as abstraction. We define its strength \(
  \strength[\Context\Shift{}]_{\mP,\pA} : (\Context\Shift \mP)\prmul \pA \to
  \Context\Shift(\mP\prmul \pA) \) by:
  \begin{multline*}
  \strength[\Context\Shift{}]_{\sort;\Context[3]}\bracks{t \in \mP_{\sort}(\Context[2]\++\Context[1]),e \in \Env{\carrier\pA}{\Context[2]}\Context[3]}_{\Context[2]}
  \definedby\\
  \bracks{
     t\in\mP_{\sort}(\Context[2]\++\Context[1])
    , \seq[{(x : \sort[1]) \in \Context[2]}]%
            {e_x\bracks{\Context[3]\xfrom{\projection_1}\Context[3]\++\Context}}
       \!\!\++\!
      \seq[{(y : \sort[2]) \in \Context[1]}]%
          {\sem y_{\pA}\bracks{[y : \sort[2]]
             \xfrom{\projection_{y[\projection_2]}} \Context[3]\++\Context}}
     \in\Env{\carrier\pA}{\Context[2]\++\Context[1]}(\Context[3]\++\Context[1])
     }_{\Context[2]\++\Context}
   \end{multline*}
   I.e., in order to propagate the environment under a scope-shift, we extend it
   by the environment that sends every variable $(y:\sort[2])\in \Context[2]$
   to its representation $\sem y$ in $\pA$, suitably weakened into the context
   $\Context[3]\++\Context[1]$.
\end{example}

When the operators in a signature do not bind variables themselves but
are merely compatible with binding, we can exhibit a strength
w.r.t.~the simpler action $(\rmul)$ thanks to the following result,
 which appears implicitly in Fiore's~\cite{fiore:soas} work (see
\appsubsecref[C.2]{pointed action from mere action} for the straightforward proof):
\begin{lemma}\lemmalabel{pointed tensorial strength}
  Let $\catMon$ be a monoidal category; $\Cat A$,$\Cat B$ two
  $\catMon$-actegories; and $F : \Cat A \to \Cat B$ a strong functor. Then
  the following morphisms exhibit $\carrier F$ as a strong functor:
  $\PointedAct F : \PointedAct{\Cat A} \to \PointedAct{\Cat B}$.
  \[
  \pstrength F_{x,a} : (\carrier F x)\PointedAct\ract a
  = (\carrier F x)\ract \carrier a
  \mathrel{\smash{\xto{\strength[F]_{x,\carrier a}}}} \carrier F(x \ract \carrier a)
  = \carrier F(x \PointedAct\ract a)
  \]
\end{lemma}

The next few examples all use this lemma to derive strengths.

\begin{example}
  Recall the restriction combinator $(\restrict{}I) :
  \Syn\gsort \to \PSh+{I\times{\RenCat{{\Fst}}}}$ and its
  specialisation that projects out a single sort $(\At\sort_0) : \Syn\gsort \to
  \PSh+{{\RenCat{{\Fst}}}}$ (where $\sort_0 \in \Sort$) that we use it to signify a sub-term
  of sort $\sort[1]_0$. Define
  $\strength[\restrict{}I] : \restrict\mP I\rmul \mQ \to
  \restrict{(\mP\rmul \mQ)}I$ by
  \(
  \strength[\restrict{}I]_{\sort,\Context}\bracks{t \in \mP_{\sort_0}\Context[2], e \in \Env{\mQ}{\Context[2]}\Context}_{\Context[2]}
  \definedby
  \bracks{t,e}_{\Context[2]}
  \), and derive a strength for the projection $\strength[\At \sort_0] : (\mP
  \At\sort_0)\rmul \mQ \to (\mP\rmul \mQ) \At\sort_0$ from it.
  Similarly, recall the \emph{extension} combinator
  $\OnlyAt{I} : \PSh{I\times{\RenCat*{\Fst[\gsort]}}}\to \Syn\gsort$
  that lets us construct nodes at a specific subset of sorts $I
  \subset \Sort\gsort$. Then $((\OnlyAt{I} \mP)\rmul \mQ)_{\sort} =
  \initial$ when $\sort \notin I$, and so we can define the
  strength $\strength[\OnlyAt I]_{\mP;\mQ} : (\OnlyAt\mP)\rmul \mQ \to \OnlyAt{\mP\rmul\mQ} $ vacuously in those sorts:
  \[
  \smash{
    \strength[\OnlyAt I]_{\mP;\mQ;\sort[1]} : ((\OnlyAt\mP)\rmul \mQ)_{\sort[1]} = \initial \xto{\coseq{}} \OnlyAt I (\mP\rmul\mQ)_{\sort[1]}
    \qquad
    \strength[\OnlyAt I]_{\mP;\mQ;\sort[2]} : ((\OnlyAt\mP)\rmul \mQ)_{\sort[2]} = (\mP\rmul \mQ)_{\sort[2]} = \OnlyAt I (\mP\rmul\mQ)_{\sort[2]}
  }
  \tag{$\sort[1] \notin I \ni \sort[2]$}
  \]
\end{example}

\begin{longexample}
  Let $\seq[i \in I]{\Cat A_i}$ be a family of $\catMon$-actegories.  The
  projection functors $\projection_j : \prod_{i \in I}\Cat A_i \to
  \Cat A_j$ have the identity as a strength
  $\strength[\projection_j]_{\seq[i]{x_i},a} = \id[x_i\ract_i a]$
  since:
  \(
  \parent{\projection_j \seq[i]{x_i}}\ract_i a
  = x_j \ract a = \projection_j{\seq[i]{x_i}\ract a}
  \).
  For a $\catMon$-actegory $\Cat B$ and a family of strong functors
  $F_i : \Cat B \to \Cat A_i$, the tupling functor $\seq[i]{\carrier F_i} :
  \Cat B \to \prod_i\Cat A_i$ has the following $\prod_i\Cat A_i$-morphism as
  strength:
  \(
  \strength[{\seq[i]{F_i}}]_{\seq[i]x,a} \definedby
  \seq[i]{(F_ix) \ract a \xto{\strength[F_i]} F_i(x \ract a)}
  \).
\end{longexample}

\begin{example}
  Recall the $I$-ary sums $\coprod_I : \Cat C^I \to \Cat C$ and $J$-ary
  products $\prod_J : \Cat C^J \to \Cat C$ that let us alternate between
  $I$-indexed operator nodes and include $J$-ary branching factor in
  categories with $I$-coproducts/$J$-products. The product has strength, and
  coproducts, if they distribute over $(\rmul)$, also have a strength:
  \[
  \strength[\prod_J] : \parent{\prod_{j \in J} x_j} \ract a \xto{\seq[j \in J]{\projection_j \ract \id}} \prod_{j \in J}(x_j \ract a)
  \qquad
  \strength[\coprod_I] : \parent{\coprod_{i \in I} x_i} \ract a
  \xto{\inv-{\coseq[i \in I]{(i:) \ract \id}}} \coprod_{i \in I}(x_i \ract a)
  \]
  When $\Cat C = \Syn\gsort$, the substitution tensor $(\rmul)$
  distributes over coproducts (cf.~\propref{heterogeneous structures right-closed}), and so the following pointwise formulae
  describe these strengths:
  \[
  \strength[\prod_J]_{\vec \mP, \mQ}\bracks{t, e}_{\Context[2]} \definedby \seq[j \in J]{\bracks{t_j, e}_{\Context[2]}}
  \qquad
  \strength[\coprod_I] : \bracks{(i : t), e}_{\Context[2]} \definedby (i : \bracks{t, e}_{\Context[2]})
  \]
\end{example}

As is well-known, strong functors compose, and their combined strength is given by
(see \applemmaref[C.2]{strengths compose}):
\[
\strength[G \compose F]_{x,a} : (\carrier{G F} x) \ract a \xto{\strength[G]} \carrier G(\carrier F
x \ract a) \xto{\carrier G\strength[F]} \carrier{GF} (x \ract a)
\]
This fact and the signature combinators covers all our
examples. E.g., cf.~the strength from \exref{cbv abst val app}.

%% file: compatibility.tex
\section{Compatible actions and structures}
\seclabel{compatible monoids} We define the substitution structure of
interpretations of the syntax that ensures they are compatible with
the operations in the signature, culminating in the Special
Representation \thmref{representation}, concluding the
tutorial. Actions also give additional perspectives
on the compatibility condition (\subsecref{new perspective}).
\begin{definition}\deflabel{compatibile structures}
  Let $\Sig$ be an $\gsort$-signature functor,
  $\Model = \triple{\carrier\Model}{-[-]_{\Model}}{\var[\Model]}$  a substitution structure.
  We say that an $\Sig$-algebra $\sem- : \carrier{\Sig\Model}\to\carrier\Model$ is \emph{compatible} with $\Model$, when:
  \diagram{compatible-monoid}
  An \emph{$\Sig$-compatible
  substitution sturucture} $\Model =
  \seq{\carrier\Model, -[-]_{\Model}, \var[\Model], \sem-_{{\Model}}}$, or \emph{$\Sig$-structure} for short,
  consists of a structure
  $\Model = \triple{\carrier\Model}{-[-]_{\Model}}{\var[\Model]}$ and an
  $\carrier\Sig$-algebra $\sem-_{\Model} : \carrier{\Sig\Model} \to
  \carrier\Model$, compatible with $\Model$.
\end{definition}

\begin{example}\exlabel{lam-compatibility}
  Recall the substitution structure $\Model$ for \CBV{} in a Cartesian-closed
  category $\Cat C$ (\exref{pure cbv
    model}). The interpretation of the \CBV{} \lamcalc{}
  equips it with a $\AbsSig$-, $\ValSig$-, and $\AppSig$-algebra structures:
  \begin{gather*}
  \sem{\sabst{x : \ty}}_{\Model}\parent{\sem{\Context, x : \ty} \xto f \sem{\ty[2]}}
  \definedby \parent{\sem{\Context} \xto{\curry f} \sem{\ty[2]}^{\sem{\ty[1]}}}
  \qquad
  \sem{\val}_{\Model}\parent{\sem{\Context}\xto f\sem{\ty[1]}}
  \definedby f
  \\
  \sem{\sapply{\parent{\sem{\Context} \xto f \sem{\ty[2]}^{\ty[1]}}}
                    {\parent{\sem{\Context} \xto a \sem{\ty[2]}}}}_{\Model}
  \definedby\parent{
  \sem{\Context} \xto{\pair fa} \sem{\ty[2]}^{\sem{\ty[1]}}\times \sem{\ty[1]} \xto{\eval} \sem{\ty[2]}}
  \end{gather*}
  The compatibility axiom for $\AbsSig$ amounts to the following equation, for each
  $\msubst  \in \Cat C(\sem\Context, \sem{\Context[2]}) \isomorphic \Env\Model{\Context[2]}\Context$:
  \begin{gather*}
  \curry\parent{\sem{\sabst{x : \ty}}f}\compose \seq[y]{\msubst{}_y} =
  \curry\parent{\sem{\Context}\times\sem\ty \xto{\seq[y]{\msubst{}_y}\times \id} \sem{\Context[2]} \xto{f} \sem{\ty[2]}}
  \end{gather*}
  It follows from the naturality $\curry$. The compatibility axioms
  for $\ValSig$ and $\AppSig$ hold immediately.
\end{example}

We combine compatibility conditions from fragments to the whole $\CBV$ language (cf.~\appsubsecref[C.3]{coprod compatible monoid} for the proof):
\begin{lemma}\lemmalabel{coprod compatible monoid}
  Let $\Monoid$ be a substitution structure and
  $\seq[i \in I]{\Sig_i}$ and $\seq-[i \in
    I]{\sem-_i : \carrier{\Sig_i\Monoid} \to \carrier\Monoid}$
  be families of $\gsort$-signature
  functors and algebras for them. The cotupled algebra:
  \(
  \coseq[i \in I]{\sem-_i} : \coprod_{i \in I}\carrier{\Sig_i\Monoid} \to \carrier\Monoid
  \)
  is compatible with $\Monoid$ iff every algebra $\Sig_i$ is
  compatible with $\Monoid$.
\end{lemma}

An $\Sig$-substitution structure \emph{homomorphism} $h : \Model[1] \to \Model[2]$ is a
morphism $h : \carrier{\Model[1]} \to \carrier{\Model[2]}$ that is
both an $\Sig$-homomorphism and an action homomorphism in the sense of \figref{sig substitution homomorphism}.
\begin{figure}
\begin{center}
  \diagram*{algebra-homo}\hfill
  \diagram*{unit-preservation}\hfill
  \diagram*{mult-preservation}
\end{center}
\caption{Definition of an $\Sig$-substitution structure homomorphism $h : \Model[1] \to \Model[2]$}
\figlabel{sig substitution homomorphism}
\end{figure}
Let $\Hole \in
\Syn\gsort$ be an $\gsort$-structure, whose elements we think of as
holes. An \emph{$\Sig$-structure over $\Hole$} is a pair
$\pair\Model\metaenv$ consisting of a $\Sig$-structure $\Model$ and a
morphism $? : \Hole \to \carrier\Model$, which we will call the \emph{metavariable interpretation} map. A morphism of
$\Sig$-structures $h : \pair-{\Model[1]}{?} \to
\pair-{\Model[2]}{{?}}$ over $\Hole$ is an $\Sig$-structure
homomorphism $h : \Model[1] \to \Model[2]$ preserving the metavariable
interpretation. We let $\kNeut \definedby \encode\Neut{\initial}$ be
the $\Syn\gsort$-structure of variables: its first-class fragment is
the presheaf of variables, and its second-class sets
are empty.

\begin{theorem}[special representation]\thmlabel{representation}
  Let $\Sig$ be an $\gsort$-signature functor, and $\Hole$ an
  $\gsort$-structure. Consider any initial algebra
  \(
  \AST[\Sig]\Hole = \encode{\AST[\Sig]_{\fst}\Hole}{\AST[\Sig]_{\snd}} \definedby \fix X. (\carrier\Sig X)\amalg \kNeut \amalg (\Hole\rmul\rest X)
  \)
  given by:
  \[
  \sem- : \carrier\Sig (\AST[\Sig]\Hole) \to \AST[\Sig]\Hole
  \qquad
  \var : \Neut \to \AST[\Sig]_{\fst}\Hole
  \qquad
  \meta-[-] : \Hole\tensor\AST[\Sig]_{\fst}\Hole \to \AST[\Sig]\Hole
  \]
  There is a unique morphism $-[-] :
  \AST[\Sig]\Hole \tensor \AST[\Sig]_{\fst}\Hole \to \AST[\Sig]\Hole$, called \emph{simultaneous substitution},
  satisfying:
  \begin{center}
    \diagram*{meta-subst-ops}\hfill
    \diagram*{meta-subst-vars}\hfill
    \diagram*{meta-subst-meta}
  \end{center}
  \noindent{}Equipping
  $\Free[\Sig]\Hole \definedby \seq{\AST[\Sig]\Hole, -[-], \var, \sem-}$ with
  $? : \Hole \xto{\runit'} \Hole\rmul\Neut \xto{\id\tensor\var} \Hole\rmul\AST[\Sig]\Hole \xto{\meta-[-]} \AST[\Sig]\Hole$,
  yields the
  free $\Sig$-substituion structure over $\Hole$.
  We call morphisms $\msubst : \Hole_1 \to \AST[\Sig]\Hole_2$ \emph{simple metavariable substitutions}, and we call their Kleisli extension $-[\msubst]: \AST[\Sig]\Hole_1 \to \AST[\Sig]\Hole_2$ \emph{metavariable substitution by $\msubst$}.
\end{theorem}

This theorem lets us prove substitution lemmata
wholesale~(cf.~\lemmaref{substitution}), concluding the tutorial.

\subsection{On pointed strengths and compatibility}
\subseclabel{new perspective}
The compatibility condition \defref{compatibile structures} is
technical, involving all \mast{} components.
It deserves to, but often does not, take the centre spot in
accounts of this theory.  In concrete examples, one can see
how these components contribute to the modularity of the theory. Each
syntactic construct, specified through a signature functor $\Sig_i$,
carries its own strength. Each model exhibits a substitution
structure, as well as an interpretation for each of these functors
through an $\Sig_i$-algebra structure. Compatibility expresses a
semantic condition on the interaction between semantic substitution
and the semantics of each $\Sig_i$ in isolation, without referring to
the rest of the syntax. We will now give a few more results, brought
about thanks to the action-based perspective, that also explain the
abstract role strength and compatibility have.

First, we relate substitution actions with the pointed monoidal
structure. The difference between monoids in a monoidal
category $\catMon$ and monoids for its associated pointed monoidal
category $\Pointed\catMon$ is the point of the monoid and the
point-preservation of the unit and multiplication. This difference
is irrelevant:
\begin{proposition}\proplabel{pointed monoids and actions coincide}
Let $\catMon$ be a monoidal category. The forgetful functor $\carrier-
: \Pointed\catMon \to \catMon$ lifts to an isomorphism between their
categories of monoids:
\[
  \MonoidCat{\Pointed\catMon}
  \mathrel{
    \raisebox{-.75ex}{$\xfrom[\Pointed{-}]{\xto{\carrier-\mspace{20mu}}\mspace{-10mu}}$}
  }
  \MonoidCat\catMon
  \qquad
  \carrier{\Monoid} \definedby
  \triple{\carrier{\carrier\Monoid}}{\carrier{-[-]_{\Monoid}}}{\carrier{\var}}
  \quad
  \carrier h\definedby h
  \qquad
  \Pointed{\Monoid} \definedby
  \triple{\pair{\carrier\Monoid}{\var}}{-[-]}{\var}
  \quad
  \Pointed h \definedby h
\]
Given a $\catMon$-actegory $\catAct$, the categories
of $\Monoid$-actions in $\catAct$ and $\Pointed\Monoid$-actions in
$\PointedAct\catAct$ are also isomorphic.
\end{proposition}
The proof in \appsubsecref[C.3]{pointed monoids and actions} is
straightforward. For monoids, we show that the pointed structure of a
pointed monoid must be its unit, and validating the point preservation
axioms. The proof for actions is immediate.

Next, we explicate the abstract role the pointed strength serves,
proved by direct calculation~(cf.~\appsubsecref[C.3]{lifting signature functors to monoids}):
\begin{proposition}\proplabel{lifting signatures to monoids}
  Let $\catMon$ be a monoidal category, $\catAct$ be
  $\catMon$-actegory, and $\Sig : \PointedAct\catAct \to \PointedAct\catAct$ a
  pointed-strong functor. For every $\catMon$-monoid, $\Sig$ lifts to the following
  functor over
  $\Monoid$-actions:\[
  \Sig_{\Monoid} : \ActionCat\Monoid\catAct \to \ActionCat\Monoid\catAct
  \quad
  \Sig_{\Monoid}\Action \definedby
  \carrier{\Sig_{\Monoid}\Action} \definedby \carrier{\Sig\Action}
  \quad
  -[-]_{\Sig_{\Monoid}\Action} : \carrier{\Sig\Action} \tensor \Monoid
  \xto{\strength_{\carrier\Action,\var}}
  \carrier{\Sig}\parent{\carrier\Action \tensor \carrier\Monoid}
  \xto{-[-]_{\Action}}
  \carrier\Action
  \]
\end{proposition}
Putting these two results together gives a new perspective on the
pointed strength and on compatible algebras. The monoidal category of
homogeneous structures $\Syn\Fst$ acts on the category of
heterogeneous structures $\Syn\gsort$, and so $\Pointed{(\Syn\Fst)}$
acts on $\PointedAct{\Syn\gsort}$. Given a substitution structure
$\SubStruct$, the following process produces an
$\Pointed{(\Syn\Fst)}$-action on $\SubStruct$ in
$\PointedAct{\Syn\gsort}$.  Moving to the first-, and second-, order
fragments, we get a substitution action
$\pair{\rest\SubStruct}{\restrict\SubStruct\snd}$. The monoid
$\rest\SubStruct$ acts on itself and on $\restrict\SubStruct\snd$, and
therefore the monoid $\Pointed-{\rest\SubStruct}$ acts on both
$\rest\SubStruct$ in $\PointedAct{\Syn\Fst}$ and
$\restrict\SubStruct\snd$ in $\PointedAct{\SndSyn\gsort}$. Pairing
these up as gives an $\Pointed-{\rest\SubStruct}$-action on
$\pair{\rest\SubStruct}{\restrict\SubStruct\snd}$ in
$\PointedAct{\Syn\Fst}\times\PointedAct{\SndSyn\gsort}$.  Combining
them through the isomorphism $\encode-- :
\Syn\Fst\times\SndSyn\gsort\xto\isomorphic\Syn\gsort$ exhibits an
$\Pointed-{\rest\SubStruct}$-action on $\SubStruct$ in
$\PointedAct{\Syn\gsort}$.
Let $\sem- : \Sig\SubStruct \to \SubStruct$ be any $\Sig$-algebra
structure. It is compatible precisely when it is an
$\rest\SubStruct$-action homomorphism $\sem- :
\Sig_{\rest\SubStruct}\SubStruct \to \SubStruct$, recasting
compatibility as substituion-preservation in a precise way.

%% file: case-studies.tex
\section{Case study: the Call-by-Value \lamcalc{}}
\seclabel{case studies}
We use algebraic signatures to modularly describe extensions to the
\CBV{} type system, and signature functors to modularly describe
extensions to its terms. Starting with a semantics based on strong
monads, we extend a basic calculus with sequential composition, functions, products, coproducts, an inductive
datatype of natural numbers, iteration,
and recursion. These require
\(
7
\) checks that each additional bunch of semantic definitions is well-defined and compatible with
the substitution structures in the corresponding algebra.
In return, the \mast{} theory lets us deduce $2^7 = 128$ different
substitution lemmata, for each language fragment. Note that not all
models can interpret each fragment, and so one cannot deduce the substitution
lemma for a fragment from the lemma for the full language.
The development is similar in spirit to Swierstra's \alacarte{}
methodology~\cite{swierstra:data-types-a-la-carte-journal,forster-stark:coq-a-la-carte},
but it also provides semantic substitution lemmata.  We use this
opportunity to also summarise the standard denotational semantics for
these features.

\subsection{The full calculus}
\figref{cbv syntax} presents the abstract syntax of all the features
we will consider without explicating their typing judgements nor their
binding structure. \figref{cbv type system} presents the typing
judgements. Our base calculus includes a construct for sequencing,
which evaluates the intermediate results in order, binding them to
variables. Extending the calculus with records adds tuples of labelled
fields, which we eliminate with a pattern matching
construct. Extending the calculus with variants adds tagged sums,
which we eliminate with a pattern matching construct. Extending the
calculus with natural numbers adds natural number literals as values,
the iso-recursive constructor $\Roll$ and deconstructor $\Unroll$, and
a bounded iteration eliminator. We further extend the calculus with
unbounded iteration $\ForLoop i\M\N$, which initialises $i$ to $\M$,
and then iterates $\N$ until it is $\Done{}$. Finally, we extend the
calculus with higher-order recursion through the $\LetRecName{}$
construct. It extends the body's ($\N$) context with $n$
mutually-recursive functions $f_1, \ldots, f_n$. Here we assume some
predefined mapping from contexts $\Context$ and tuple-types
$\sRecord\Context$, identifying each position $(x : \ty) \in \Context$
with a label $\syntax x$. For example, sending its position to its
corresponding numeral.

\subseclabel{cbv-development}
\begin{figure}
  \input{cbv-type-system}
  \caption{Type system of \CBV{}, omitting analogous rules for value records and variants.}
  \figlabel{cbv type system}
\end{figure}

\subsection{Simple types \alacarte}

To develop these calculi and their models fully modularly, we
first need to treat their sets of types modularly.  We will use the
classical \alacarte{} methodology, for initial algebra semantics in
$\Set$, to mix and choose the collection of simple types we work with
in each case. We will typically work with respect to an ordinary signature
functor $\TySig : \Set \to \Set$ that has an initial algebra $\TySig
(\fix\TySig) \to \fix \TySig$.  This functor specifies the signature
for the simple types given by $\Type_{\TySig} \definedby \fix \TySig$.  For any
set $\Type$, whether inductively given by such a signature functor or
not, define the sorting system $\CBVSort[\Type]$ as the coproduct diagram:
\[
\Fst[{\CBVSort[\Type]}] \definedby
\Type \subset
\Sort[{\CBVSort[\Type]}] \definedby \set{\ty, \Comp \ty \suchthat \ty \in \Type}
\xfrom{\Comp}
\Type \defines \Snd[{\CBVSort[\Type]}]
\]
We will also work with respect to a $\mast{}$
$\CBVSort[\Type]$-signature functor $\Sig$, and define an
$\Sig$-monoid, which by the Special Representation
\thmref{representation} satisfies the Substitution \lemmaref{substitution}.

\begin{example}
In the simplest case, all we have are base types. The signature
functor for types is the constant functor \( \BaseTySig : \Set \to
\Set \), \( \BaseTySig X \definedby \Base \). It has the identity
function $\id : \Base \to \Base$ as its initial algebra, thus $\Base =
\fix\BaseTySig \defines \Type$. Summarising, the set of types
is the set of base types.
\end{example}

\begin{example}\exlabel{stlc types initiality}
  To accommodate function types, take the functor $\FunTySig \definedby X \mapsto X\times X : \Set
  \to \Set$. Then:
  \[
  \BaseTySig \amalg \FunTySig X \isomorphic
  \Variant{\base \suchthat \base \in \Base}\amalg\Variant{(\FUN{}{}) : X\times X}
  \]
  Its initial algebra is $\Type$, with
  $\sem- : (\BaseTySig \amalg \FunTySig)\Type \to \Type$
  where
  \(
  \sem{\base} \definedby \base
  \); \(
  \sem{(\FUN{}{})}\pair{\ty[1]}{\ty[2]} \definedby \FUN{\ty[1]}{\ty[2]}
  \).
\end{example}

\begin{example}\exlabel{simple rows}
  Next, we deal with records and variants uniformly.
  Let $X$ be a set. A \emph{row} in $X$ is a function $\seq[i \in
    I]{\Cons{C_i} \mapsto x_i}$ from a finite set of \emph{field/constructor labels}
  $\set{\Cons{C_i} \suchthat i \in I}$ to $X$.
  Letting $\FieldLabel$ be the set of field labels, define $\RowSig : \Set \to \Set$ by \(\RowSig X \definedby
  \coprod_{I \finsubset \FieldLabel} X^I
 \).
  The functors for record and variant types are:
  \(
  \RecordTySig, \VariantTySig \definedby \RowSig : \Set \to \Set
  \).
\end{example}

The collection of types $\CbvType[\Base]$ defined inductively in
\figref{cbv syntax} is the initial algebra for the functor:
\[
\FullCBVTySig \definedby
\BaseTySig
\amalg((\FUN{}{}):\FunTySig)
\amalg(\sRecord-:\RecordTySig)
\amalg(\sVariant-:\VariantTySig)
\amalg(\Nat : \NatTySig)
: \Set \to \Set
\]
We want to work with more than $2^4 = 16$ collections
of types, given by various restrictions of this collection, depending
on the typing needs of each language fragment.  As a running example,
consider the term former for higher-order recursive definitions. For
each function signature $f[x_1:\ty_1, \ldots, x_n : \ty_n] \sof
\ty[2]$, we will need to identify the function type
$\FUN{\sRecord{x_1:\ty_1, \ldots, x_n : \ty_n}}{\ty[2]}$.  In fragments
that contain both records and all function types, this construct will
need to identify the compound type $\FUN{\sRecord{x_1:\ty_1, \ldots,
    x_n : \ty_n}}{\ty[2]}$. But in fragments that only contain
higher-order recursion without records, we will fuse function
application with record creation.  To allow such flexibility, we use
the following concepts.  We define a \emph{typing need} to be a
signature functor $\TySig[2] : \Set \to \Set$.  For higher-order
recursion, we use the typing need $\RecReq X \definedby
\RenCat X\times X$.  A \emph{fulfillment} of a
typing need $\TySig[2]$ in a set $\Type$ is then a relation $(\models)
: \TySig[2] \Type \rto \Type$, which we will write as $k \models \ty :
\TySig[2]$ for every $k \in \TySig[2](\Type)$ and $\ty \in \Type$ satisfying
$\pair k\ty \in (\models)$.
A fulfillment explains which types
satisfy the typing need. For example, the fullfillment of the typing need
for higher-order recursion $\RecReq$ in the set of types for the full
calculus is given by the relation defined by:
\[
\pair{[x_1 : \ty[1]_1, \ldots, x_n : \ty[1]_n]}{\ty[2]}\models
\parent{\FUN{\sRecord{\Cons{x_1} \sof \ty_1, \ldots, \Cons{x_n}\sof \ty_n}}{\ty[2]}}
: \RecReq
\tag{$\ty[1]_1, \ldots, \ty[1]_n, \ty[2] \in \CbvType[\Base]$}
\]
Note that we apply two type constructors in this fulfillment: the
function type constructor $(\FUN{}{})$ and the record type constructor
$(\sRecord-)$. In the fulfillment for fragments that include
higher-order recursion but not records, we will fulfill this need by
the single type-constructor for $n$-ary functions
$(\FUN{\sRecord{-}}-)$ that these fragments include.
We will use the following typing needs, and only them, in our case study:
\begin{itemize}
\item When we need
  the unit type, we impose the typing need
  $\EmptyRecordTySig \definedby \terminal \definedby \set{\star}$, fulfilled by
  the empty record type, when the fragment includes records,
  and by the unit type otherwise:
  $\star \models \sRecord{\mathvisiblespace} : \EmptyRecordTySig$.
\item When we need to deconstruct natural numbers, we impose
  the typing need:
  \[
  \NatConstTySig \definedby (\Nat : \terminal)\amalg
  (\Cons\Maybe : \Id)
  \]
  We will fulfill it in fragments with the natural numbers and bounded
  iteration over them by specifying the natural number type and an
  option type for each type:\[
  \Nat \star \models \Nat : \NatConstTySig
  \qquad
  \Maybe\ty \models \sVariant{\Cons 0 \sof \sRecord\mathvisiblespace, \Cons{(1+)}
    \sof \ty} : \NatConstTySig
  \tag{$\ty \in \Type$}
  \]
  When the fragment doesn't include all variant or record types, we
  will ensure it includes the single type-constructor $\sVariant{\Cons
    0 \sof \sRecord\mathvisiblespace, \Cons{(1+)}\sof-}$.
\item To type the bodies of while-loops, we impose the need
  $\NatConstTySig \definedby \Record{\Done: \Id, \Cont : \Id}$
  fulfilled by a binary variant type:
  \[
  \pair{\Done : \ty[1]}{\Cont : \ty[2]} \models
  \sVariant{\Done \sof \ty[1], \Cont \sof \ty[2]} :
  \NatConstTySig
  \tag{$\ty[1], \ty[2] \in \Type$}
  \]
\item To type recursive function definitions, we impose $\RecReq X
  \definedby \RenCat X\times X$, our running example.
\end{itemize}

\subsection{The substitution structures}
\subseclabel{cbv subst structure main}
Let $\Type$ be a set whose elements represent types. A \emph{strong-monad
model} $\seq{\Cat C, \sem-, \Monad}$ consists of:
\begin{itemize}
\item A locally-small Cartesian category $\Cat C$ with chosen finite products.
\item An interpretation function $\sem- : \Type \to \Cat C$.
\item A strong monad~\cite{kock:monads-on-sym-mon-closed-cats}
  $\Monad$ over $\Cat C$, i.e.,
  an assignment of:
  \begin{itemize}
  \item an object $\Monad x$ to every $x \in \Cat C$;
  \item a morphism $\return_x : x \to \Monad x$ to every $x \in \Cat C$;
  \item a morphism $\bind[a,x,y] f : a \times \Monad x \to \Monad y$ to every $a,x,y \in \Cat C$ and
    $f : a \times x \to \Monad y$;
  \end{itemize}
  satisfying the four equations~\cite{mcdermott-uustalu:strong-monads}
  in \figref{strong monad laws}, w.r.t~the cartesian monoidal
  structure $\seq{\Cat C, (\times), \terminal, \assoc, \ellunit,
    \runit}$.
  \begin{figure}
    \begin{center}
    \diagram*{strong-monad-monoidal-unit}
    \quad
    \diagram*{strong-monad-naturality}
    \quad
    \diagram*{strong-monad-monadic-unit}
    \\
    \diagram*{strong-monad-assoc}
    \end{center}
    \caption{Strong monad laws}
    \figlabel{strong monad laws}
  \end{figure}
\end{itemize}
Each such strong-monad model induces a substitution structure $\Model$ in
$\CBVSort[\Type]$-structures. First, let:
\begin{gather*}
  \sem\Context \definedby \Env{\sem-}\Context \definedby \prod_{(x : \ty) \in \Context} \sem \ty
\qquad
\sem{\Context[1] \xto\ren \Context[2]}
\definedby \Env{\sem-}\ren : \sem{\Context[1]} \xto{\seq[{(y : \ty[2]) \in \Context[2]}]{\projection_y}}
\sem{\Context[2]}
\end{gather*}
I.e., interpret contexts as products and their renamings as tupled projections.
The $\CBVSort[\Type]$-structure $\Model$:
\begin{gather*}
  \sem{\Comp \ty} \definedby \Monad\sem\ty
\qquad
\carrier\Model_{\sort}\Context \definedby \Cat C(\sem\Context, \sem \sort)
\qquad
\carrier\Model_{\sort}\parent{\Context \xfrom\ren \Context[2]} :
\parent{\sem{\Context} \xto f \sem\sort} \mapsto
\parent{\sem{\Context[2]} \xto{\sem\ren} \sem\Context \xto f \sem\sort}
\end{gather*}
I.e.,
$\carrier\Model_{\ty}\Context \definedby \Cat C(\sem\Context, \sem \ty)$ and
$\carrier\Model_{\Comp\ty}\Context \definedby \Cat C(\sem\Context, \Monad\sem \ty)$ so
the semantics of computations are Kleisli arrows for the monad $\Monad$.
The monoid's unit interprets variables by projecting the appropriate component:
\[
\var : \Neut \to \rest{\carrier\Model}
\qquad
\var[{\ty[2],\Context}]y \definedby \parent{\sem\Context =
  \parent{\prod\nolimits_{(x : \ty) \in \Context}\sem \ty} \xto{\projection_y} \sem{\ty[2]}}
\]
To define substitution, we use the isomorphism which internalises tupling:
\[
\overline{(-)} \definedby \seq[{(y : \ty[2]) \in \Context[2]}]{\projection_y \compose (-)}
: \Cat C(\sem\Context, \sem{\Context[2]})
\xto\isomorphic
\parent{\prod_{(y : \ty[2]) \in \Context[2]}\Cat C(\sem\Context, \sem{\ty[2]})}
=
\Env{\carrier\Model}{\Context[2]}\Context
\]
We express the functorial action of $\Env{\carrier\Model}{}$
through this isomorphism as follows:
\[
\Env{\carrier\Model}{\Context[2]_1\xto\ren \Context[2]_2,\Context} :
\parent{\overline{\sem\Context \xto\msubst \sem{\Context[2]_1}}}
\mapsto
\parent{\overline{\sem\Context \xto\msubst \sem{\Context[2]_1} \xto{\sem\ren} \sem{\Context[2]_2}}}
\]
We then define substitution \(-[-]_{\Model} : \Model \tensor \Model \to \Model\) by
pre-composition:
\begin{gather*}
  \parent{\sem{\Context[2]} \xto f \sem\sort}
  \bracks{\overline{\sem\Context \xto{\msubst} \sem{\Context[2]}}}_{\Model, \sort,\Context}
  {}\definedby \parent{
  \sem\Context \xto{\msubst}
  \sem{\Context[2]} \xto{f}
  \sem\sort}
\end{gather*}
\appref[A]{case-studies} details the proof that this definition forms a
substitution structure.

\subsection{The \CBV{} customisation menu}
We will now consider some fragments of the full $\CBV{}$
calculus. \figref{cbv menu} list the constructs in each fragment,
which fragments of the type system they require, and what model
structure they need. We will treat the base, sequential, and
functional fragments. The other fragments admit a similar treatment.
We define each fragment, and then explain
how to combine fragments together into one calculus and its model
class of interest. Together, these combinations
describe $2^7 = 128$ different calculi, and
their denotational semantics.
\begin{figure}
  \begin{center}
    \input{cbv-a-la-carte}
  \end{center}
\caption{A customisation menu of \CBV{} fragments}
\figlabel{cbv menu}
\end{figure}
Thanks to \mast{} and the Special Representation
\thmref{representation}, the substitution operation and denotational
semantics for each combination satisfy a substitution lemma. Formally,
let $\Extension$ be the set of $7$ extensions listed in the `name'
column. For each fragment $\frag \subset \Extension$, we specify:
\begin{itemize}
\item a simple signature $\TySig_{\frag} : \Set \to \Set$, which we
  define \alacarte{} as $\TySig_{\frag} \definedby \BaseTySig\amalg
  \coprod_{\ext \in \frag}\TySig_{\frag}^{\ext} $, inducing the set
  $\Type_{\frag} \definedby \fix\TySig_{\frag}$ and sorting system
  $\CBVSort[\frag] \definedby \CBVSort[\Type_{\frag}]$;
\item a
  fulfillment $(\models_{\frag}^{\ext}) :
  \TySig[2]_{\ext}\Type_{\frag} \rto \Type_{\frag}$
  for each typing need $\TySig[2]_{\ext} : \Set \to \Set$
  and extension $\ext \in \frag$;
\item a $\CBVSort[\Type]$-signature functor $\Sig_{\frag} \definedby
  \BaseSig \amalg \coprod_{\ext \in \frag}\Sig^{\ext}_{\frag}$
  describing the syntactic constructs in this fragment;
\item additional structure or properties we require of the
  substitution structure $\Model$ for a strong-monad model
  $\triple{\Cat C}{\sem-}{\Monad}$.  These requirements
  impose structure and properties of the type interpretation function
  $\sem- : \Type_{\frag} \to
  \Cat C$, typically via universal properties involving
  categorical structures over the model.
\end{itemize}
We then fix such a substitution structure $\Model$, and further
specify:
\begin{itemize}
\item a compatible $\Sig$-algebra for $\Model$ making it an $\Sig$-structure.
\end{itemize}

\subsubsection*{Base fragment}
All our fragments include the base fragment: the signature functor
$\TySig_{\frag}$ includes the set of base types and so $\Type_{\frag}$
includes them. There are no typing need for the base fragment, and it contributes
no tuples to the fulfillment relation.  The
base calculus has, for each type $\ty \in \Type_{\frag}$, one
operator coercing $\ty$-values to $\ty$-computations. Its binding
signature functor and its derived strength are:
\[
\BaseSig X \definedby \coprod_{\ty \in \Type_{\TySig}}
  (\val_{\ty} : \OnlyAt{\Comp\ty} (X\At \ty))
\qquad
\strength[\BaseSig]_{P,A,\Context}
\bracks{\val_{\ty}(p \in P_{\ty}\Context[2]),
  \msubst \in \Env A{\Context[2]}\Context} \definedby
\val_{\ty}[p,\msubst]
\]
The base requirement of a model is for it to be a strong-monad model
$\Model \definedby \seq{\Cat C, \sem-, \Monad}$. Define its
$\BaseSig$-algebra structure as follows, and see calculation
\appeqref[A.1]{eq:val-compatibility} in the appendix for the compatibility
condition:
\[
\sem{\val_{\ty}{\sem\Context \xto f \sem\ty}}_{\Model}
\definedby \parent{\sem\Context \xto f \sem\ty \xto{\return}\Monad \sem A}
\]

\subsubsection*{Sequential fragment}
This fragment does not extend the language with new types nor does it
impose any typing requirements, and so $\TySig[1]_{\frag}^{\seqFrag}
\definedby \TySig[2]_{\seqFrag} \definedby \initial$. The fulfillment
relation is then the empty relation $(\models_{frag}^{\seqFrag}) :
\TySig[2]_{\seqFrag}\Type_{\frag} \definedby \emptyset \rto
\Type_{\frag}$.  The sequential fragment has, for every non-empty
context $[x_0 : \ty[1]_0, \ldots, x_n : \ty[1]_n] \in
\RenCat*{\Type_{\frag}}$ and type $\ty[2] \in \Type_{\frag}$, one
sequencing operator $\Let*{x_0\:=\M_0; \ldots; x_n\:=\M_n}\N$:
\[
\Sig_{\frag}^{\seqFrag} X \definedby
  \coprod_{n\in\naturals}
  \coprod_{[x_0: \ty[1]_0, \ldots, x_n : \ty[1]_n] \in \RenCat*{\Type_{\frag}}}
  \coprod_{\ty[2] \in \Type_{\frag}}
  \parent{\begin{array}{@{}>{\displaystyle}l@{}}
      \parent{\Let*{x_0 : \ty_0\:=\_; \ldots; x_n : \ty_n \:=\_}{\_ : \ty[2]}} :
      \\\qquad
      \OnlyAt{\Comp \ty[2]}\parent{
        \begin{array}{@{}>{\displaystyle}l@{}}
        \parent{\prod_{i=0}^n
          [x_0 : \ty_1, \ldots, x_{i-1}: \ty_{i-1}]\Shift X\At{\Comp \ty[1]_i}}
          \\\qquad\times
          ([x_0 : \ty_1, \ldots, x_{n}: \ty_{n}]
          \Shift X \At{\Comp \ty[2]})
      \end{array}}
    \end{array}}
\]
with its induced tensorial strength given in \figref{let strength}.
\begin{figure}
\begin{multline*}
\strength[\Sig_{\frag}^{\seqFrag}]_{P,A,\Context}
\bracks{
  \begin{aligned}
    &\Let*{
  \begin{aligned}[t]
    x_0 : A_0 & \:= (p_0 \in P_{\Comp\ty[1]_0}\Context[2])\\
      x_1 : A_1 & \:= (p_1 \in P_{\Comp\ty[1]_1}(\Context[2], x_0:\ty[1]_0))\\
      &\vdots\\
      x_n : A_n & \:= (p_n \in P_{\Comp\ty[1]_n}(\Context[2], x_0:\ty[1]_0, \ldots, x_{n-1}:\ty[1]_{n-1}))
      \\
  \end{aligned}\\&
  }{(q \in P_{\Comp\ty[2]}(\Context[2], x_0:\ty[1]_0, \ldots, x_n:\ty[1]_n))}
  , \msubst \in \Env A{\Context[2]}\Context
  \end{aligned}
}_{\Context[2]}
\\\definedby
\parent{
  \begin{aligned}
    &\Let*{
  \begin{aligned}[t]
    x_0 : A_0 & \:= \bracks{p_0, \msubst}_{\Context[2]}\\
    x_1 : A_1 & \:= \bracks{p_1, \msubst\++\seq{x_0 : \var[A] x_0}}_{\Context[2], x_0 : A_0}\\
      &\vdots\\
      x_n : A_n & \:= \bracks{p_n, \msubst\++\seq{x_0 : \var[A] x_0, \ldots, x_{n-1} : \var[A] x_{n-1}}}_{\Context[2], x_0 : \ty_0, \ldots, x_{n-1} : \ty_{n-1}}
  \end{aligned}\\&
    }{\bracks{q \in P_{\Comp\ty[2]}, \msubst\++\seq{x_0 : \var[A] x_0, \ldots, x_{n} : \var[A] x_{n}}}_{\Context[2], x_0 : \ty_0, \ldots, x_{n} : \ty_{n}}}
  \end{aligned}
}
\end{multline*}
\caption{Pointed strength for the sequential signature functor}
\figlabel{let strength}
\end{figure}
Given a strong-monad model $\Model \definedby
\seq{\Cat C, \sem-, \Monad}$, we require no
additional semantic structure of it. We use
the monadic bind to interpret the $\LetName{}$ construct, and to
ease dealing with the intermediate results bound to $x_1, \ldots, x_n$,
we use the following derived semantic structure. Consider the \emph{Cartesian strength}
of the monad:
\[
\strength_{a,b} : a \times \Monad b \xto{\bind[{a,b,a\times b}]\id} \Monad (a \times b)
\]
We use it to define, for every $f : a \times x \to \Monad y$ a morphism that keeps the intermediate result:
\[
\keep f : a \times x \xto{\pair{\projection_2}f} x \times \Monad y \xto\strength
          \Monad (x \times y) \qquad
\qquad
\bindkeep f : a \times \Monad x \xto{\bind(\keep f)} \Monad (x \times y)
\]
Define the $\Sig_{\frag}^{\seqFrag}$-algebra structure
as follows. Given $\ty_0, \ldots, \ty_n, \ty[2]$, let:
\[
\Context[2]   \definedby [x_0 : \ty_0, \ldots, x_{n  } : \ty_{n  }]
\qquad
\Context[2]_i \definedby [x_0 : \ty_0, \ldots, x_{i-1} : \ty_{i-1}]
\tag*{for all $i = 1, \ldots, n$}
\]
and then define, suppressing canonical isomorphisms such as
  $\Context[2]_i\times\ty_i \isomorphic \Context[2]_{i+1}$:
\begin{multline*}
  \bigsem{\begin{array}{@{}>{\displaystyle}l@{}}
    \begin{aligned}[t]
    \Let*{
        x_0 : \ty_0&\:={\sem\Context \xto{f_0} \Monad\sem{\ty_0}}\\
        x_1 : \ty_1&\:={\sem\Context\times\sem{\Context[2]_1} \xto{f_1} \Monad\sem{\ty_1}}\\
        &\vdots\\[-7pt]
        x_n : \ty_n &\:={\sem\Context\times\sem{\Context[2]_n} \xto{f_n} \Monad\sem{\ty_n}}
    }{\sem{\Context\++\Context[2]} \xto{g} \Monad\sem{\ty[2]}} : \ty[2]
    \end{aligned}
    \end{array}
  }\definedby\\
  \sem\Context
    \xto{\pair{\id}{f_0}}
  \sem{\Context}\times\Monad\sem{\Context[2]_1}
  \xto{\pair\id{\bindkeep{f_1}}}
  \sem{\Context}\times\Monad\sem{\Context[2]_2}
  \to \cdots
  \to \sem{\Context}\times\Monad\sem{\Context[2]_n}
  \xto{\pair\id{\bindkeep{f_n}}}
  \sem{\Context}\times\Monad\sem{\Context[2]}
  \xto{\bind g}
  \Monad\sem{\ty[2]}
\end{multline*}
We prove that this $\Sig_{\frag}^{\seqFrag}$-algebra is compatible
with substitution in \appsubsecref[A.3]{seq-fragment-proof}.

\subsubsection*{Functional fragment}
For this fragment we extend the set of types---\alacarte---using
the following simple signature functor for this fragment to add function types:
\(
\TySig_{\frag}^{\funFrag}X \definedby \set{\FUN xy \suchthat x,y \in X}
\isomorphic X\times X
\). We impose no typing typing requirements for this extension
($\TySig[2]_{\funFrag} = \initial$) and the fulfillment relation for
these fragments is empty
($(\models_{\frag}^{\funFrag}) : \TySig[2]_{\funFrag}\Type_{frag} =
\emptyset \rto \Type_{\frag}$).
These fragments extend the base language with function abstraction and application,
which we add using the following signature functor:
\[
\Sig_{\frag}^{\funFrag} X \definedby \coprod_{\mathclap{\ty,\ty[2] \in \Type_{\TySig}}}
\begin{array}[t]{@{}>{\displaystyle}l@{}}
\parent{
  (\sabst{x:\ty}{}) : \OnlyAt{\FUN\ty{\ty[2]}}[x : \ty]\Shift X \At {\Comp \ty[2]}
}\amalg \parent{
  (\sapply{}{}) : \OnlyAt{\ty[2]} (X \At {\Comp (\FUN{\ty[1]}{\ty[2]}}))\times(X \At {\Comp\ty[1]})
}
\end{array}
\]
Its derived strength is given in \figref{fun strength}.
\begin{figure}
\[
\begin{array}{@{}>{\displaystyle}l@{{}\definedby{}}l@{}}
\strength[\Sig_{\frag}^{\funFrag}]_{P,C,\FUN\ty{\ty[2]},\Context}\bracks{
\sabst{x : \ty}\parent{p \in P_{\Comp \ty[2]}(\Context[2], x : \ty)}
,\msubst \in \Env C{\Context[2]}\Context}_{\Context[2]}
&
\sabst{x : \ty}\bracks{p, (\msubst, x : \var x)}_{\Context[2], x : \ty}
\\
\strength[\FunSig]_{P,C,{\ty[2]},\Context}\bracks{
  \sapply{\parent{p \in P_{\Comp \FUN\ty{\ty[2]}}\Context[2]}}
       {\parent{q \in P_{\Comp \ty[1]}\Context[2]}}
,\msubst \in \Env C{\Context[2]}\Context}_{\Context[2]}
&
\sapply{\bracks{p, \msubst}_{\Context[2]}}
       {\bracks{q, \msubst}_{\Context[2]}}
\end{array}
\]
\caption{Pointed strength for the functional-fragment signature functor}
\figlabel{fun strength}
\end{figure}
We require models $\Model$
for fragments with $\funFrag \in \frag$ must be equipped with a choice
of \emph{Klesili exponentials} $\pair{(\Monad y)^x}{\eval : (\Monad
  y)^x \times x \to \Monad y}$, for every pair of objects $x,y \in \Cat C$.
Define the $\Sig_{\frag}^{\funFrag}$-algebra structure by:
\[
\begin{array}{@{}>{\displaystyle}l@{}}
\sem{{\sabst{x : \ty}\parent{\sem\Context \times \sem\ty \xto{f} \Monad\sem{\ty[2]}}}}
  \definedby \parent{\sem\Context\xto{\curry f} \sem{\FUN\ty{\ty[2]}}}
  \\\\
  \begin{array}{@{}>{\displaystyle}l@{}}
  \sem{\sapply{\parent{\sem\Context \xto{f} \Monad\sem{\FUN\ty{\ty[2]}}}}
               {\parent{\sem\Context \xto{a} \Monad\sem{\ty}}}}
  \definedby\\\qquad\qquad
  \sem\Context
  \xto{\pair{\id}{f}}
  \sem\Context\times\Monad\sem{\FUN\ty{\ty[2]}}
  \xto{\bindkeep\parent{\sem\Context\times\sem{\FUN\ty{\ty[2]} }\xto{\projection_1} \sem\Context \xto a \Monad\sem\ty}}
  \Monad\parent{\sem{\FUN\ty{\ty[2]}}\times\sem\ty}
  \xto{\bind\eval}
  \Monad\sem{\ty[2]}
  \end{array}
\end{array}
\]
We prove that this $\Sig_{\frag}^{\seqFrag}$-algebra is compatible
with substitution in \appsubsecref[A.4]{fun-fragment-proof}.

We demonstrated the various moving parts in using \mast{} to define
syntax and semantics \alacarte{}. The other fragments follow a similar
treatment, defining signature functors $\RecordSig$, $\VariantSig$,
$\NatSig$, $\WhileSig$, and $\RecSig$, each with their typing needs
and model structure and properties, as in \figref{cbv menu}.  We omit
these details. We conclude by reaping the fruit of our labour:
the standard substitution lemma for denotational
semantics. While substitution lemmata are not hard to prove, they are tedious to
establish formally. The Special Representation \thmref{representation} justifies
omitting them from most technical developments:
\begin{lem*}[substitution]\lemmalabel{substitution}
  For every term $\Context[2] \types \M : \ty$ and substitution
  $\seq[{(y:\ty[2]) \in \Context[2]}]{\Context \types \msubst_y :
    \ty[2]}$, we have:
  \[
  \sem{\M\Subst\msubst} = \sem\M\compose \seq[y]{\sem{\msubst_y}}
  \]
\end{lem*}
\begin{proofNoQED}
  By the homomorphism property of the denotational semantics:
  \[
  \sem{\M\Subst\msubst} =
  \sem{(-[-])[\M, \msubst]_{\Context[2]}}
  =
  (-[-])\bracks{\sem{\M}, \seq[{y \in \Context[2]}]{\sem\msubst_y}}_{\Context[2]}
  =
  \sem\M\compose \seq[y]{\sem{\msubst_y}}
  \tag*{\qed}
  \]
\end{proofNoQED}

%% file: cbv-type-system.tex
\begin{gather*}
  \begin{prooftree}[template={$\scriptstyle\inserttext$}]
    \hypo{(x : \ty) \in \Context}
    \infer1{\Context \types x : \ty}
  \end{prooftree}
  \qquad
  \begin{prooftree}[template={$\scriptstyle\inserttext$}]
    \hypo{\Context \types \V : \ty}
    \infer1{\Context \types \val\V : \Comp\ty}
  \end{prooftree}
  \qquad
  \begin{prooftree}[template={$\scriptstyle\inserttext$}]
    \hypo{\Context, x_1 : \ty_1, \ldots, x_n : \ty_n \types \N : \Comp\ty[2]}
    \hypo{\text{for all $i < n$: } \Context, x_1 : \ty_1, \ldots, x_i : \ty_i \types
      \M_{i+1} : \Comp\ty_{i+1}}
    \infer2{\Context \types \Let*{x_1\:=\M_1; \ldots; x_n\:=\M_n}\N : \Comp\ty[2]}
  \end{prooftree}
  \\[\baselineskip]
  \begin{prooftree}[template={$\scriptstyle\inserttext$}]
    \hypo{\Context, x : \ty  \types \M : \Comp\ty[2]}
    \infer1{\Context \types \sabst{x : \ty}\M : \FUN\ty{\ty[2]}}
  \end{prooftree}
  \qquad
  \begin{prooftree}[template={$\scriptstyle\inserttext$}]
    \hypo{\Context \types \M[1] : \Comp(\FUN\ty\ty[2])}
    \hypo{\Context \types \M[2] : \Comp\ty}
    \infer2{\Context \types \sapply{\M[1]}{\M[2]} : \Comp\ty[2]}
  \end{prooftree}
  \\[\baselineskip]
  \begin{prooftree}[template={$\scriptstyle\inserttext$}]
    \hypo{\text{for all $1 \leq i \leq n$: }
      \Context \types \M_i : \ty_i
    }
    \infer1{\Context \types
      \sseq{\Cons{C_1} \sof \M_1, \ldots, \Cons{C_n}\sof \M_n} :
      \Comp{\sRecord{\Cons{C_1} \sof \ty_1, \ldots, \Cons{C_n}\sof \ty_n}}}
  \end{prooftree}
  \qquad
    \begin{prooftree}[template={$\scriptstyle\inserttext$}]
    \hypo{\Context \types \M :
      \Comp{\sRecord{\Cons{C_1} \sof \ty_1, \ldots, \Cons{C_n}\sof \ty_n}}
    }
    \hypo{\Context, x_1 : \ty_1, \ldots, x_n : \ty_n \types \N : \Comp\ty[2]}
    \infer2{\Context \types
      \RecordCase\M{\Cons{C_1}x_1, \ldots, \Cons{C_n}x_n}\N : \Comp\ty[2]}
  \end{prooftree}
  \\[\baselineskip]
  \begin{prooftree}[template={$\scriptstyle\inserttext$}]
    \hypo{\ty = \sVariant{\Cons{C_i} \sof \ty_i \suchthat* i \in I}}
    \hypo{\Context \types \M : \Comp\ty_i}
    \infer2{\Context \types \ty.\Cons{C_i} \M: \Comp\ty}
  \end{prooftree}
  \qquad
  \begin{prooftree}[template={$\scriptstyle\inserttext$}]
    \hypo{\Context \types \M : \sVariant{\Cons{C_i} \sof \ty_i \suchthat* i \in I}}
    \hypo{\text{for all $1 \leq i \leq n$: }
      \Context, x_i : \ty_i \types
      \M_i
      : \Comp\ty[2]
    }
    \infer2{\Context \types \VariantCase\M{\Clause{\Cons{C_i}x_i} \M_i \suchthat* i \in I}\N: \Comp\ty[2]}
  \end{prooftree}
  \\[\baselineskip]
  \begin{prooftree}[template={$\scriptstyle\inserttext$}]
    \infer0{\Context \types \literal n : \Nat}
  \end{prooftree}
  \qquad
  \begin{prooftree}[template={$\scriptstyle\inserttext$}]
    \hypo{\Context \types \M : \Comp\Nat}
    \infer1{\Context \types \Unroll \M :
      \sVariant{\literal 0\sof \sRecord{\mathvisiblespace},\ (\Suc{})\sof \Nat}}
  \end{prooftree}
  \qquad
  \begin{prooftree}[template={$\scriptstyle\inserttext$}]
    \hypo{\Context \types \M :
      \Comp \sVariant{\literal 0\sof \sRecord{\mathvisiblespace},\ (\Suc{})\sof \Nat}}
    \infer1{\Context \types \Roll\M : \Comp\Nat}
  \end{prooftree}
  \quad
  \begin{prooftree}[template={$\scriptstyle\inserttext$}]
    \hypo{\Context \types \M : \Comp\Nat}
    \hypo{\Context, x : \sVariant{\literal 0\sof \sRecord{\mathvisiblespace},\ (\Suc{})\sof \ty}
    \types \M[2] : \Comp \ty}
    \infer2{\Context \types \Fold* \M x{\M[2]} : \Comp \ty}
  \end{prooftree}
  \\[\baselineskip]%
  \begin{prooftree}[template={$\scriptstyle\inserttext$}]
    \hypo{\Context \types \M : \Comp\ty}
    \hypo{\Context, i : \ty \types \N : \Comp\sVariant{\Done\sof \ty[2],  \Cont\sof \ty[1]}}
    \infer2{\Context \types \ForLoop i\M\N : \Comp\ty[2]}
  \end{prooftree}
  \quad
  \begin{prooftree}[template={$\scriptstyle\inserttext$}]
    \hypo{\smash{\overbrace{\scriptstyle\Context,
      f_1 : \FUN{\sRecord{\Context_1}}{\ty_1},
      \ldots,
      f_n : \FUN{\sRecord{\Context_n}}{\ty_n}}^{\Context[2]\definedby}}
      \types
       \N: \ty[2]}
    \hypo{\text{for all $1 \leq i \leq n$: }
      \Context[2]\++\Context_n \types \M_n : \ty_n}
    \infer2{\Context \types \LetRec{
     f_1 \Context_1 \sof \ty_1 \:= \M_1;
     \ldots;
     f_n \Context_n \sof \ty_n \:= \M_n}
       \N: \ty[2]}
  \end{prooftree}
\end{gather*}

%% file: cbv-a-la-carte.tex
  \begin{tabular}{@{}p{.1\textwidth}>{\centering}p{.37\textwidth}>{\centering}p{.16\textwidth}p{.30\textwidth}@{}}
    name & syntactic constructs & typing needs & additional model needs
    \\\hline
    base & returning a value: \(
    \val{}
    \)
    &
    &
    strong monad over a Cartesian category
    \\
    sequential & sequencing: \(
    \LetName{}
    \)
    &
    &
    \\
    functions &
    abstraction and application\[
    (\sabst x:\ty), (\sapply{}{})
    \]
    &
    function\[
    (\FUN{}{})
    \]
    &
    Kleisli exponentials
    \\
    records
    &
    constructors and pattern match
    \begin{gather*}
    \sseq{\Cons{C_1} \sof -, \ldots, \Cons{C_n}\sof -}\\
    \RecordCase-{\Cons{C_1}x_1, \ldots, \Cons{C_n}x_n}-
    \end{gather*}
    &record\[
    \sRecord{C_i : - \suchthat* i \in I}
    \]&\\
    variants&
    constructors and pattern match
    \begin{gather*}
    \ty.\Cons{C_i} -,\
    \VariantCase-{\Clause{\Cons{C_i}x_i} - \suchthat* i \in I}
    \end{gather*}
    &
    variant
    \[
    \sVariant{C_i : - \suchthat* i \in I}
    \]
    &
    distributive category\\
    natural numbers &
    the zero and successor constructors,
    literals, empty record, (de)constructors, and bounded iteration,
    the pattern matching
    \begin{gather*}
      \Cons{0}, (\Cons{1+}),
      \literal n, \sseq{\mathvisiblespace},\Unroll, \Roll\\
      \Fold* - x{-}\\
      \VariantCase-{\Clause{\Cons{0}}-, \Clause{\Cons{1+}x}-}
    \end{gather*}
    &
    naturals,
    empty record, and the variants
    \begin{gather*}
    \Nat\\
    \sVariant{\begin{array}{@{}l@{}}
        \literal 0\sof \sRecord{\mathvisiblespace},\\(\Suc{})\sof -
      \end{array}
    }
    \end{gather*}
    &
    binary coproducts distributed over by the products, and
    a natural numbers object\\
    while
    &
    the constructors, unbounded iteration
    \[
    \Done, \Cont,\ForLoop i--
    \]
    &
    the variants
    \begin{gather*}
      \sVariant{\begin{array}{@{}l@{}}
          \Done\sof -,\\
          \Cont\sof -
        \end{array}
      }
    \end{gather*}
    &
    binary coproducts, distributive products,
    and the monad has a complete Elgot structure~\cite{adamek-et-al:iterative-monads,adamek-et-al:elgot-algebras,aczel-et-al:completely-iterative-theories,elgot:monadic-computation,bloom-esik:iteration-theories,goncharov-et-al:unguarded-recursion}
    \\
    recursion
    &
    relevant record constructor, function application, recursion
    \[
    \sseq{-}, (\sapply{}{}),\LetRecName{}
    \]
    &
    the functions
    \[
    (\FUN{\sRecord{x_i : - \suchthat* i \in I}})
    \]
    & uniform parameterised monadic fixed-points~\cite{hasegawa-kakutani:axioms-for-recursion,simpson-plotkin:complete-axioms-for-cat-fixed-points}, Kleisli exponentials
  \end{tabular}

%% file: technical-development.tex
\section{Technical development outline}
\seclabel{technical development} Our development is relatively
straightforward thanks to several abstractions: bi-categories and the
right-closed actegorical structure of substitution (\subsecref{bicat
  dev}) and the General Representation \thmref{general representation}
(\subsecref{general representation theorem main}) which abstracts from
the concrete details of presheaf categories and their tensors. We also
relate this development to the pre-proceedings manuscript, which used
\emph{skew} monoidal structures (\subsecref{skew dev main}).  In this
section, we summarise how these abstract components intertwine, and
relegate the remaining details to Appendices~\appref*[B]{skew
  bicats}--\appref*[D]{representation}.

\input{bicat-main}
\pagebreak
\input{representation-main}
\input{skew}

%% file: bicat-main.tex
\subsection{Bicategorical development}
\subseclabel{bicat dev}

Fiore, Gambino, Hyland, and
Winskel~\cite{fiore-et-al:relative-pseudomonads-etc} package the
sophistication involved in the classical substitution tensor product
in a bicategory we denote by $\RenCat\Prof$.  Its vertices/$0$-cells
are small categories $\SC[1], \SC[2], \SC[3], \ldots$.  The category
$\RenCat S$ of $S$-sorted contexts is the finite-product
completion of $S$, and extends to categories.  The arrows/$1$-cells $P
: \SC[1] \rTo \SC[2]$ in $\RenCat\Prof$ are Kleisli profunctors $P : \SC[1]
\rto \RenCat{\SC[2]}$, i.e. presheaves $P \in \PSh{(\opposite{\SC[1]}
  \times \RenCat{\SC[2]})}$, equivalently functors $P : \SC[1]\times
\opposite-{\RenCat{\SC[2]}} \to \Set$. Its faces/$2$-cells $\alpha :
\mP \To \mQ$ are natural transformations, with their usual vertical and
horizontal composition.
Arrows $\mP : \SC[1] \rTo \SC[2]$
and $\mQ : \SC[2] \rTo \SC[3]$ compose diagrammatically using the
same formula as the substitution tensor product, and the profunctor of
variables, suitably generalised, is the identity $\Vars : \SC \rTo
\SC$. One recovers the classical setting by restricting to the full
sub-bicategory over a set of sorts $\gsort$, since a
 one vertex bicategory is a monoidal category.

\begin{proposition}\proplabel{actions from two 0-cells}
  Let $\Cat B$ be a bicategory. Every two $0$-cells $\SmallCat A,
  \SmallCat B \in \Cat B$ induce an actegory $\pair{\Cat B(\SmallCat
    A, \SmallCat A)}{\Cat B(\SmallCat B, \SmallCat A)}$. Its monoidal
  category is given by the endo-$1$-cells $X : \SmallCat A \to
  \SmallCat A$ and their $2$-cells. It acts on the $1$-cells $P :
  \SmallCat B \to \SmallCat A$ and their $2$-cells through $1$-cell
  post-composition and horizontal composition.
\end{proposition}

The proof is in \appsubsecref[B.2]{actions from bicategories}.  We obtain
\thmref{structures are actions} from \propref{actions from two
  0-cells} by taking $\Cat B \definedby \RenCat\Prof$, $\SmallCat A
\definedby \Fst$, and $\SmallCat B \definedby \Snd$.

%% file: representation-main.tex
\subsection{The representation theorem}
\subseclabel{general representation theorem main}
\begin{wrapfigure}[5]{r}{3.1cm}
  \hspace{-.75cm}
  \diagram*{right-exponential}
\end{wrapfigure}
Adapting the classical proof for the representation theorem is
straightforward thanks to its level of generality.  Let $\pair{\catMon
  }{\catAct}$ be an actegory, and $a \in \catAct$ and $x \in \catMon$. Recall that a
\emph{right exponential of $a$ by $x$} is an object $\aexp ax$
equipped with a universal morphism $\eval : \parent{(\aexp ax)\ract
  x} \to a$, i.e., for every arrow $f : b \ract x \to a$ there is a
unique arrow $\curry f : b \to (\aexp a x)$ making the diagram on the
right commute. To distinguish the special case when the monoidal
category acts on itself, we'll use the notation $\texp xy$ for a right
exponential of $x$ by $y$. An actegory $\pair{\catMon}{\Cat A}$ is
\emph{closed} when all right-exponentials exist. In that case,
$(\ract)$ distributes over coproducts in $\Cat A$.
We prove the following result in
\appsubsecref[B.3]{exponentiation with kan lifts}.
\begin{proposition}\proplabel{heterogeneous structures right-closed}
  The actegory of $\gsort$-structures is right-closed:
  \(
  (\aexp{\restrict\mP\snd}{\rest\mQ})_{\sort}\Context \definedby
  \int_{\Context[2]\in \RenCat{{\Fst}}} (\mP_{\sort}\Context[2])^{\rEnv\mQ{\Context[2]}\Context}
  \).
\end{proposition}

We prove the following generalisation of the classical theory in
\appref[D]{representation}.  It implies the Special Representation
\thmref{representation}.  Like its classical counterpart, it abstracts
from the technical details of $\PointedAct{\Syn\gsort}$.
\begin{theorem}[general representation]\thmlabel{general representation}
  Let $\pair{\catMon}{\catAct}$ be a closed actegory with finite coproducts.
  Letting $\catSkew \definedby \catMon\times\catAct$ be the product $\catMon$-actegory
  and $\kNeut \definedby \pair\mNeut\initial \in \catSkew$, take any pointed-strong functor  $\Sig : \PointedAct{\catSkew} \to
  \PointedAct{\catSkew}$,  object $\ghole \in
  \catSkew$, and initial algebra structure
  $\seq{\sem- : \carrier\Sig \AST \to \AST, \var : \mNeut \to \AST_1, \meta-[-] : \ghole\tensor \AST_1 \to \AST}$
  over the object
  $\AST = \pair-{\AST_1}{\AST_2} \definedby \fix
  \pair-{x_1}{x_2}. (\carrier\Sig\pair-{x_1}{x_2}%
  \amalg \kNeut \amalg (\ghole \rmul x_1)$.
  There is a unique $\catSkew$-morphism, called
  \emph{simultaneous substitution},
  $-[-] : \AST \tensor \AST_1 \to \AST$
  satisfying analogous equations to
  \thmref{representation}.  The free $\Sig$-action over
  $\ghole$ is then $\Free[\Sig]\ghole \definedby
  \seq{\AST, -[-], \var, \sem-}$ equipped with the arrow $? : \ghole
  \xto{\runit'} \ghole\rmul\mNeut \xto{\id\tensor\var} \ghole\tensor
  \AST_1 \xto{\meta-[-]} \AST$.
\end{theorem}

%% file: skew.tex
\subsection{Skew-monoidal categories}
\subseclabel{skew dev main}
In an earlier version of this article, we developed \mast{} by
recourse to the following structure:
\[
(\bmul) : \Syn\gsort \times \Syn\gsort \to
\Syn\gsort
\quad
P\bmul Q \definedby P \ract \rest Q
\quad
\kNeut \in \Syn\gsort
\quad
\kNeut[\sort]\Context \definedby \set{x \suchthat (x:\sort) \in \Context}
\]
This structure has been used, e.g., by Goncharov et al.~\cite{goncharov-et-al:cbpv}
and by Greg Brown as
the universe of syntax for \CBPV{}. It falls short from being a
monoidal category: while we can define the mediator maps, the putative
left unitor $\ellunit : \kNeut \bmul P \to P$ is not invertible in the
presence of second-class sorts. For example, for each $\sort \in
\Snd$, we have $\kNeut[\snd\sort]\Context = \emptyset$, and so:
\(
(\kNeut\bmul \terminal)_{\snd \sort}\Context = \int^{\Context[2]} \kNeut[\snd\sort]\Context[2] \times \terminal
= \int^{\Context[2]}\emptyset \times \terminal = \emptyset \not\isomorphic \terminal
\).
Therefore, even though $\Syn\gsort$ fails to be a monoidal category,
this structure exhibits it as a \emph{skew} monoidal
category~(cf.~\appsubsecref[C.4]{skew monoidal def}).
Moreover, both the associator $(\assoc : (P \bmul Q)\bmul R
\to P \bmul (Q\bmul R))$ and right unitor ($\runit' : P \to P \bmul
\kNeut$) are invertible, in a situation we call an \emph{associative}
and \emph{right-unital} skew monoidal category. To our surprise, the
general representation theorem of the classical theory remains true
even under the weaker assumptions that the left unitor is not
invertible. That development took advantage of results by Fiore and
Szamozvancev~\cite{fiore-szamozvancev:soas-theory} about skew monoidal
categories for the classical theory. We show
how to recover the skew situation, and its special representation
theorem, from our actegorical perspective.

Let $\pair{\catMon}{\catAct}$ be an actegory. Assume $\catAct$ has an
initial object $\initial$ preserved by the action $(\ract) : \catAct
\times \catMon \to \catAct$: the unique morphisms
$\coseq{} : \initial \to \initial\ract a$ are invertible for each $a \in
\catMon$. Tensors of closed actegories preserve $\initial$.%
\begin{proposition}\proplabel{skew from action}
Let $\pair{\catMon}{\catAct}$ be an actegory. If $\catAct$ has an
initial object $\initial$ and the action $(\ract) : \catAct \times
\catMon \to \catAct$ preserves it, then $\catSkew \definedby
\seq{\carrier\catMon\times\carrier\catAct, (\bmul), \kNeut, \assoc, \runit', \ellunit}$ is
an associative right-unital skew monoidal category, where:
\[
(\bmul) : \carrier \catSkew \times \carrier \catSkew \to \carrier\catSkew
\quad
\pair ax \bmul \pair by \definedby \pair{a \tensor b}{x \ract b}
\quad
\kNeut \definedby \pair{\mNeut}{\initial}
\quad
\assoc \definedby \pair\assoc\assoc
\quad
\runit' \definedby \pair{\inv\runit}{\inv\runit}
\quad
\ellunit \definedby \pair\ellunit{\coseq{}}
\]
We have a carrier-preserving isomorphism between the categories of
$\pair{\catMon}{\catAct}$-actions and $\catSkew$-monoids. Moreover, if
$\pair{\catMon}{\catAct}$ are (right-)closed, then $\catSkew$ is
right-closed.
\end{proposition}
See \appsubsecref[C.4]{skew monoidal proofs} for the proof.
The general representation theorem for the skew setting does not
follow from \propref{skew from action} and \thmref{general representation}.
The actegorical angle is natural, however, and extends
previous work~\cite{fiore-turi:name-and-value-passing}.

%% file: related-work.tex
\section{Related work}
\seclabel{related work}

The two POPL-Mark challenges~\cite{poplmark1,poplmark2} galvanise
the programming-language community to hard
problems in formalisation. Both challenges emphasise representation and
manipulation of syntax with binding and substitution. We do
not target mechnisation and computational realisation of abstract
syntax with binding and substitution, but we note the ample work of
this nature. All modern mechanisation systems support libraries or
features for abstract syntax with
binding~\cite[e.g.]{schafer-et-al:autosubst,aydemir-et-al:lngen,urban:nominal-isabelle/hol,forster-stark:coq-a-la-carte,kathrin-stark:thesis,stark-et-al:autosubst-2,pientka:beluga,gacek:thesis,gacek-et-al:abella-journal,sewell-et-al:ott}. These
generate specialised functions, lemmata, and proofs given a
description of the syntax, instead of proving
a general theory of syntax and substitution.  We direct
the reader to Allais et al.'s related work
section~\cite{allais-et-al:type-and-scope-safe-syntax-journal} which
surveys: non-de Bruijn approaches to
binding~\cite{gabbay-pitts:nominal-sets,chlipala:parametric-hoas,chargueraud:locally-nameless};
alternative binding
structure~\cite{weirich-et-al:binders-unbound,poulsen-et-al:definitional-interpreters,cheney:towards-names,ghani-et-al:cyclic-structures,hamana:initial-semantics-for-cyclic-sharing};
more work on
automation~\cite{polonowski:automatically-generated-syntaxes,keuchel-et-al:needle-and-knot};
universes of syntax and generic
programming~\cite{keuchel-jeuring:generic-conversions,keuchel:msc-thesis,benton-et-al:strongly-typed-terms,lee-et-al:gmeta,erdi:well-scoped-syntax,copello-et-al:variable-convention,morris-et-al:regular-tree-types}.
To those we add
HOAS~\cite{hofmann:semantical-analysis-of-ho-abstract-syntax,pfenning-elliott:hoas},
monadic~\cite{altenkirch-reus} and
functorial~\cite{blanchette:bindings-as-bounded-natural-functors}
representations.

The presheaf
approach~\cite{fiore:soas,arkor-mcdermott:abstract-clones-for-abstract-syntax,fiore-plotkin-turi:ast,fiore-hamana:mpats,sterling-morrison:syntax-and-semantics-of-abstract-binding-trees}
lends itself to mathematical operational
semantics~\cite{turi-plotkin:mathematical-os,goncharov-et-al:ho-mos}
and G.~structural operational semantics
(GSOS)~\cite{goncharov-et-al:cbpv}.
Fiore and Turi~\cite{fiore-turi:name-and-value-passing}
considered a special case of our heterogeneous situation in which the
constructs for first-class sorts can be described
independently from the second-class sorts in terms of actions. In our situation,
we have mutual dependency between terms of first-class and
second-class sorts, e.g., \CBV{} terms include values, and functions,
which are values, are abstracted terms.
McBride varies the indexing
category e.g., thinnings/order-preserving embeddings to implement
co-de Bruijn representations~\cite{mcbride:thinning}.  In a different
direction, Fiore and Saville reduce the free $\Sig$-monoid to the
existence of certain list objects~\cite{fiore-saville:list-objects}.

Fiore and Szamozvancev
reformulate~\cite{fiore-szamozvancev:soas-theory,szamozvancev:thesis}
and implement~\cite{fiore-szamozvancev:formal-soas} the classical
theory in terms of homogeneous \emph{families}, i.e., sort-and-context
indexed sets without a given functorial action. These families have an
associated tensor without a quotient \( (P \zmul Q)_{\sort}\Context
\definedby \coprod\nolimits_{\Context[2]} P_{\sort}\Context[2] \times
\prod\nolimits_{(x : \sort[2]) \in \Context[2]} Q_{\sort[2]}\Context
\).  The theory compensates for the missing quotient by demanding
additional axioms for substitution. Their tensor product $(\zmul)$
is skew, and it is neither unital
nor associative.

Close to the presheaf approach is the familial approach of Hirschowitz
et
al.~\cite{hirschowitz-et-al:nameless-dummies,borthelle-et-al:cellular-howe-theorem,hirschowitz-et-al:modules-over-monads,hirschowitz-maggesi:modules-over-monads},
which also uses a skew tensor for technical reasons involving
operational semantics and bisimilarity.  Ahrens also uses families
rather than presheaves~\cite{ahrens:modules-over-relative-monads}
including an
implementation~\cite{ahrens-et-al:implementing-typed-abstract-syntax}
in UniMath~\cite{unitmath}. The introductory text by Lamiaux and
Ahrens~\cite{lamiaux-ahrens:survey} provides many connections between these
approaches and related work.

Fiore, Gambino, Hyland, and Winskel's Kleisli
bicategories~\cite{fiore-et-al:relative-pseudomonads-etc} follow their
earlier work~\cite{fiore-gambino-hyland-winskel:species-bicat}
generalising Joyal's species of structure~\cite{joyal:species}.
Olimpieri et al.~\cite{olimpieri:thesis,%
clairambault-et-al:games-and-species} used these, e.g., when
studying intersection type systems.  More recently, Fiore, Galal, and
Paquet introduce a bicategory of stable
species~\cite{galal-et-al:stable-species}. Ahrens et al's
aforementioned
implementation~\cite{ahrens-et-al:implementing-typed-abstract-syntax}
also uses such bicategorical ideas.

The Expression Problem concerns the design of an abstract syntax tree
datatype for simply-typed expression language without binding and an
implementation of an evaluator. Wadler attributes the problem to
\cite{reynolds:types-as-procedural-data-structures}. In Wadler's
formualtion, the challenge is to extend the language with new term
constructions without recompiling the code for the previous versions.
Wadler reviewed several existing solutions, including object oriented
ones by Cook~\cite{cook:oop-vs-abstract-data-types}; Krishnamurthi et
al.~\cite{krishnamurthi-et-al:synthesizing-oo-and-functional-design};
and Zenger and
Odersky~\cite{zenger:msc-thesis,zenger-odersky:extensible-datatypes,zenger-odersky:mixins-workshop},
before presenting a solution based on recursive generics in Java.

%% file: conclusion.tex
\section{Conclusion}
\seclabel{conclusion}%
We have extended the classical theory of abstract syntax with binding
and substitution with second-class sorts through actegorical structure.  This
extension is straightforward thanks to the existing bicategorical
perspective. Our tutorial separates the concepts needed to employ the theory
from its underlying technical machinery and development.  We expect similar results
from an analogous case study on Levy's \CBPV{}. There the basic
calculus comprises of returners and sequencing, and one can extend it
with value products and coproducts, thunks, computation products, and
functions. Thus, with \mast{} one could deduce $2^6 = 64$ different
substitution lemmata by checking the compatibility conditions for $6$
signatures.  We want to develop a corresponding algorithmic theory
using familial skew actions~\cite{fiore-szamozvancev:soas-theory}
and implement it. We are also interested in developing a
multi-categorical perspective. It may avoid the need for
quotients, arising through the tensor product and represent the
multi-category~\cite{hermida:representable-multicategories,burroni:t-categories,leinster:fcmulticategories,leinster:higher-operads,dawson-pare-pronk:the-span-construction,cruttwell-shulman:generalized-multicategories}.
This perspective might simplify the algorithmic theory.

%% file: acks.tex
This work was supported by the Air Force Office of Scientific Research
under award number FA9550-21-1-0038, ERC Grant BLAST, two ARIA SGAI
TA1.1 grants, a Royal Society University Research Fellowship, as well
as FWF Project AUTOSARD P~36623.  For the purpose of Open Access the
authors have applied a CC BY public copyright licence to any Author
Accepted Manuscript version arising from this submission. We thank the
chairs and editor for their patience, and the anonymous referees and
these people for interesting and useful discussions and
suggestions:
Nathanael Arkor; %
Bob Atkey; %
Greg Brown; %
Vikraman Choudhury; %
Yotam Dvir; %
Nicola Gambino; %
Sergey Goncharov; %
Nick Hu; %
M.~Codrin Iftode; %
Meven Lennon-Bertrand; %
Paul B. Levy; %
Cristina Matache; %
Justus Matthiesen; %
Conor McBride; %
Sean K.~Moss; %
Filip Sieczkowski; %
Peter M.~Sewell; %
Stelios Tsampas; %
Zoe Stafford; %
Dmitrij Szamozvancev; and
Jacob Walters.

%% file: case-studies-appendix.tex
\section{Further details on the Call-by-Value case study}
\applabel{case-studies}

We include the calculations we omitted from \secref{case studies}:
showing that the monoid multiplication and its action are well-defined
and validate the action axioms; and the compatibility conditions for
the base, sequential, and functional fragments.


\subsection{Substitution structure}
We defined in \subsecref{cbv subst structure main} the substitution
structure on a strong-monad model $\seq{\Cat C, \sem-, \Monad}$ by
pre-composition:
\begin{gather*}
  \parent{\sem{\Context[2]} \xto f \sem\sort}
  \bracks{\overline{\sem\Context \xto{\msubst} \sem{\Context[2]}}}_{\Model, \sort,\Context}
  {}\definedby \parent{
  \sem\Context \xto{\msubst}
  \sem{\Context[2]} \xto{f}
  \sem\sort}
\end{gather*}
It respects the coend quotient. Indeed, for $\ren : \Context[2]_1
\to \Context[2]_2$, chase a generic element
$\pair{f}{\msubst} \in \carrier\Model_{\sort}\Context[2]_2 \times
\Env{\carrier\Model}{\Context[2]_1}\Context$ as in
\figref{substitution respects coend}.
\begin{figure}
  \diagram{monadic-monoid-subst-well-defined}
  \caption{Well-definedness of semantic substitution}
  \figlabel{substitution respects coend}
\end{figure}
It is natural in $\Context$ since $\sem\Context$ is natural in $\Context$,
and pre-/post-composition is natural. Thus we have
construct the data of a substitution structure:
\[
\Neut \xto\var \rest{\carrier\Model}
\qquad
\carrier\Model\xfrom{-[-]_{\Model}} \carrier\Model\tensor \rest{\carrier\Model}
\]
This structure satisfies the action axioms.
Straightforward calculation shows the action axioms.
\begin{figure}
  \diagram{monadic-monoid-subst-lft-unit}
  \caption{Substitution monoid left neutrality}
  \figlabel{monadic-monoid-subst-lft-unit}
\end{figure}
\figref{monadic-monoid-subst-lft-unit} establishes the monoid's left
unitality axiom. Figures~\figref*{monadic-monoid-subst-rgt-unit} and~%
\figref*{monadic-monoid-subst-assoc}%
establish the right unitality and associativity axioms for both the
monoid and its action simultaneously. The right unitality argument, for clarity,
identifies environments of variables
with renamings, and the following presentation of the right unitor:
\[
\overline{(-)} \definedby \seq[{(y : \ty[2]) \in \Context[2]}]{-[y]}
: \RenCat{{\Fst}}(\Context, \Context[2])
\xto\isomorphic
\parent{\prod_{(y : \ty[2]) \in \Context[2]}\Neut[{\ty[2]}]\Context}
=
\Env{{\Neut}}{\Context[2]}\Context
\quad
\runit_{P} : (P \tensor \Neut)_s\Context \ni \coseq[{\Context[2]}]{P_s\Context[2]\ni t,\overline\ren}
\mapsto t[\ren]_P \in P_s\Context
\]
\begin{figure}
  \diagram{monadic-monoid-subst-rgt-unit}
  \caption{Substitution structure right neutrality}
  \figlabel{monadic-monoid-subst-rgt-unit}
\end{figure}
\begin{figure}
  \diagram{monadic-monoid-subst-assoc}
  \caption{Substitution structure associativity}
  \figlabel{monadic-monoid-subst-assoc}
\end{figure}

\subsection{Base fragment}
\subseclabel{base-fragment-proof}
We  validate the compatibility axiom:
\begin{equation}\label{eq:val-compatibility}
  \diagram*{val-compatibility}
\end{equation}

\subsection{Sequential fragment}
\subseclabel{seq-fragment-proof}
To show the compatibility condition for this algebra with the substitution
monoid structure, take any:
\[
\bracks{
  \begin{aligned}
    &\Let*{
  \begin{aligned}[t]
      x_0 : A_0 & \:= \sem{\Context[3]} \xto{f_0} \Monad\sem{\ty[1]_0} \\
      x_1 : A_1 & \:= \sem{\Context[3]}\times\sem{\Context[2]_1}
                        \xto{f_1} \Monad\sem{\ty[1]_1}\\
      &\vdots\\
      x_n : A_n & \:= \sem{\Context[3]}\times\sem{\Context[2]_n}
                        \xto{f_n} \Monad\sem{\ty[1]_n}
      \\
  \end{aligned}\\&
  }{\sem{\Context[3]}\times\sem{\Context[2]} \xto{g} \Monad\sem{\ty[2]}}
  , \sem\Context \xto\msubst \sem{\Context[3]}
  \end{aligned}
}_{\Context[3]} \in ((\SeqSig\carrier\Model)\tensor\rest{\carrier\Model})_{\Comp\ty[2]}\Context
\]
The compatibily condition then amounts to the following diagram:
\diagram{seq-compatibility}
The remaining proof obligation \SeqCompatProofObligationB{} follows by
applying $\pair{\id}{\bind-}$ to the following equation:
\[
\diagram*{seq-compatibility-aux}
\tag*{\SeqCompatProofObligationB{}}
\label{SeqCompatProofObligationB}
\]

\subsection{Functional fragment}
\subseclabel{fun-fragment-proof}

This algebra structure is compatible with the substitution monoid
structure. For abstraction, it amounts to the following equation (cf.~\exref{lam-compatibility}):
\diagram{lam-compatibility}
For application, it amounts to the following calculation:
\diagram{app-compatibility}
The remaining proof obligation \AppCompatProofObligation{} is similar to
\ref{SeqCompatProofObligationB} in the sequential case.

%% file: bicat-app.tex
\section{Bicategorical development}
\applabel{skew bicats}
We recall the definition of a bicategory.
Readers already familiar with this definition may skip to
\subsecref{end of bicat definition}.

\subsection{Definition}
A bicategory generalises the notion of a monoidal category. Formally, a \emph{bicategory}~\cite{benabou:intro-to-bicats}
$\Cat B$
consists of:
  \begin{itemize}
    \item{} \emph{Vertices/$0$-cells} $a, b, c, \ldots \in \Cat B$
    \item{} For every pair of vertices $a,b$, a category $\Cat B(a,b)$
      whose:
      \begin{itemize}
      \item objects are the \emph{arrows/$1$-cells} $f,g,h, \ldots : a \to b$;
        and whose
      \item morphisms are the \emph{faces/$2$-cells} $\alpha,\beta,\gamma,\ldots : f \To g$;
      \item identities are the \emph{strictly commuting faces/identity $2$-cells} $\id : f = g$;
      \item composition operations sending two \emph{vertically
        composable} $2$-cells to their \emph{vertical composition}:
        \[
        \begin{tikzcd}[ampersand replacement=\&]
	        a \&\&\& b
	  \ToArrow{ 0}{f}{height=-24pt}{1-1}{1-4}
          \ToArrow[pos=0.4]{1}{g}{height= 0pt}{1-1}{1-4}
          \ToArrow'{2}{h}{height= 24pt}{1-1}{1-4}
          \arrow["\,\alpha", shorten <=5pt, shorten >=5pt, Rightarrow, from=0, to=1]
          \arrow["\,\beta" , shorten <=5pt, shorten >=5pt, Rightarrow, from=1, to=2]
        \end{tikzcd}
        \mapsto
        \begin{tikzcd}[ampersand replacement=\&]
	        a \&\&\& b
	  \ToArrow{ 0}{f}{height=-24pt}{1-1}{1-4}
          \ToArrow'{2}{h}{height= 24pt}{1-1}{1-4}
          \arrow["\,\beta\compose\alpha", shorten <=5pt, shorten >=5pt, Rightarrow, from=0, to=2]
        \end{tikzcd}
        \]
        The category axioms amount to requiring vertical
        composition to be associative and the identity $2$-cells
        to compose neutrally on both sides.
      \end{itemize}
    \item{} For every $a,b,c \in \Cat B$, a functor $\Cat B(b,c)\times\Cat B(a,b) \to \Cat B(a,c)$ whose action on:
      \begin{itemize}
      \item objects sends every pair of \emph{composable}
        arrows/$1$-cells $a \xto f b \xto g c$ to their
        \emph{arrow/$1$-cell composition} which we write in
        \emph{application order} as $g \compose f : a \to c$ and in
        \emph{diagram order} as $f\tensor g : a \to c$.
      \item morphisms sends every pair of \emph{horizontally
        composable} $2$-cells to their \emph{horizontal composition}:
        \[
        \begin{tikzcd}[ampersand replacement=\&]
	        a \&\& b \&\& c
	  \ToArrow{ 0}{f}{height=-12pt}{1-1}{1-3}
          \ToArrow'{1}{g}{height= 12pt}{1-1}{1-3}
          \ToArrow{ 2}{h}{height=-12pt}{1-3}{1-5}
          \ToArrow'{3}{k}{height= 12pt}{1-3}{1-5}
          \arrow["\,\alpha", shorten <=5pt, shorten >=5pt, Rightarrow, from=0, to=1]
          \arrow["\,\beta ", shorten <=5pt, shorten >=5pt, Rightarrow, from=2, to=3]
        \end{tikzcd}
        \mapsto
        \begin{tikzcd}[ampersand replacement=\&]
	        a \&\& c
	  \ToArrow{ 0}{f\tensor h}{height=-12pt}{1-1}{1-3}
          \ToArrow'{1}{g\tensor k}{height= 12pt}{1-1}{1-3}
          \arrow["\,\alpha\tensor\beta", shorten <=5pt, shorten >=5pt, Rightarrow, from=0, to=1]
        \end{tikzcd}
        \]
        The functoriality axioms amount to the so called \emph{interchange laws},
        meaning $\id[f] \tensor \id[g] = \id[f\tensor g]$ so that:
        \[
        \begin{tikzcd}[ampersand replacement=\&]
	        a \&\& b \&\& c
	  \ToArrow{ 0}{f}{height=-12pt}{1-1}{1-3}
          \ToArrow'{1}{f'}{height= 12pt}{1-1}{1-3}
          \ToArrow{ 2}{g}{height=-12pt}{1-3}{1-5}
          \ToArrow'{3}{g'}{height= 12pt}{1-3}{1-5}
          \Commute 01
          \Commute 23
        \end{tikzcd}
        \implies
        \begin{tikzcd}[ampersand replacement=\&]
	        a \&\& c
	  \ToArrow{ 0}{f\tensor g}{height=-12pt}{1-1}{1-3}
          \ToArrow'{1}{f'\tensor g'}{height= 12pt}{1-1}{1-3}
          \Commute 01
        \end{tikzcd}
        \]
        and $(\beta \compose \alpha)\tensor (\delta \compose \gamma) =
             (\beta \tensor \gamma)\compose (\alpha \tensor \gamma)$, so that:
        \[
        \begin{tikzcd}[ampersand replacement=\&]
	        a \&\& b \&\& c
	  \ToArrow{ 0}{f}{height=-24pt}{1-1}{1-3}
          \ToArrow'[pos=0.4]{1}{g}{height=  0pt}{1-1}{1-3}
          \ToArrow'{2}{h}{height= 24pt}{1-1}{1-3}
          \ToArrow{ 3}{k}{height=-24pt}{1-3}{1-5}
          \ToArrow'[pos=0.4]{4}{p}{height=  0pt}{1-3}{1-5}
          \ToArrow'{5}{q}{height= 24pt}{1-3}{1-5}
          \arrow["\,\alpha", shorten <=5pt, shorten >=5pt, Rightarrow, from=0, to=1]
          \arrow["\,\beta ", shorten <=5pt, shorten >=5pt, Rightarrow, from=1, to=2]
          \arrow["\,\gamma", shorten <=5pt, shorten >=5pt, Rightarrow, from=3, to=4]
          \arrow["\,\delta ", shorten <=5pt, shorten >=5pt, Rightarrow, from=4, to=5]
        \end{tikzcd}
        \mapsto
        \begin{tikzcd}[ampersand replacement=\&]
	        a \&\&\& c
	  \ToArrow{ 0}{f\tensor k}{height=-24pt}{1-1}{1-4}
          \ToArrow'{1}{h\tensor q}{height= 24pt}{1-1}{1-4}
          \arrow["\,(\beta \tensor \gamma)\compose (\alpha \tensor \gamma)", shorten <=5pt,
            shorten >=5pt, Rightarrow, from=0, to=1,
            curve={height=10pt}, shift left=-1em]
        \end{tikzcd}
        =
        \begin{tikzcd}[ampersand replacement=\&]
	        a \&\&\& c
	  \ToArrow{ 0}{f\tensor k}{height=-24pt}{1-1}{1-4}
          \ToArrow'{1}{h\tensor q}{height= 24pt}{1-1}{1-4}
          \arrow["{(\beta \compose \alpha)\tensor (\delta \compose \gamma)}",
            shorten <=5pt, shorten >=5pt, Rightarrow, from=0, to=1,
            curve={height=10pt}, shift left=-1em]
        \end{tikzcd}
        \]
      \end{itemize}
    \item For every $a \in \Cat B$, a specified \emph{identity} $1$-cell $\mNeut : a \to a$.
    \item For every $a \xto f b \xto g c \xto h d$, an
      \emph{associator} $2$-isomorphism $\assoc : (f \tensor g) \tensor h \To
      f \tensor (g \tensor h)$, invertible and natural in $f$, $g$, and $h$ in the
      sense that:%
      \refstepcounter{junk}\label{associator naturality}
      \begin{multline*}
        \begin{tikzcd}[ampersand replacement=\&]
	  a \&\& b \&\& c \&\& d
	  \ToArrow{ 0}{f}{height=-12pt}{1-1}{1-3}
          \ToArrow'{1}{f'}{height= 12pt}{1-1}{1-3}
          \ToArrow{ 2}{g}{height=-12pt}{1-3}{1-5}
          \ToArrow'{3}{g'}{height= 12pt}{1-3}{1-5}
          \ToArrow{ 4}{h }{height=-12pt}{1-5}{1-7}
          \ToArrow'{5}{h'}{height= 12pt}{1-5}{1-7}
          \arrow["{\alpha}",shorten <=5pt,shorten >=5pt, Rightarrow, from=0, to=1]
          \arrow["{\beta}",shorten <=5pt,shorten >=5pt, Rightarrow, from=2, to=3]
          \arrow["{\gamma}",shorten <=5pt,shorten >=5pt, Rightarrow, from=4, to=5]
        \end{tikzcd}
        \\\implies
        \begin{tikzcd}[ampersand replacement=\&]
	  a \&\&\&\& d
	  \ToArrow{ 0}{(f\tensor g)\tensor h}{height=-32pt}{1-1}{1-5}
          \ToArrow[pos=.25]{1}{f\tensor (g\tensor h)}{height= 0pt}{1-1}{1-5}
          \ToArrow'{2}{f'\tensor (g'\tensor h')}{height= 32pt}{1-1}{1-5}
          \arrow["\assoc",
            shorten <=5pt,shorten >=5pt, Rightarrow, from=0, to=1]
          \arrow["{\alpha\tensor(\beta\tensor \gamma)}",
            shorten <=5pt,shorten >=5pt, Rightarrow, from=1, to=2]
        \end{tikzcd}
        =
        \begin{tikzcd}[ampersand replacement=\&]
	  a \&\&\&\& d
	  \ToArrow{ 0}{(f\tensor g)\tensor h}{height=-32pt}{1-1}{1-5}
          \ToArrow[pos=.25]{1}{(f'\tensor g')\tensor h'}{height= 0pt}{1-1}{1-5}
          \ToArrow'{2}{f'\tensor (g'\tensor h')}{height= 32pt}{1-1}{1-5}
          \arrow["{(\alpha\tensor \beta)\tensor \gamma}",
            shorten <=5pt,shorten >=5pt, Rightarrow, from=0, to=1]
          \arrow["\assoc",
            shorten <=5pt,shorten >=5pt, Rightarrow,from=1, to=2]
        \end{tikzcd}
      \end{multline*}
    \item For every $a \xto f b$,
      \emph{left} and \emph{right} unitor $2$-isomorphisms $\ellunit : \mNeut \tensor f \To
      f$ and $\runit' : f\To f \tensor \mNeut $, invertible and natural in $f$ in the
      sense that:
      \begin{align*}
        \begin{tikzcd}[ampersand replacement=\&]
	  a \&\& b
	  \ToArrow{ 0}{f}{height=-12pt}{1-1}{1-3}
          \ToArrow'{1}{f'}{height= 12pt}{1-1}{1-3}
          \arrow["{\alpha}",shorten <=5pt,shorten >=5pt, Rightarrow, from=0, to=1]
        \end{tikzcd}
        \implies&
        \begin{tikzcd}[ampersand replacement=\&]
	  a \&\& b
	  \ToArrow{ 0}{\mNeut \tensor f}{height=-24pt}{1-1}{1-3}
          \ToArrow[pos=.25]{1}{f}{height= 0pt}{1-1}{1-3}
          \ToArrow'{2}{f'}{height= 24pt}{1-1}{1-3}
          \arrow["\ellunit",
            shorten <=5pt,shorten >=5pt, Rightarrow, from=0, to=1]
          \arrow["\alpha",
            shorten <=5pt,shorten >=5pt, Rightarrow, from=1, to=2]
        \end{tikzcd}
        =
        \begin{tikzcd}[ampersand replacement=\&]
	  a \&\& b
	  \ToArrow{ 0}{\mNeut\tensor f}{height=-24pt}{1-1}{1-3}
          \ToArrow[pos=.25]{1}{\mNeut\tensor f'}{height= 0pt}{1-1}{1-3}
          \ToArrow'{2}{f'}{height= 24pt}{1-1}{1-3}
          \arrow["\mNeut\tensor\alpha",
            shorten <=5pt,shorten >=5pt, Rightarrow, from=0, to=1]
          \arrow["\ellunit",
            shorten <=5pt,shorten >=5pt, Rightarrow, from=1, to=2]
        \end{tikzcd}
        \text{ and }
        \begin{tikzcd}[ampersand replacement=\&]
	  a \&\& b
	  \ToArrow{ 0}{f}{height=-24pt}{1-1}{1-3}
          \ToArrow[pos=.25]{1}{f\tensor\mNeut}{height= 0pt}{1-1}{1-3}
          \ToArrow'{2}{f'\tensor\mNeut}{height= 24pt}{1-1}{1-3}
          \arrow["\runit'",
            shorten <=5pt,shorten >=5pt, Rightarrow, from=0, to=1]
          \arrow["\alpha\tensor\mNeut",
            shorten <=5pt,shorten >=5pt, Rightarrow, from=1, to=2]
        \end{tikzcd}
        =
        \begin{tikzcd}[ampersand replacement=\&]
	  a \&\& b
          \ToArrow{0}{f}{height=-24pt}{1-1}{1-3}
          \ToArrow[pos=.25]{1}{f'}{height= 0pt}{1-1}{1-3}
	  \ToArrow'{2}{f'\tensor\mNeut}{height= 24pt}{1-1}{1-3}
          \arrow["\alpha",
            shorten <=5pt,shorten >=5pt, Rightarrow, from=0, to=1]
          \arrow["\runit'",
            shorten <=5pt,shorten >=5pt, Rightarrow, from=1, to=2]
        \end{tikzcd}
      \end{align*}
    \item such that the following $2$-cell equations hold:
      \begin{center}
        \diagram*{bicategorical-pentagon}\qquad\diagram*{bicategorical-triangle}
      \end{center}
  \end{itemize}
  \refstepcounter{junk}

\subsection{Monoidal actions from bicategories}
\subseclabel{end of bicat definition}
\subseclabel{actions from bicategories}
A bicategory lets us refine the tensor so that not every $a$ and $b$
can be tensored, merely composable ones. Famously, when there is only one
$0$-cell, every pair of $1$-cells is composable, recovering the notion of
a monoidal category. When we restrict attention to two $0$-cells
and ignore the identities in one of them, we obtain an actegory:

\begin{proposition@}{\propref*{actions from two 0-cells}}
  Let $\Cat B$ be a bicategory. Every two $0$-cells $\SmallCat A,
  \SmallCat B \in \Cat B$ induce an actegory $\pair{\Cat B(\SmallCat
    A, \SmallCat A)}{\Cat B(\SmallCat B, \SmallCat A)}$. Its monoidal
  category is given by the endo-$1$-cells $X : \SmallCat A \to
  \SmallCat A$ and their $2$-cells. It acts on the $1$-cells $P :
  \SmallCat B \to \SmallCat A$ and their $2$-cells through $1$-cell
  post-composition and horizontal composition.
\end{proposition@}
\begin{proof}
  By the famous result, $\catMon \definedby \seq{\Cat B(\SmallCat A,
    \SmallCat A), (\tensor), \mNeut[\SmallCat A], \assoc, \ellunit,
    \runit}$ is a monoidal category. Define its actegory by
  $\catAct \definedby \seq{\Cat B(\SmallCat B, \SmallCat A),
    (\tensor), \assoc, \runit}$. The action pentagon and triangle
  are the bicategorical ones, specialised.
\end{proof}

\subsection{Exponentiation qua right Kan lifts}
\subseclabel{exponentiation with kan lifts}
In a bicategory $\Cat B$, every $1$-cell $g : b \to c$ induces
the \emph{post-composition} functor $(-\tensor g) : \Cat B(a, b) \to
\Cat B(a, c)$. Given any $1$-cell $f : a \to c$, a \emph{right Kan
lift} of $f$ along $g$ is a universal arrow from $(-\tensor g)$ to $f$, i.e.,
a $1$-cell $(f \tfrom g) : a \to b$ and a $2$-cell $\eval : (f \tfrom g) \tensor g \To f$
that is terminal among these $2$-cells:
\[
  \begin{tikzcd}[ampersand replacement=\&]
    a \&\& b \\
      \&\& c
    \arrow["f \tfrom g", from={1-1}, to={1-3}]
    \arrow["f"'{name=0}, from={1-1}, to={2-3}]
    \arrow["g",from={1-3}, to={2-3}]
    \arrow["\eval", Rightarrow,  shorten >=5pt, shorten <=5pt, from={1-3}, to=0]
  \end{tikzcd}
  \qquad
  \forall
    \begin{tikzcd}[ampersand replacement=\&]
    a \&\& b \\
      \&\& c
    \arrow["h", from={1-1}, to={1-3}]
    \arrow["f"'{name=0}, from={1-1}, to={2-3}]
    \arrow["g",from={1-3}, to={2-3}]
    \arrow["\alpha", Rightarrow,  shorten >=5pt, shorten <=5pt, from={1-3}, to=0]
  \end{tikzcd}
  \exists!
  \begin{tikzcd}[ampersand replacement=\&]
    a \&\& b
    \arrow["h"{name = 1}, from={1-1}, to={1-3}, curve={height=-12pt}]
    \arrow["f \tfrom g"'{name = 2}, from={1-1}, to={1-3}, curve={height=12pt}]
    \arrow["\beta", Rightarrow,  shorten >=5pt, shorten <=5pt, from=1, to=2]
  \end{tikzcd}
  .\quad
  \alpha =
  \begin{tikzcd}[ampersand replacement=\&]
    a \&\& b \\\\
    \&\& c
    \arrow["h"{name=1}, from={1-1}, to={1-3}, curve={height=-12pt}]
    \arrow["f \tfrom g"'{name=2,pos=.45}, from={1-1}, to={1-3}, curve={height=12pt}]
    \arrow["f"'{name=0}, from={1-1}, to={3-3}, curve={height=12pt}]
    \arrow["g",from={1-3}, to={3-3}]
    \arrow["\eval", Rightarrow,  shorten >=5pt, shorten <=5pt, from={1-3}, to=0]
    \arrow["\beta"{pos=.25}, Rightarrow,  shorten >=5pt, shorten <=5pt, from=1, to=2]
  \end{tikzcd}
\]
A bicategory is \emph{(right)-liftal} when it has all
right Kan lifts along all $1$-cells. This terminology is not
standard. It follows Street~\cite{street:elementary-cosmoi-i}
who calls a bicategory \emph{extensional}
when it has all right Kan extensions.

The bicategory of profunctors $\Prof \definedby \seq{\Category{Cat},
  (\rto), (\To), (\odot), \terminal, \assoc, \ellunit, \runit}$ has as
$0$-cells the small categories $\SmallCat A, \SmallCat B, \SmallCat C,
\ldots$. As $1$-cells $P : \SmallCat A \rto \SmallCat B$ the
\emph{profunctors}, i.e., presheaves $P \in \PSh+{\opposite{\SmallCat
    A} \times \SmallCat B}$, i.e., functors $P : \SmallCat A \times
\opposite{\SmallCat B} \to \Set$. Its $2$-cells $\alpha : P \To Q :
\SmallCat A \rto \SmallCat B$ are natural transformations $\alpha : P
\to Q$. We compose the $1$-cells $\SmallCat A \overset P\rto \SmallCat
B \overset Q\rto \SmallCat C$ through the coend formula $(P \odot
Q)_ac \definedby \int^{b \in \SmallCat B} P_ab\times Q_bc$. The
mediators reassociate brackets and project the left/right component.
The bicategory $\Prof$ is well-known to be liftal and extensional.
\figref{right-kan-lifts-def-of-bimodules} presents the right Kan lifts
explicitly.  \newcommand\PlaceLift[4]{
    \begin{array}[t]{@{}l@{}}
    #1:\\\\
    {{\begin{tikzcd}[ampersand replacement=\&]
        #2
      \end{tikzcd}}
    }
    \end{array}
    &
    \begin{array}[t]{@{}c@{}}
      #3
    \\\\
    {{\begin{tikzcd}[ampersand replacement=\&]
        #4
      \end{tikzcd}}
    }
    \end{array}
}
\begin{figure}
  \[
  \begin{array}{@{}l@{\quad}l@{}}
  \PlaceLift\Prof{
        \SC[1] \&\& \SC[2] \\
        \&\& \SC[3]
        \rtoArrow{}{P \dfrom Q}{}{1-1}{1-3}
        \rtoArrow'{0}{P}{}{1-1}{2-3}
        \rtoArrow{}{Q}{}{1-3}{2-3}
        \arrow["\eval"{name=a}, Rightarrow,  shorten >=5pt, shorten <=5pt, from={1-3}, to=0]
  }{
    (P \dfrom Q)_ab \definedby
    \int_{c \in \SC[3]}(P_ac)^{Q_bc}
  }{
        {\parent{\int_{c'\in\SC[3]} (P_ac')^{Q_bc'}}\times Q_bc}
                                   \&\& ((P \dfrom Q)\odot Q)_ac \\
        {(P_ac)^{Q_bc}\times Q_bc} \&\& {Q_ac}
        \ToArrow{toplab}{[-]_b}{}{1-1}{1-3}
        \ToArrow{}{\eval}{}{1-3}{2-3}
        \ToArrow'{}{\projection_c\times \id}{}{1-1}{2-1}
        \ToArrow'{botlab}{\eval}{}{2-1}{2-3}
        \CommuteDefBy{toplab}{botlab}
  }
  \end{array}
  \]
  \caption{Right Kan lifts in the bicategory of profunctors.}
  \figlabel{right-kan-lifts-def-of-bimodules}
\end{figure}
We derive the right Kan lifts for the Kleisli bicategory from the
lifts of profunctors.  For
completeness's sake, we prove this well-known result.
\begin{theorem}
  The Kleisli bicategory
  $\RenCat\Prof$ is liftal.
  \figref{right-kan-lifts-def} describes its
  right Kan lifts.
\end{theorem}
\begin{figure}
  \[
  \begin{array}{@{}l@{\quad}l@{}}
  \PlaceLift{\RenCat\Prof}{
        \SC[1] \&\& \SC[2] \\
        \&\& \SC[3]
        \rToArrow{}{P \tfrom Q}{}{1-1}{1-3}
        \rToArrow'{0}{P}{}{1-1}{2-3}
        \rToArrow{}{Q}{}{1-3}{2-3}
        \arrow["\eval"{name=a}, Rightarrow,  shorten >=5pt, shorten <=5pt, from={1-3}, to=0]
  }{
    {(P \tfrom Q)} \definedby (P \dfrom \Env Q{})
    \quad\text{i.e.: }
    {(P \tfrom Q)_a\Context} \definedby
    \int_{\Context[2] \in \RenCat{\SC[3]} }(P_a\Context[2])^{\Env Q{\Context[2]} \Context}
  }{
        {\parent{\int_{\Context[3]\in\RenCat{\SC[3]}} (
           P_a\Context[3])^{\Env Q{\Context[2]}\Context[3]}}\times \Env Q{\Context[2]}\Context}
                                   \&\& ((P \tfrom Q)\tensor Q)_a\Context \\
        {{(P_a\Context[1])^{\Env Q{\Context[2]}\Context[1]}}\times \Env Q{\Context[2]}\Context} \&\& {Q_a\Context[1]}
        \ToArrow{toplab}{[-]_{\Context[2]}}{}{1-1}{1-3}
        \ToArrow{}{\eval}{}{1-3}{2-3}
        \ToArrow'{}{\projection_{\Context[1]}\times \id}{}{1-1}{2-1}
        \ToArrow'{botlab}{\eval}{}{2-1}{2-3}
        \CommuteDefBy{toplab}{botlab}
  }
  \end{array}
  \]
  \caption{Deriving right Kan lifts in the Kleisli bicategory
    of profunctors..}
  \figlabel{right-kan-lifts-def}
\end{figure}
\begin{proof}
  To define the situation as on the
  left, use the situation on the right:
  \[
  \begin{tikzcd}[ampersand replacement=\&]
        \SC[1] \&\& \SC[2] \\
        \&\& \SC[3]
        \rToArrow{}{P \tfrom Q}{}{1-1}{1-3}
        \rToArrow'{0}{P}{}{1-1}{2-3}
        \rToArrow{}{Q}{}{1-3}{2-3}
        \arrow["\eval"{name=a}, Rightarrow,  shorten >=5pt, shorten <=5pt, from={1-3}, to=0]
  \end{tikzcd}
  \qquad\text{take:}
  \begin{tikzcd}[ampersand replacement=\&]
        \SC[1] \&\& \RenCat{\SC[2]} \\
        \&\& \RenCat{\SC[3]}
        \rtoArrow{}{P \dfrom \Env Q{}}{}{1-1}{1-3}
        \rtoArrow'{0}{P}{}{1-1}{2-3}
        \rtoArrow{}{\Env Q{}}{}{1-3}{2-3}
        \arrow["\eval"{name=a}, Rightarrow,  shorten >=5pt, shorten <=5pt, from={1-3}, to=0]
  \end{tikzcd}
  \]
  To show that this lift is universal, consider any other lift:
  \[
  \begin{tikzcd}[ampersand replacement=\&]
        \SC[1] \&\& \SC[2] \\
        \&\& \SC[3]
        \rToArrow{}{L}{}{1-1}{1-3}
        \rToArrow'{0}{P}{}{1-1}{2-3}
        \rToArrow{}{Q}{}{1-3}{2-3}
        \arrow["\alpha"{name=a}, Rightarrow,  shorten >=5pt, shorten <=5pt, from={1-3}, to=0]
  \end{tikzcd}
  \text{ i.e.:}
  \begin{tikzcd}[ampersand replacement=\&]
        \SC[1] \&\& \RenCat{\SC[2]} \\
        \&\& \RenCat{\SC[3]}
        \rtoArrow{}{L}{}{1-1}{1-3}
        \rtoArrow'{0}{P}{}{1-1}{2-3}
        \rtoArrow{}{\Env Q{}}{}{1-3}{2-3}
        \arrow["\alpha"{name=a}, Rightarrow,  shorten >=5pt, shorten <=5pt, from={1-3}, to=0]
  \end{tikzcd}
  \implies \exists!
  \begin{tikzcd}[ampersand replacement=\&]
        \SC[1] \& \RenCat{\SC[2]}
        \rtoArrow{ 0}{L}{height=-12pt}{1-1}{1-2}
        \rtoArrow'{1}{P\tfrom Q}{height=+12pt}{1-1}{1-2}
        \arrow["\beta"{name=a}, Rightarrow,  shorten >=5pt, shorten <=5pt, from={0}, to=1]
  \end{tikzcd}
  .\alpha =
  \begin{tikzcd}[ampersand replacement=\&]
        \SC[1] \&\& \RenCat{\SC[2]} \\\\
        \&\& \RenCat{\SC[3]}
        \rtoArrow{0}{L}{height=-12pt}{1-1}{1-3}
        \rtoArrow'{1}{P\dfrom Q}{height= 12pt}{1-1}{1-3}
        \rtoArrow'{2}{P}{height=+12pt}{1-1}{3-3}
        \rtoArrow{}{\Env Q{}}{}{1-3}{3-3}
        \arrow["\beta"{name=a}, Rightarrow,  shorten >=5pt, shorten <=5pt, from={0}, to=1]
        \arrow["\eval"{name=a,pos=.7}, Rightarrow,  shorten >=5pt,
          shorten <=5pt, from={1-3}, to=2, curve={height=-12pt}]
  \end{tikzcd}
  \]
  For such a $\beta$ we therefore have:
  \[
  \begin{aligned}
  \alpha &= \eval \compose (\beta \odot \Env Q{}) \\
         &= \eval \compose (\beta \tensor Q)
  \end{aligned}\qquad
  \text{ i.e.: }
  \alpha =
  \begin{tikzcd}[ampersand replacement=\&]
        \SC[1] \&\& \RenCat{\SC[2]} \\\\
        \&\& \RenCat{\SC[3]}
        \rToArrow{0}{L}{height=-12pt}{1-1}{1-3}
        \rToArrow'{1}{P\tfrom Q}{height= 12pt}{1-1}{1-3}
        \rToArrow'{2}{P}{height=+12pt}{1-1}{3-3}
        \rToArrow{}{\Env Q{}}{}{1-3}{3-3}
        \arrow["\beta"{name=a}, Rightarrow,  shorten >=5pt, shorten <=5pt, from={0}, to=1]
        \arrow["\eval"{name=a,pos=.7}, Rightarrow,  shorten >=5pt,
          shorten <=5pt, from={1-3}, to=2, curve={height=-12pt}]
  \end{tikzcd}
  \]
  and so $\pair{P \tfrom Q}{\eval}$ is a right Kan lift in $\RenCat\Prof$.
  The formula in \figref{right-kan-lifts-def} spells this definition out.
\end{proof}

Unfolding the definition of right-closed monoidal categories and
closed actions, we deduce \propref{heterogeneous structures
  right-closed}.

A similar argument shows that the pointed action inherits any closed structure:
\begin{proposition}
  Let $\catAct$ be a closed $\catMon$-actegory. Then the actegory
  $\PointedAct\catAct$ is closed: for every object $x \in
  \carrier\catAct$ and pointed object $a \in \Pointed\catMon$, the
  right exponential is given by:
  \[
  (x \pfrom A) \definedby x \tfrom \carrier A
  \qquad
  \eval : (x\pfrom A) \pract A
  = (x\tfrom \carrier A) \ract \carrier A
  \xto \eval x
  \] The currying bijection is given by
  \(
  \curry \parent{y \pract A \xto g x} \definedby
  \curry \parent{y \ract \carrier A = y \pract A \xto g x}
  \).
\end{proposition}

%% file: monoidal-cat.tex
\section{Monoidal categories}
\applabel{skew monoidal cats}
We prove some technical results concerning monoidal categories and
actegories: the well-definedness of pointed tensors, results for
deriving strengths for functors, and medley technical results
concerning monoids and actions. We then recall the full definition of
skew monoidal categories and show how an actegory gives rise to a skew
monoidal category.

\subsection{Pointed tensors}
\subseclabel{monoidal proofs: pointed tensors}
For completeness, we include the calculations required by Fiore's
\propref{pointed tensor} and \exref{pointed action}:
\begin{proposition@}{\propref*{pointed tensor} and \exref{pointed action}}[Fiore~\cite{fiore:soas}]
  Every monoidal category $\catMon$ yields a
  monoidal category of pointed objects
  \(
  \Pointed\catMon \definedby \seq{\Pointed{\carrier\catMon}, (\ptensor), \pinitial[\mNeut], \assoc, \ellunit, \runit}
  \).
  I.e., the mediators preserve points, hence lift to
  $\Pointed\catMon$.
  Let $\catAct$ be a $\catMon$-actegory. Define the \emph{pointed action}
  $\Pointed\catMon$-actegory $\PointedAct\catAct =
  \seq{\carrier\catAct, (\pract), \assoc, \runit}$ by \( (a
  \mathbin{\PointedAct\ract} A) \definedby (a \ract \carrier A) \).
\end{proposition@}
As a consequence, the pointed monoidal category $\Pointed\catMon$ acts
on $\catMon$ as $\PointedAct\catMon$.
\begin{proof}
  The functorial action of the pointed tensor $(\ptensor)$ from
\subsecref{signature functors}, and the initial map out of $\pinitial$
are well-defined by the calculations in \figref{pointed tensor well-defined}.
\begin{figure}
\begin{center}
  \diagram*{well-defined-pointed-tensor}
  \qquad
  \diagram*{well-defined-pinitial-map}
\end{center}
\caption{Well-definedness of the pointed monoidal structure}
\figlabel{pointed tensor well-defined}
\end{figure}
  \figref{pointed lifting skew structure} shows the mediators lift
  to the category of pointed objects.
  \begin{figure}
  \[
  \diagram*{pointed-lift-left-unitor}
  \quad
  \diagram*{pointed-lift-right-unitor}
  \quad
  \diagram*{pointed-lift-associator}
  \]
  \caption{Lifting the mediators from a monoidal category to its pointed objects}
  \figlabel{pointed lifting skew structure}
  \end{figure}
  Consider the functor $\carrier- : \Pointed{\carrier{\Cat C}} \to
  \carrier{\Cat C}$.  An invertible pointed arrow is a pointed
  isomorphism, i.e., this functor reflects isomorphisms.  Moreover, by
  appealing to faithfulness, the monoidal axioms for $\Pointed{\Cat
    C}$ follow from those of $\Cat C$.
  The $\Pointed\catMon$-actegory axioms follow from those of
  $\PointedAct\catAct$.
\end{proof}

\subsection{Strong functors}
\subseclabel{pointed action from mere action}
In \subsecref{signature functors} we claimed \lemmaref{pointed
  tensorial strength} has a straightforward proof:
\begin{lemma}
  Let $\Cat C$ be a skew monoidal category; $\Cat A$,$\Cat B$ two
  $\Cat C$-actions; and $F : \Cat A \to \Cat B$ a strong functor. Then
  the following exhibits $\carrier F$ as a strong functor:
  $\PointedAct F : \PointedAct{\Cat A} \to \PointedAct{\Cat B}$.
  \[
  \pstrength F_{x,a} : (\carrier F x)\PointedAct\ract a
  = (\carrier F x)\ract \carrier a
  \mathrel{\smash{\xto{\strength[F]_{x,\carrier a}}}} \carrier F(x \ract \carrier a)
  = \carrier F(x \PointedAct\ract a)
  \]
\end{lemma}
\begin{proof}
  Consider a strong functor $F : \Cat A \to \Cat B$ and define a
  putative $\pstrength F$ as in the lemma statement.
  For naturality, take any $f : x \to x$ in $\Cat
  C$ and $g : a \to b$ in
  $\Pointed{\Cat C}$, and calculate as in \figref{pointed strength naturality}.
  \begin{figure}
    \diagram{pointed-strength-naturality}
    \caption{Naturality of the pointed strength}
    \figlabel{pointed strength naturality}
  \end{figure}
  For the strength pentagon and rectangle, calculate as in \figref{pointed strength axioms}.
  \begin{figure}
  \diagram{pointed-strength-pentagon}
  \diagram{pointed-strength-triangle}
  \caption{Strength axioms for the pointed strength}
  \figlabel{pointed strength axioms}
  \end{figure}
\end{proof}

The fact that strong functors compose is well established in some form of
another~\cite[e.g.]{eilenberg-kelly:closed-categories}:
\begin{lemma}[well-known]
\lemmalabel{strengths compose} Let $\Cat{A}_1 \xto F \Cat{A}_2 \xto
G \Cat{A}_3$ be $\catMon$-strong functors. Then composing the strengths
yields a strong functor $G \compose F : \Cat{A}_1 \to \Cat{A}_3$:
\[
\strength[G \compose F]_{x,a} : (G F x) \ract a \xto{\strength[G]} G(F x \ract a) \xto{G\strength[F]} GF (x \ract a)
\]
\end{lemma}
\begin{proof}
  Straightforward calculation as in \figref{composite strength}.
  \begin{figure}
  \diagram{composite-strength-triangle}
  \diagram{composite-strength-pentagon}
  \caption{Strength axioms for composite strengths}
  \figlabel{composite strength}
  \end{figure}
\end{proof}

\subsection{Monoids and actions}
\subseclabel{coprod compatible monoid}
\subseclabel{pointed monoids and actions}
\subseclabel{lifting signature functors to monoids}

First, we deal with compatible monoids and discharge \lemmaref{coprod compatible monoid}:
\begin{lemma@}{\lemmaref*{coprod compatible monoid}}
  Let $\Monoid$ be a substitution structure and
  $\seq[i \in I]{\Sig_i}$ and $\seq[i \in
    I]{\sem-_i : \carrier{\Sig_i\Monoid} \to \carrier\Monoid}$
  be families of $\gsort$-signature
  functors and algebras for them. The cotupled algebra:
  \(
  \coseq[i \in I]{\sem-_i} : \coprod_{i \in I}\carrier{\Sig_i\Monoid} \to \carrier\Monoid
  \)
  is compatible with $\Monoid$ iff every algebra $\Sig_i$ is
  compatible with $\Monoid$.
\end{lemma@}
\begin{proof}
  For each $j \in I$, consider the diagram in \figref{compatible monoid coprod}.
  \begin{figure}
    \diagram{compatible-monoid-coprod}
    \caption{Compatibility for coproduct signature functors}
    \figlabel{compatible monoid coprod}
  \end{figure}
  The outer face is the compatibility condition for the cotupled
  algebra, and the inner face is the compatibility condition for the
  $j$-th algebra. Since $\Syn\gsort$ is distributive, the outer face
  commutes iff all inner faces commute, as we wanted.
\end{proof}

Next, we show that substitution monoids and pointed substitution
monoids coincide:
\begin{proposition@}{\propref*{pointed monoids and actions coincide}}
Let $\catMon$ be a monoidal category. The forgetful functor $\carrier-
: \Pointed\catMon \to \catMon$ lifts to an isomorphism between their
categories of monoids:
\[
  \MonoidCat{\Pointed\catMon}
  \mathrel{
    \raisebox{-.75ex}{$\xfrom[\Pointed{-}]{\xto{\carrier-\mspace{20mu}}\mspace{-10mu}}$}
  }
  \MonoidCat\catMon
  \qquad
  \carrier{\Monoid} \definedby
  \triple{\carrier{\carrier\Monoid}}{\carrier{-[-]_{\Monoid}}}{\carrier{\var}}
  \quad
  \carrier h\definedby h
  \qquad
  \Pointed{\Monoid} \definedby
  \triple{\pair{\carrier\Monoid}{\var}}{-[-]}{\var}
  \quad
  \Pointed h \definedby h
\]
Given a $\catMon$-actegory $\catAct$, the categories
of $\Monoid$-actions in $\catAct$ and $\Pointed\Monoid$-actions in
$\PointedAct\catAct$ are also isomorphic.
\end{proposition@}
\begin{proof}
For monoids, $\carrier- : \MonoidCat{\Pointed\catMon} \to
\MonoidCat\catMon$ is well-defined because $\carrier- : \Pointed\catMon
\to \catMon$ preserves the monoidal structure strictly. The inverse
functor $\Pointed{-} : \MonoidCat{\Pointed\catMon} \from
\MonoidCat\catMon$ is well defined. Indeed, the unit lifts to a pointed
object morphism $\var : \pinitial[\mNeut] \to
\pair{\carrier\Monoid}\var = \carrier{\Pointed\Monoid}$ by fiat.
Multiplication is a pointed map:
\diagram{monoid-multiplication-lifts} and faithfulness of $\carrier{-}
: \Pointed\catMon \to \catMon$ derives the monoid axioms and that the
functorial action is well-defined. Most of the proof these functors
are each other's inverses is immediate. The only twist is to show that
$\Monoid = \Pointed{(\carrier\Monoid)}$ for $\Monoid \in
\MonoidCat{\Pointed\catMon}$. The fact that the unit is a pointed
object morphism $\var[\Monoid] : \pair{\mNeut}{\id} =
\pinitial[\mNeut] \to \carrier{\Pointed\Monoid}$ ensures
$\var[\Monoid] = \var[\carrier\Monoid]$, i.e., the pointed structure
of a pointed monoid must be its unit.
Turning to actions, the same data
$\pair{\carrier\Action}{-[-]}$ is an $\Monoid$-action iff it is an
$\Pointed\Monoid$-action, and similarly for their homomorphisms.
\end{proof}

We conclude this treatment by showing how a pointed-strong functor
lets us lift actions:
\begin{proposition@}{\propref*{lifting signatures to monoids}}
  Let $\catMon$ be a monoidal category, $\catAct$ be
  $\catMon$-actegory, and $\Sig : \PointedAct\catAct \to \PointedAct\catAct$ a
  pointed-strong functor. For every $\catMon$-monoid, $\Sig$ lifts to the following
  functor over
  $\Monoid$-actions:\[
  \Sig_{\Monoid} : \ActionCat\Monoid\catAct \to \ActionCat\Monoid\catAct
  \quad
  \Sig_{\Monoid}\Action \definedby
  \carrier{\Sig_{\Monoid}\Action} \definedby \carrier{\Sig\Action}
  \quad
  -[-]_{\Sig_{\Monoid}\Action} : \carrier{\Sig\Action} \tensor \Monoid
  \xto{\strength_{\carrier\Action,\var}}
  \carrier{\Sig}\parent{\carrier\Action \tensor \carrier\Monoid}
  \xto{-[-]_{\Action}}
  \carrier\Action
  \]
\end{proposition@}
\begin{proof}
  \Figref{action lifts pentagon} proves the action pentagon axiom, and
  \Figref{action lifts triangle} the triangle axiom.
  \begin{figure}
    \diagram{action-lifts-pentagon}
    \caption{Validating the action pentagon axiom for $\Sig\Action$.}
    \figlabel{action lifts pentagon}
  \end{figure}
  \begin{figure}
    \diagram{action-lifts-triangle}
    \caption{Validating the action triangle axiom for $\Sig\Action$.}
    \figlabel{action lifts triangle}
  \end{figure}
\end{proof}

\subsection{Skew monoidal categories}
\subseclabel{skew monoidal def}
\subseclabel{skew monoidal proofs}
Recall that a \emph{left-skew monoidal
category}~\cite{szlachanyi:skew-monoidal-categories-and-bialgebroids}
$\seq{\Cat C, (\tensor), \Neut, \assoc, \ellunit, \runit'}$ consists
of a category $\Cat C$ equipped with a functor $(\tensor) : \Cat C
\times \Cat C \to \Cat C$, an object $\mNeut \in \Cat C$, and three
transformations called the \emph{associator}, natural in $a,b,c \in
\Cat C$ and the \emph{left/right unitors}, natural in $a \in \Cat C$:
\[
\assoc : (a \tensor b) \tensor c \to
a \tensor (b \tensor c)
\qquad
\ellunit : \Neut \tensor a \to a
\qquad
\runit' : a \to a \tensor \Neut
\]
satisfying the equations in \figref{skew monoidal cat}.
\begin{figure}
\begin{center}
  \begin{tabular}{@{}cc@{}}
    \begin{tabular}{@{}c@{}}
      \diagram*{skew-monoidal-pentagon}\\~\\
      \diagram*{skew-monoidal-rectangle}
    \end{tabular}
    &
    \begin{tabular}{@{}c@{}}
      \diagram*{skew-monoidal-left}\\~\\
      \diagram*{skew-monoidal-right}\\~\\
      \diagram*{skew-monoidal-triangle}
    \end{tabular}
  \end{tabular}
\end{center}
\caption{Axioms for a skew monoidal category}
\figlabel{skew monoidal cat}
\end{figure}
The two left axioms generalise the monoidal pentagon and triangle
axioms. We say that a skew monoidal category is \emph{associative}
when the associator is invertible, and \emph{right-unital} when the
right unitor is invertible.

\begin{proposition@}{\propref*{skew from action}}
Let $\pair{\catMon}{\catAct}$ be an actegory. If $\catAct$ has an
initial object $\initial$ and the action $(\ract) : \catAct \times
\catMon \to \catAct$ preserves it, then $\catSkew \definedby
\seq{\carrier\catMon\times\carrier\catAct, (\bmul), \kNeut, \assoc, \runit', \ellunit}$ is
an associative right-unital skew monoidal category, where:
\[
(\bmul) : \carrier \catSkew \times \carrier \catSkew \to \carrier\catSkew
\quad
\pair ax \bmul \pair by \definedby \pair{a \tensor b}{x \ract b}
\quad
\kNeut \definedby \pair{\mNeut}{\initial}
\quad
\assoc \definedby \pair\assoc\assoc
\quad
\runit' \definedby \pair{\inv\runit}{\inv\runit}
\quad
\ellunit \definedby \pair\ellunit{\coseq{}}
\]
We have a carrier-preserving isomorphism between the categories of
$\pair{\catMon}{\catAct}$-actions and $\catSkew$-monoids. Moreover, if
$\pair{\catMon}{\catAct}$ are (right-)closed, then $\catSkew$ is
right-closed.
\end{proposition@}
\begin{proof}
  We need to validate both the $\catMon$-component and the
  $\catAct$-component for each of the $5$ diagrams. We explain how to
  discharge the $10$ diagrams systematically.

  For the $\catMon$-components, the projection functor
  $\projection_1 : \carrier\catSkew \to \carrier\catMon$ preserves the
  monoidal structure strictly:
  \[
  \begin{array}{c@{\qquad}c@{\qquad}c}
  \diagram*{mon-projection-preserves-associator}
  &
  \diagram*{mon-projection-preserves-lft-unitor}
  &
  \diagram*{mon-projection-preserves-rgt-unitor}
  \\
  (p = (a,x), q = (b,y), r = (c,z))
  &
  \multicolumn{2}{c}{(p = (a,x))}
  \end{array}
  \]
  and so the $\catMon$-component of each diagram reduces by pasting
  these squares to the corresponding diagram for $\catMon$, which
  commute by coherence.

  For the $\catAct$-components, the monoidal pentagon reduces to the
  action pentagon and the monoidal rectangle reduces to the action
  triangle, similarly. The skew triangle reduces to a diagram whose
  source is the initial object $\initial$, and so commutes. The skew
  left diagram reduces, by the assumed preservation of the initial
  object, to a diagram whose source is $(\initial\ract a)\ract b
  \isomorphic \initial \ract b \isomorphic \initial$, and so commutes
  as well.

  The $\catAct$-component of the skew right diagram reduces to the
  diagram on the left ($*$). Since the right unitor $\runit$ is
  invertible, it suffices to prove the diagram on the right ($**$):
  \[
  \diagram*{skew-mon-goal} \impliedby
  \diagram*{skew-mon-goal-modified}
  \]
  The diagram $(**)$ commutes by the following argument due to
  Kelly~\cite[Thm~7]{kelly:on-maclanes-conditions-for-coherence}. Calculate:
  \diagram{kellys-pentagon}
  The outer face is the action pentagon axiom, and so the diagram
  marked with (I) commutes. Calculate:
  \diagram{skew-mon-goal-proof}
  proving diagram ($**$).
\end{proof}

%% file: representation-appendix.tex
\section{Proving the representation theorem}
\applabel{representation}
We dedicate this section to proving the General Representation
\thmref{general representation}. The proof becomes relatively
straightforward once we develop the appropriate technique to define and
reason about parameterised homomorphisms $\AST \tensor A \to x$.
Szamozvancev~\cite[Sec.~2.2]{szamozvancev:thesis} describes this
technique. In \subsecref{parameterised initiality}, we generalise this
technique to the actegorical setting.  In \subsecref{abstract syntax},
we define the compatible action of abstract syntax by defining
capture-avoiding substitution generically and establishing its
properties.  We conclude, in \subsecref{universality}, by showing the
abstract syntax is universal among all compatible monoids. These prove
the desired theorem:

\begin{theorem@}{\thmref*{general representation}}[general representation]
  Let $\pair{\catMon}{\catAct}$ be a closed actegory with finite coproducts.
  Letting $\catSkew \definedby \catMon\times\catAct$ be the product $\catMon$-actegory
  and $\kNeut \definedby \pair\mNeut\initial \in \catSkew$, take any pointed-strong functor  $\Sig : \PointedAct{\catSkew} \to
  \PointedAct{\catSkew}$,  object $\ghole \in
  \catSkew$, and initial algebra structure
  $\seq{\sem- : \carrier\Sig \AST \to \AST, \var : \mNeut \to \AST_1, \meta-[-] : \ghole\tensor \AST_1 \to \AST}$
  over the object
  $\AST = \pair-{\AST_1}{\AST_2} \definedby \fix
  \pair-{x_1}{x_2}. (\carrier\Sig\pair-{x_1}{x_2}%
  \amalg \kNeut \amalg (\ghole \rmul x_1)$.
  There is a unique $\catSkew$-morphism, called
  \emph{simultaneous substitution},
  $-[-] : \AST \tensor \AST_1 \to \AST$
  satisfying analogous equations to
  \thmref{representation}.  The free $\Sig$-action over
  $\ghole$ is then $\Free[\Sig]\ghole \definedby
  \seq{\AST, -[-], \var, \sem-}$ equipped with the arrow $? : \ghole
  \xto{\runit'} \ghole\rmul\mNeut \xto{\id\tensor\var} \ghole\tensor
  \AST_1 \xto{\meta-[-]} \AST$.
\end{theorem@}

In the remainder of the section, take $\pair{\catMon}{\catAct}$,
$\catSkew$, $\kNeut$, $\Sig$, $\ghole$, $\sem-$, $\AST =
\pair-{\AST_1}{\AST_2}$, $\var$, and $\meta-[-]$ as in \thmref{general representation}.

\subsection{Parameterised initiality and its induction principle}
\subseclabel{parameterised initiality}
A pointed $\gsort$-structure $A \in \Pointed{\Syn\gsort}$ represents a
presheaf whose sets may contain variables. We will be traversing the
object $\AST$, representing the abstract syntax,
manipulating environments over pointed objects. Defining and reasoning
these traversals becomes straightforward once we axiomatise their
properties.
\begin{lemma}\lemmalabel{parameterised initiality}
  Let $F : \PointedAct\catSkew \to \PointedAct\catSkew$ be a pointed-strong functor,
  and $\pair{\sem- : \carrier F\gast \to \gast}{\var : \Neut \to \gast_1}$ be an initial algebra
  structure over
  $\gast = \pair{\gast_1}{\gast_2} \definedby \fix (\carrier F \amalg \kNeut)$.
  For all pointed object $A \in \carrier{\Pointed\catMon}$,
  object $x \in \carrier\catSkew$, and morphisms:
  \[
  f : \carrier F x \to x
  \qquad
  e : \carrier A \to x \qquad
  \]
  there is a unique morphism $\fold_F f\,e : \gast \tensor
  \carrier A \to x$ satisfying:
  \[
  \diagram*{general-var-preservation}
  \qquad
  \diagram*{general-ops-preservation}
  \]
\end{lemma}
\begin{proof}
  Note that the exponential in $\PointedAct\catSkew$
  is given by:
  \(
  (x \pfrom A) = \pair{x_1 \pfrom A}{x_2 \pfrom A}
  \) and
  \[
  \parent{(x \pfrom A)\paction A \xto{\eval_{x,A}} x}
  = \pair{
    \parent{(x_1 \pfrom A)\paction A \xto{\eval_{x,A}} x_1}
  }{
    \parent{(x_2 \pfrom A)\paction A \xto{\eval_{x,A}} x_2}
  }
  \]
  Define $g_{\mNeut} : \mNeut \to (x_1 \pfrom A)$
  and $g_{\ops} : \carrier F(x \pfrom A) \to x \pfrom A$
  as the unique
  morphisms satisfying:
  \[
  \diagram*{general-param-initiality-var-map-def}
  \qquad
  \diagram*{general-param-initiality-ops-map-def}
  \]
  By initiality of $\gast = \fix(F \amalg \kNeut)$,
  there is a unique morphism $h = \pair{h_1}{h_2} : \gast \to (x
  \pfrom A)$ satisfying:
  \[
  \diagram*{general-param-initiality-var-case}
  \qquad
  \diagram*{general-param-initiality-ops-case}
  \]
  Define $\fold_F f\,e \definedby \uncurry\, h : \gast \paction A \xto{h \paction \id}
  (x \pfrom A)\paction A \xto\eval x$.
  It satisfies the two required properties:
  \[
  \diagram*{general-var-preservation-proof}
  \qquad
  \diagram*{general-ops-preservation-proof}
  \]
  To show it is the unique such morphism, take any other $\phi =
  \pair{\phi_1}{\phi_2} : \gast \paction A \to x$ satisfying:
  \[
  \diagram*{general-var-preservation-unique}
  \qquad
  \diagram*{general-ops-preservation-unique}
  \]
  Take $h' = \pair{h'_1}{h'_2} : \gast \xto{\curry \phi} (x\pfrom A)$
  noting $h'_i : \gast_i  \xto{\curry \phi_i} (x_i\pfrom A)$. Calculate:
  \[
  \diagram*{general-param-initiality-var-case-unique}
  \quad
  \diagram*{general-param-initiality-ops-case-unique}
  \]
  By the universal property of the exponentials it follows that $h'$
  also satisfies the defining equations for $h$, and so $h' =
  h$. Therefore $\fold_F f\,e = \uncurry h = \uncurry h' = \phi$.
\end{proof}

We use the the special case where $F \definedby \Sig \amalg
(\ghole\tensor-_1)$, noting that $\ghole\tensor-_1$ has $\assoc$ as
its pointed strength:
\begin{theorem}[parameterised initiality]\thmlabel{parameterised initiality}
  For all pointed object $A \in \carrier{\Pointed\catMon}$,
  object $x \in \carrier\catSkew$, and morphisms:
  \[
  f : \carrier F x \to x
  \qquad
  g : \ghole\tensor x_1 \to x
  \qquad
  e : \carrier A \to x \qquad
  \]
  there is a unique morphism $\fold f\,g\,e : \gast \tensor
  \carrier A \to x$ satisfying:
  \[
  \diagram*{foldr-var-preservation}
  \ \ %
  \diagram*{foldr-ops-preservation}
  \ \ %
  \diagram*{foldr-hole-preservation}
  \]
\end{theorem}
Parameterised initiality also provides the following induction
principle. To prove $h_1 = h_2$ for two morphisms $h_1,h_2 :
\gast\paction A \to x$, we commute the following six diagrams for some $f$,$g$,$e$.
We mark the internal faces that
we do not commute as part of the proof as the Induction Hypothesis~(IH):
\[
\diagram*{parameterised-induction-var}\quad
\diagram*{parameterised-induction-ops}\quad
\diagram*{parameterised-induction-hole}
\implies
\diagram*{parameterised-induction-conc}
\]
since then $h_1 = \fold\,f\,g\,e = h_2$.  This notation lets us retain
the diagram shape when applying the principle. We will
the Parameterised Initiality \lemmaref{parameterised initiality}
and its induction principle,
since we can treat both operations and metavariables uniformly.
The Parameterised Initiality \thmref{parameterised initiality} can deal with
cases in which they require different treatment.

\subsection{Abstract syntax}
\subseclabel{abstract syntax}
In this section we define capture avoiding substitution and show it
forms a compatible structure over $\ghole$. The three analogous to
the equations  `op case', `var case', and `metavariable case' in
\thmref{representation} specify
$-[-]_{\AST} : \AST \paction \var \to \AST$
through parameterised initiality as:
\[
-[-]_{\AST} = \fold\,\parent{\carrier\Sig \AST \xto{\sem-}\AST}
                  \,\parent{\ghole \tensor \AST_1 \xto{?-[-]}\AST}
                  \,\parent{\carrier{\var} = \AST_1 \xto{\id[\AST]} \AST_1}
\]
We let $\gSig\definedby \Sig\amalg(\ghole\tensor-_1)$ be the
pointed-strong endofunctor as in the proof of the Parameterised
Initiality \thmref{parameterised initiality}. By setting $\gsem- :
\coseq{\sem-, ?-[-]} : \gSig \AST \to \AST$, we can treat both the op
and metavariable cases uniformly. We will refer to these as the `joint
case'.

The equations `op case' and `var case' are the compatibility and
monoid left unit axioms. We need to show the action's right unit and
associativity axioms:
\[
\diagram*{subst-right}
\qquad
\diagram*{subst-assoc}
\]
These two equations amount to the syntactic substitution lemma, which
we prove by induction. \Figref{subst right} proves the right unit
axioms and \figref{subst assoc} proves associativity.  The
associativity proof, in the face marked with \AssocProofObligation,
relies on the following lemma---we need to know substitution is a
pointed morphism in order to appeal to the naturality of the pointed
strength.
\begin{lemma}
  Simultaneous substitution preserves points, i.e.,
  $-[-] : \var \to
  \var\Pointed\tensor \var$ in $\Pointed{\catMon}$.
\end{lemma}
\begin{proof}
  Calculate:
  \\\hphantom{a}\hspace{.125\textwidth}
  \raisebox{1cm}
  {\diagram*{subst-point-preserving}}
\end{proof}

Only the associativity proof makes essential use of parameterised
induction. In the right unit proof, we can
invert the right unitor, and appeal instead to non-parameterised
induction over $\AST$.
\begin{figure}
  \begin{gather*}
  \diagram*{subst-right-var-case}
  \quad
  \diagram*{subst-right-ops-case}
  \end{gather*}
  \caption{Inductive proof for the substitution monoid right unit axiom}
  \figlabel{subst right}
\end{figure}

\begin{figure}
  \begin{gather*}
  \diagram*{subst-assoc-var-case}
  \\
  \diagram*{subst-assoc-ops-case}
  \end{gather*}
  \caption{Inductive proof for the syntactic substitution lemma}
  \figlabel{subst assoc}
\end{figure}

To show $\Free[\Sig]\ghole$ is initial over $\ghole$,
define the unit for this adjunction by
\(
? : x
  \xto{\runit'} x\rmul\Neut \xto{\id\tensor\var} x\tensor
  \AST[\Sig]x \xto{\meta} \AST[\Sig]x
\).
\begin{lemma}
  We can recover $?-[-]$ from $?$ and substitution:
  \(
  ?-[-] : \ghole\tensor\AST_1 \xto{? \tensor \id} \AST \tensor \AST_1 \xto{-[-]} \AST
  \).
\end{lemma}
\begin{proof}
  Calculate as in \figref{recover-meta-from-metaenv-proof}.
  \begin{figure}
    \diagram{recover-meta-from-metaenv-proof}
    \caption{Reconstructing the structure map $?-[-]$ from the unit $?$.}
    \figlabel{recover-meta-from-metaenv-proof}
  \end{figure}
\end{proof}

\subsection{Universality among compatible monoids}
\subseclabel{universality}
To complete the proof, take any compatible structure $\Model$
equipped with an arrow $?_{\Model} : \ghole \to \carrier\Model$.
Define:
\[
h : \AST \to \carrier\Model
\qquad
h \definedby \fold
  \,\parent{\carrier{\Sig\Model} \xto{\sem-_{\Model}} \carrier\Model}
  \,\parent{?-[-]_{\Model} : x \tensor \carrier\Model
            \xto{?_{\Model}\tensor \id} \carrier\Model \tensor \carrier\Model
            \xto{-[-]_{\Model}} \carrier\Model}
  \,\var[{\Model}]
\]
I.e., $h$ is the unique arrow satisfying:
\begin{equation}\label{init-mon-def}
    \diagram*{init-mon-ops}\quad
    \diagram*{init-mon-vars}\quad
    \diagram*{init-mon-meta}
\end{equation}
We will show it is the unique substitution $\Sig$-structure
homomorphism preserving $?$. First, we express $?$:
\begin{lemma}\lemmalabel{? from ?-[-] in any substitution structure}
  We recover $?_{\Model}$ from $?-[-]_{\Model}$
  and $?-[-]_{\Model}$ satisfies an analogue to the metavariable case:
  \[
  \diagram*{recover-metaenv-from-meta}
  \qquad
  \diagram*{monoid-meta-case}
  \]
\end{lemma}
\begin{proof}
  By straightforward calculation, see \figref{recover metaenv from meta proof}.
  \begin{figure}
    \[
    \diagram*{recover-metaenv-from-meta-proof}
    \quad
    \diagram*{monoid-meta-case-proof}
    \]
    \caption{Proving \lemmaref{? from ?-[-] in any substitution structure}}
    \figlabel{recover metaenv from meta proof}
  \end{figure}
\end{proof}
Next, we show the uniqueness of $h$:
\begin{lemma}\lemmalabel{uniqueness of den sem}
  If $k : \AST \to \Model$ is an $\Sig$-structure homomorphism
  extending $?_{\Model}$ along $?_{\AST}$, then $k = h$.
\end{lemma}
\begin{proof}
  Since $k$ is an $\Sig$-homomorphism, the equation `op preservation' holds.
  Since $k_1$ is a monoid-homomorphism, the equation `var preservation' holds.
  For the equation `metavariable preservation':
  \diagram{init-mon-meta-proof}
  Thus by definition $k = h$.
\end{proof}

\begin{lemma}\lemmalabel{metaenv-extension}
  The arrow $h$ preserves the metavariable inclusion:
  \diagram{metaenv-extension}
\end{lemma}
\begin{proof}
  Calculate:
  \diagram{metaenv-extension-proof}
  as we wanted.
\end{proof}

We complete the final step in the proof:

\begin{lemma}\lemmalabel{den sem monoid homo}
  The arrow $h$ is a substitution structure homomorphism.
\end{lemma}
\begin{proof}
  By the `var preservation' equation, $h$ preserves the unit of the
  monoids. It remains to show it preserves substitution,
  amounting to the semantic substitution lemma:
  \diagram{den-sem-homo}

  Recall the pointed-strong functor $\gSig \definedby \Sig \amalg
  (\ghole \tensor -_1)$.  Thanks to the `metavariable case' equation
  of \lemmaref{? from ?-[-] in any substitution structure}, we can
  equip $\carrier\Model$ with an $\gSig$-algebra structure
  $\gsem-_{\Model} : \carrier{\gSig\Model}
  \xto{\coseq{\sem-_{\Model,}, ?-[-]_{\Model}}} \carrier \Model$. We
  then combine the compatibility axiom for $\Model$ with the
  `metavariable case' from \lemmaref{? from ?-[-] in any substitution
    structure} into one equation, and `op preservation' and
  `metavariable preservation' into the other equation:
  \[
  \diagram*{compatible-joint}
  \qquad
  \diagram*{init-mon-join}
  \]
  We now appeal to \lemmaref{parameterised initiality}, through
  the calculations in
  \figref{substitution preservation}.
  \begin{figure}
    \diagram{den-sem-homo-var-case}
    \diagram{den-sem-homo-ops-case}
    \caption{Proving the semantic substitution lemma by parametrised induction}
    \figlabel{substitution preservation}
  \end{figure}
\end{proof}